\documentstyle[graphicx,lscape,times]{mn}

\newif\ifAMStwofonts

\ifoldfss
  \ifCUPmtlplainloaded \else
    \NewTextAlphabet{textbfit} {cmbxti10} {}
    \NewTextAlphabet{textbfss} {cmssbx10} {}
    \NewMathAlphabet{mathbfit} {cmbxti10} {} 
    \NewMathAlphabet{mathbfss} {cmssbx10} {} 
  \fi
  \ifAMStwofonts
    \ifCUPmtlplainloaded \else
      \NewSymbolFont{upmath} {eurm10}
      \NewSymbolFont{AMSa} {msam10}
      \NewMathSymbol{\upi}     {0}{upmath}{19}
      \NewMathSymbol{\umu}     {0}{upmath}{16}
      \NewMathSymbol{\upartial}{0}{upmath}{40}
      \NewMathSymbol{\leqslant}{3}{AMSa}{36}
      \NewMathSymbol{\geqslant}{3}{AMSa}{3E}

      \let\leq=\leqslant \let\le=\leqslant
      \let\geq=\geqslant \let\ge=\geqslant
    \fi
  \fi
\fi 

\ifnfssone
  \newmathalphabet{\mathit}
  \addtoversion{normal}{\mathit}{cmr}{m}{it}
  \addtoversion{bold}{\mathit}{cmr}{bx}{it}
  \newmathalphabet{\mathbfit} 
  \addtoversion{normal}{\mathbfit}{cmr}{bx}{it}
  \addtoversion{bold}{\mathbfit}{cmr}{bx}{it}
  \newmathalphabet{\mathbfss} 
  \addtoversion{normal}{\mathbfss}{cmss}{bx}{n}
  \addtoversion{bold}{\mathbfss}{cmss}{bx}{n}
  \ifAMStwofonts
    \ifCUPmtlplainloaded \else
      \UseAMStwoboldmath
      \makeatletter
      \new@mathgroup\upmath@group
      \define@mathgroup\mv@normal\upmath@group{eur}{m}{n}
      \define@mathgroup\mv@bold\upmath@group{eur}{b}{n}
      \edef\UPM{\hexnumber\upmath@group}
      \new@mathgroup\amsa@group
      \define@mathgroup\mv@normal\amsa@group{msa}{m}{n}
      \define@mathgroup\mv@bold\amsa@group{msa}{m}{n}
      \edef\AMSa{\hexnumber\amsa@group}
      \makeatother
      \mathchardef\upi="0\UPM19
      \mathchardef\umu="0\UPM16
      \mathchardef\upartial="0\UPM40
      \mathchardef\leqslant="3\AMSa36
      \mathchardef\geqslant="3\AMSa3E

      \let\leq=\leqslant \let\le=\leqslant
      \let\geq=\geqslant \let\ge=\geqslant
    \fi
  \fi
\fi 

\ifnfsstwo
  \DeclareMathAlphabet{\mathbfit}{OT1}{cmr}{bx}{it}
  \SetMathAlphabet\mathbfit{bold}{OT1}{cmr}{bx}{it}
  \DeclareMathAlphabet{\mathbfss}{OT1}{cmss}{bx}{n}
  \SetMathAlphabet\mathbfss{bold}{OT1}{cmss}{bx}{n}
  \ifAMStwofonts
    \ifCUPmtlplainloaded \else
      \DeclareSymbolFont{UPM}{U}{eur}{m}{n}
      \SetSymbolFont{UPM}{bold}{U}{eur}{b}{n}
      \DeclareSymbolFont{AMSa}{U}{msa}{m}{n}
      \DeclareMathSymbol{\upi}{0}{UPM}{"19}
      \DeclareMathSymbol{\umu}{0}{UPM}{"16}
      \DeclareMathSymbol{\upartial}{0}{UPM}{"40}
      \DeclareMathSymbol{\leqslant}{3}{AMSa}{"36}
      \DeclareMathSymbol{\geqslant}{3}{AMSa}{"3E}

      \let\leq=\leqslant \let\le=\leqslant
      \let\geq=\geqslant \let\ge=\geqslant
    \fi
  \fi
\fi 

\ifCUPmtlplainloaded \else
  \ifAMStwofonts \else 
    \def\upi{\pi}
    \def\umu{\mu}
    \def\upartial{\partial}
  \fi
\fi

\title[Comparing local globular cluster systems]
  {Comparing the properties of local globular cluster systems: implications for the formation of the Galactic halo}
\author[A.~D.~Mackey \& G.~F.~Gilmore]
  {A.~D.~Mackey$^1$\thanks{E-mail: dmackey@ast.cam.ac.uk}
  and G.~F.~Gilmore$^1$\\
  $^1$Institute of Astronomy, University of Cambridge, Madingley Road,
  Cambridge CB3 0HA}
\date{Accepted --. Received --}
\pagerange{000--000}
\pubyear{2004}

\def\LaTeX{L\kern-.36em\raise.3ex\hbox{a}\kern-.15em
    T\kern-.1667em\lower.7ex\hbox{E}\kern-.125emX}

\begin{document}

\label{firstpage}

\maketitle

\begin{abstract}
We investigate the hypothesis that some fraction of the globular clusters presently observed in the 
Galactic halo formed in external dwarf galaxies. This is done by means of a detailed comparison between the 
``old halo'', ``young halo,'' and ``bulge/disk'' sub-systems defined by Zinn \shortcite{zinn:93a} and the 
globular clusters in the LMC, SMC, and Fornax and Sagittarius dwarf spheroidal galaxies. 
We first use high quality photometry from {\em Hubble Space Telescope} images to derive 
a complete set of uniform measurements of horizontal branch (HB) morphology in the external clusters. We also 
compile structural and metallicity measurements for these objects and update the database of such measurements
for the Galactic globular clusters, including new calculations of HB morphology for
$11$ objects. Using these data together with recent measurements of globular cluster kinematics and ages
we examine the characteristics of the three Galactic cluster sub-systems. Each is quite distinct in terms
of their spatial and age distributions, age-metallicity relationships, and typical orbital parameters,
although we observe some old halo clusters with ages and orbits more similar to those of young halo objects.
In addition, almost all of the Galactic globular clusters with large core radii fall into the young halo
sub-system, while the old halo and bulge/disk ensembles are characterized by compact clusters.
We demonstrate that the majority of the external globular clusters are essentially indistinguishable from the 
Galactic young halo objects in terms of HB morphology, but $\sim 20-30$ per cent of 
external clusters have HB morphologies most similar to the Galactic old halo clusters.
We further show that the external clusters have a distribution of core radii which very closely matches
that for the young halo objects. The old halo distribution of core radii can be very well represented
by a composite distribution formed from $\sim 83-85$ per cent of objects with structures typical of bulge/disk
clusters, and $\sim 15-17$ per cent of objects with structures typical of external clusters. Taken together
our results fully support the accretion hypothesis. We conclude that all $30$ young halo clusters and 
$15-17$ per cent of the old halo clusters ($10-12$ objects) are of external origin. Based on cluster number
counts, we estimate that the Galaxy may have experienced $\sim 7$ merger events with cluster-bearing 
dwarf-spheroidal-type galaxies during its lifetime, building up $\sim 45-50$ per cent of the mass of
the Galactic stellar halo. Finally, we identify a number of old halo objects which have properties 
characteristic of accreted clusters. Several of the clusters associated with the recently proposed dwarf 
galaxy in Canis Major fall into this category.
\end{abstract}


\section{Introduction}
\label{s:intro}
Understanding in detail the formation of the Galaxy has long been considered one of the great challenges
of modern astronomy. Eggen, Lynden-Bell \& Sandage \shortcite{eggen:62} observed that for high velocity 
stars, those with lower abundances possess more halo-like orbits (i.e., higher eccentricities and orbital
energies, and lower orbital angular momenta). They argued that this implied that the oldest stars were formed 
in a rapid collapse of gas towards the Galactic plane. In contrast, Searle \& Zinn \shortcite{searle:78}
examined the abundances of a sample of Galactic globular clusters, and found no correlation between
metallicity and Galactocentric distance ($R_{\rm{gc}}$) for objects more distant than $R_{\rm{gc}} = 8$ kpc. 
They argued that this fact, along with other anomalies in the colour-magnitude diagrams (CMDs) of distant
globular clusters, implied that the outer halo of the Galaxy must have been formed in a hierarchical 
fashion, via the accretion and merger of protogalactic fragments over an extended period of time.
Over the past two decades, the general consensus has formed that a composite model, consisting of a dissipative
collapse together with numerous merger events, provides the best explanation of the presently available data.

As evidenced by the paper of Searle \& Zinn \shortcite{searle:78}, the $\sim 150$ Galactic globular 
clusters are important tools for both tracing the properties of our 
Galaxy in the present epoch and piecing together its formation history. These objects are generally 
bright (at least compared to most field stars) and so can be observed to large Galactocentric radii 
($> 100$ kpc) or through regions of high extinction, plus many of their parameters (such as distances, 
abundances, and ages) are relatively simple to measure. Their kinematics can provide information about 
different regions of the Galaxy (see e.g., Armandroff \shortcite{armandroff:89}; van den Bergh 
\shortcite{vdb:93}; Zinn \shortcite{zinn:93a}; Dinescu et al. \shortcite{dinescu:99,dinescu:03}), while 
their structures allow the gravitational potential of the Galaxy, or its enclosed mass, to be inferred 
at different positions (see e.g., Odenkirchen et al. \shortcite{odenkirchen:03}; Bellazzini 
\shortcite{bellazzini:04}). 

Consideration of the many parameters relevant to the Galactic globular clusters reveals them to be a
very inhomogeneous group. It is therefore often useful to split the full sample into various different
sub-systems. One of the most successful classification schemes is that of Zinn \shortcite{zinn:93a}
(see also van den Bergh \shortcite{vdb:93} and the references therein), who defined three sub-systems 
according to cluster metallicity and horizontal branch (HB) morphology. He found that the metal-rich 
globular clusters ($[$Fe$/$H$] > -0.8$) are confined to the bulge and disk of the Galaxy, while the 
metal-poor globular clusters ($[$Fe$/$H$] < -0.8$) are generally located in the Galactic halo. These 
halo objects exhibit a clear dispersion in their HB morphology at given metallicity, with most of the 
inner halo clusters defining a tight relationship between $[$Fe$/$H$]$ and the HB index of Lee, 
Demarque \& Zinn \shortcite{lee:94}, but many of the clusters in the more remote outer regions of the 
halo possessing much redder HB morphologies for a given abundance. This phenomenon is known in the 
literature as the second parameter effect, since a second parameter in addition to metallicity is evidently 
governing HB morphology in these clusters. There is good evidence that at least some of the effect can be 
explained by taking cluster age as the second parameter, so that younger clusters have redder HB morphology at
constant $[$Fe$/$H$]$ (see e.g., Lee et al. \shortcite{lee:94}). This led Zinn \shortcite{zinn:93a} to
define two halo sub-systems with blue and red HB types at given metallicity -- the so-called ``old halo''
and ``young halo'' systems, respectively. 

Zinn \shortcite{zinn:93a} found that the Galactic globular cluster sample for which he possessed suitable
data ($87$ clusters) split relatively cleanly into these three groups. Moreover, he showed that these 
systems are quite distinct in terms of a number of other parameters -- for example, their typical ages, 
spatial distributions and mean dynamical characteristics. He argued that the observed trends suggest
that the bulge/disk clusters and the majority of the old halo objects formed during a dissipative
collapse of the type envisioned by Eggen et al. \shortcite{eggen:62} (which ultimately formed the
Galactic disk), while the young halo objects (and possibly a number of the metal poor old halo objects) 
are not native to the Galaxy at all, but rather were formed in external dwarf galaxies and later accreted 
into the Galactic halo when these parent galaxies merged with our own and were destroyed, in the
manner suggested by Searle \& Zinn \shortcite{searle:78}. Van den Bergh \shortcite{vdb:93} arrived at a 
similar conclusion via and equivalent classification scheme. We also note that the results of Zinn 
\shortcite{zinn:96}, using a slightly modified classification scheme, support this conclusion.

Since Zinn's \shortcite{zinn:93a} paper, there have been a number of developments supporting the above
scenario. Most significant is the discovery of the Sagittarius dwarf spheroidal (dSph) galaxy 
\cite{ibata:94,ibata:95}, which is deep into the process of merging with, and being disrupted by, the Galaxy.
This object has brought with it a full retinue of globular clusters, with properties similar to those
of Zinn's young halo ensemble (see e.g., Da Costa \& Armandroff \shortcite{dacosta:95}). 
More recently has come the realization that the Galactic halo exhibits a large amount of sub-structure,
visible with a number of new large surveys (e.g., SDSS and 2MASS). For example the Sagittarius dSph itself
has left a debris stream fully encircling the Galaxy (e.g., Ibata et al. \shortcite{ibata:01}; 
Mart\'{\i}nez-Delgado et al. \shortcite{md:01,md:04}; Newberg et al. \shortcite{newberg:02,newberg:03}),
while the ``Monoceros Ring'' identified by Newberg et al. \shortcite{newberg:02} and Yanny et al. 
\shortcite{yanny:03} (sometimes known as the 
Galactic anti-centre stellar structure -- GASS) is probably the tidal debris of a destroyed dwarf 
galaxy whose nucleus lies towards Canis Major \cite{martin:04}. Again, there are a number of globular
clusters possibly associated with this structure. 

Several studies have also demonstrated that the merger scenario is not inconsistent with the observed
properties of field halo stars, although a number of constraints have been placed. For example, Unavane, 
Wyse \& Gilmore \shortcite{unavane:96} examined the fraction of stars in the halo which have colours 
consistent with a metal-poor, intermediate-age population matching those typically observed in Local Group 
dwarf spheroidal galaxies. They conclude that the star counts imply an upper limit of $\sim 60$ mergers with 
low luminosity dwarf spheroidals (i.e., Carina-like objects), or $\leq 6$ mergers with more luminous 
Fornax-like objects within the last $\leq 10$ Gyr. 

Shetrone, C\^ot\'e \& Sargent \shortcite{shetrone:01}
examined abundance patterns in the low luminosity Draco, Sextans, and Ursa Minor dSph galaxies, and found
strong consistency between these three sets of measurements. Moreover, their measured $[\alpha /$Fe$]$ ratios 
in the low luminosity dwarf galaxies are $\sim 0.2$ dex lower than those of halo field stars in the same
metallicity range. This observation, along with several additional mismatches in elemental abundance ratios
between the three dSph galaxies and halo field stars of equivalent metallicity, led these authors to conclude
that the Galactic halo cannot have been assembled entirely through the accretion of low luminosity dwarf
galaxies similar to Draco, Sextans, and Ursa Minor; however, they place no constraints on the accretion of
more luminous and complex dwarf galaxies such as Fornax and Sagittarius. 

Most recently, Venn et al. 
\shortcite{venn:04} have extended this idea, compiling a large amount of data concerning the chemistries of 
stars in several different regions of the Galaxy and a number of nearby low mass dSph systems. Like Shetrone
et al. \shortcite{shetrone:01}, these authors have demonstrated that the chemical signatures of most of the 
dSph stars are distinct from the stars in each of the components of the Galaxy. This implies that no Galactic 
component can have been formed primarily via the accretion of low mass dSph systems at late epochs, although 
the authors do not rule out very early mergers with such objects before significant chemical enrichment can 
have occurred, nor mergers with larger systems like the Sagittarius dwarf or the LMC.

In addition to these discoveries, the past decade has witnessed strong development in the field
of observational technology, which has led to a large amount of new data concerning the both the
Galactic globular clusters and those in external galaxies. For example, Dinescu et al. \shortcite{dinescu:99}
have determined orbital parameters for a sample of $\sim 40$ globular clusters from a proper motion
study, while large samples of clusters have been subjected to relative dating studies (e.g., Rosenberg
et al. \shortcite{rosenberg:98}; Salaris \& Weiss \shortcite{salaris:02}). With all this in mind, it is a 
good time to re-examine Zinn's \shortcite{zinn:93a} classification, to see what new conclusions can
be drawn about the formation of the Galaxy and the properties of the Galactic globular cluster sub-systems.

As Zinn \shortcite{zinn:93a} noted, for the accretion scenario described above to be viable, the 
young halo clusters must bear at least some resemblance to the globular clusters which are observed in
the dwarf galaxies associated with the Milky Way. A number of authors have shown this to be true
for various properties of different groups of external clusters. Da Costa \shortcite{dacosta:03}
demonstrates it for cluster structures and luminosities (clusters in Local Group dwarf 
galaxies and in the outer halo tend to be less luminous and of lower central concentration than those found 
at small Galactocentric radii), ages (clusters in dwarf galaxies are often relatively young compared with 
the oldest Galactic globulars, similar to those objects in the young halo sub-system), and chemical abundance 
(a combined sample of clusters in Local Group dwarf galaxies has a distribution of abundances matching that
of the outer halo clusters). In terms of HB morphology, Zinn \shortcite{zinn:93b} showed that some of the
globular clusters in the LMC and Fornax dwarf do lie to the red of the old halo clusters on a plot of
$[$Fe$/$H$]$ versus HB-type, but not necessarily in the region occupied by the majority of young halo
clusters. Similarly, Smith, Rich \& Neill \shortcite{smith:98} place the clusters in the Fornax and Sagittarius
dwarf galaxies on an $[$Fe$/$H$]$ versus HB-type diagram, concluding that while these clusters are generally
redder at given abundance than the Galactic old halo clusters, they do not occupy the same region on the plot
as the majority of the young halo clusters. Johnson et al. \shortcite{johnson:99} look at the position
of ten LMC globular clusters on an $[$Fe$/$H$]$ versus HB-type diagram and conclude that approximately
half of these objects resemble the Galactic old halo clusters in terms of HB morphology, while the other
half again have redder HB-types. Finally, we remark that the {\it field} populations of dwarf spheroidals
generally appear to exhibit HB morphologies that are more similar to those of young halo clusters than
old halo clusters -- that is, relatively red at given metallicity (e.g., Harbeck et al. 
\shortcite{harbeck:01}).

None of these HB morphology studies has examined the complete sample of local external clusters, and the 
measurements of HB morphology are by no means uniform both in terms of original observations and reduction
and measurement procedures. However, given the presence of a large number of archival {\it Hubble Space
Telescope} observations of local external clusters, we are now in a position to undertake a uniform
study of these objects. It is the aim of the present paper to construct a uniform set of measurements
of HB morphology for the local external globular clusters, and directly compare these with the old and
young halo clusters to assess the viability of the accretion hypothesis. In addition, while both
Da Costa \shortcite{dacosta:03} and Mackey \& Gilmore \shortcite{nhgc:03} examine the typical structures
of these external clusters, with the recent high quality radial brightness profiles of Mackey \& Gilmore 
\shortcite{sbp1,sbp2,sbp3} now available for a large number of local external clusters, we are in a position
to compare thoroughly the typical structures of external clusters with those belonging to the Galaxy.

Let us briefly review the globular cluster systems of the dwarf galaxies associated with the Milky Way.
Forbes et al. \shortcite{forbes:00} list $13$ dwarf companions to the Milky Way. Of these, only the
four most massive -- the LMC, SMC, and Fornax and Sagittarius dwarf spheroidals -- contain globular cluster
systems. The largest of these four galaxies is the LMC, at a distance of $\sim 50$ kpc, followed by the
SMC at $\sim 60$ kpc. Both these galaxies contain extensive systems of massive clusters, of ages ranging 
from the very newly formed (i.e., a few Myr old) to essentially coeval with the oldest Galactic globulars. 
The LMC cluster system is interesting in that it exhibits an age gap covering $\sim 4-13$ Gyr (where the
upper edge is set by the assumed age of the globular clusters coeval with the Galactic globulars).
Only one cluster -- the intriguing ESO121-SC03 (with an age of $\sim 9$ Gyr) -- lies in the age gap. 
The ensemble of very old clusters in this galaxy numbers $15$, with the most recent additions being
NGC 1928 and 1939, which Mackey \& Gilmore \shortcite{acs1} have shown to be coeval with the oldest
globular clusters in both the LMC and the Galaxy. The SMC contains fewer very old clusters (in fact, only
one cluster, NGC 121, which might be considered coeval with the majority of the Galactic halo objects)
but does possess a number of massive clusters with ages of $\sim 6-9$ Gyr (so matching the youngest
globular clusters observed in the Galactic system). 

The Fornax dwarf spheroidal, at a distance of $\sim 140$ kpc, contains $5$ {\it bona fide} globular clusters, 
all of which are relatively well studied. It is more difficult to define the cluster system of the (almost
disrupted) Sagittarius dwarf, because of the strong possibility that is has already lost some objects 
into the Galactic halo. The four well established members of this galaxy are M54, Terzan 7 and 8, and 
Arp 2 (e.g., Da Costa \& Armandroff \shortcite{dacosta:95}), although numerous additional previous members 
have been postulated. Even so, only two globular clusters have been observed so far to be surrounded
by Sagittarius debris -- these are Pal. 12 \cite{md:02} and NGC 4147 \cite{bellazzini:03a}. Dinescu et al.
\shortcite{dinescu:00} have measured a proper motion for Pal. 12 and find its kinematics also to be 
consistent with capture from the Sagittarius dwarf. For the moment, we set this as the cluster system
associated with this galaxy; however we note that it is likely in future that tidal debris will be found
around more globular clusters, therefore unambiguously associating them with Sagittarius.

We also briefly remark on the recent discovery of a probable disrupted dwarf galaxy in Canis Major
\cite{martin:04}. At least four globular clusters, plus a number of old open clusters, have been hypothesized
as previous members of this system \cite{martin:04,bellazzini:03b,frinchaboy:04}. We do not include this 
galaxy in the present external sample for a number of reasons. First, the reality of the detection still
needs to be proven unambiguously. A number of authors have postulated explanations for the observed stellar 
over-density other than a disrupted dwarf galaxy -- for example, Momany et al. \shortcite{momany:04}
assign the over-density to the warp and flare in the outer Galactic disk. Second, the number and identity
of the globular clusters associated with this over-density is certainly not yet clear enough to assemble
a complete sample, as we have done with the other four external galaxies. Finally, if the observed 
over-density {\it is} associated with a disrupted dwarf galaxy, then it would seem that this object
has already donated its globular cluster system to the Galactic halo. It is therefore most
sensible (at least at this stage) to consider these objects as members of one of the Galactic sub-systems.

We proceed as follows. In the next Section we present observations and data reduction procedures for the 
calculation of HB morphologies for the external clusters. We also compile metallicity and structural
measurements for this sample, since these are vital to the subsequent discussion. In Section \ref{s:ggc},
we assemble similar data for the Galactic globular clusters, with the aim of compiling as complete a
sample as possible. This includes the derivation of HB morphologies for $11$ clusters with no previous
measurements. With this complete, we classify all the clusters into systems defined similarly to those
of Zinn \shortcite{zinn:93a} and examine the properties of each, in particular taking into account recent
data concerning the ages and kinematics of significant samples of Galactic globular clusters.
In Section \ref{s:comparison}, we compare the properties of the external clusters with those of the
Galactic sub-systems -- specifically concentrating on HB morphology and cluster structure. We consider a
thought experiment in which the four external galaxies have been completely absorbed by the Galaxy
in merger events, and we try to answer the question as to what type of clusters would now reside in the
Galactic halo. Finally, in Section \ref{s:discussion}, we consider the implications of our results for 
the merger scenario. 

\section{New Measurements of Horizontal Branch Morphology in External Clusters}
\label{s:hbm}
Recently, Mackey \& Gilmore \shortcite{rrlyr} used archival WFPC2 images to identify and
study $197$ new RR Lyrae stars in four of the five globular clusters in the Fornax dSph galaxy.
The discovery of these stars, along with the high quality colour-magnitude diagrams
(CMDs) derived from the WFPC2 images, allowed the HB morphologies for the four clusters to be 
accurately measured (the results are listed in Table \ref{t:hbresults}, below). In addition, 
because all four clusters possess well populated horizontal branches, the $V-I$ colours of the 
edges of the instability strip at the level of the horizontal branch were derived for each cluster. 

These results open the way for similar work on other external clusters. Specifically, as described
above, we aim to investigate in a uniform manner the HB morphologies of as many local external globular
clusters as possible. HB types have previously been derived for many external clusters; however these 
measurements are widely scattered, and a large variety of observations, reduction procedures, and 
counting techniques employed to obtain them. Therefore, it is highly desirable to obtain a set of 
parameters from as uniform an observation group, and via as uniform a reduction and calculation 
procedure as possible.

\subsection{Observations}
\label{ss:obs}

\begin{table*}
\begin{minipage}{130mm}
\caption{Available ($V,\,I$) WFPC2 observations of external clusters for HB calculations.}
\begin{tabular}{@{}lcllcclcc}
\hline \hline
Cluster & Program & Data-group & Reference & $N_V$ & Exposure & Data-group & $N_I$ & Exposure \\
Name & ID & (F555W) & Image & & Range (s) & (F814W) & & Range (s) \\
\hline
NGC 1466 & 5897 & u2xj0105b & u2xj0105t & 5 & 260--1000 & u2xj0108b & 6 & 260--1000 \\
NGC 1754 & 5916 & u2xq0103b & u2xq0103t & 5 & 20--500 & u2xq0109b & 6 & 20--600 \\
NGC 1786 & 5897 & u2xj0205b & u2xj0205t & 5 & 260--1000 & u2xj0208b & 6 & 260--1000 \\
NGC 1835 & 5916 & u2xq0203b & u2xq0203t & 5 & 20--500 & u2xq0209b & 6 & 20--600 \\
NGC 1841 & 5897 & u2xj0708b & u2xj0708t & 3 & 700--800 & u2xj0707b & 1 & 800 \\
NGC 1898 & 5916 & u2xq0303b & u2xq0303t & 5 & 20--500 & u2xq0309b & 6 & 20--600 \\
NGC 1916 & 5916 & u2xq0403b & u2xq0403t & 5 & 20--500 & u2xq0409b & 6 & 20--600 \\
NGC 2005 & 5916 & u2xq0503b & u2xq0503t & 5 & 20--500 & u2xq0509b & 6 & 20--600 \\
NGC 2019 & 5916 & u2xq0603b & u2xq0603t & 5 & 20--500 & u2xq0609b & 6 & 20--600 \\
NGC 2210 & 5897 & u2xj0405b & u2xj0405t & 5 & 260--1000 & u2xj0408b & 6 & 260--1000 \\
NGC 2257 & 5897 & u2xj0505b & u2xj0505t & 5 & 260--1000 & u2xj0508b & 6 & 260--1000 \\
Hodge 11 & 5897 & u2xj0305b & u2xj0305t & 5 & 260--1000 & u2xj0308b & 6 & 260--1000 \\
Reticulum & 5897 & u2xj0605b & u2xj0605t & 5 & 260--1000 & u2xj0608b & 6 & 260--1000 \\
NGC 121 & 6604 & u3770501b & u3770501r & 10 & 40--400 & u377050bb & 8 & 20--500 \\
Fornax 4 & 5637 & u2lb0205b & u2lb0205t & 3 & 200--1100 & u2lb0203b & 3 & 200--1100 \\
M54 & 6701 & u37ga40cb & u37ga40cr & 6 & 300--350 & u37ga401b & 6 & 260--300 \\
Terzan 7 & 6701 & u37g020cb & u37g020cr & 6 & 300--350 & u37g0201b & 6 & 260--300 \\
Terzan 8 & 6701 & u37g130cb & u37g130cr & 6 & 300--350 & u37g1306b & 1 & 260 \\
Arp 2 & 6701 & u37g010cb &u37g010cm  & 6 & 300--350 & u37g0101b & 6 & 260--300 \\
\hline
\label{t:hbobs}
\end{tabular}
\end{minipage}
\end{table*}

The first stage in this process is the assembly of a set of cluster images suitable for photometry
of high resolution and accuracy. Given this requirement and the availability of the photometry 
pipeline established for the Fornax RR Lyrae study, WFPC2 imaging is the most appropriate.
As described earlier, the LMC possesses $16$ clusters which are known to be comparable in age to 
the Galactic globular clusters -- that is, older than the LMC age gap (or in the case of the unique
cluster ESO121-SC03, within the age gap). Detailed F555W and F814W WFPC2 observations exist in 
the {\em HST} archive for $13$ of these clusters, while the remaining three (NGC 1928, 1939, and
ESO121-SC03) have been observed recently with the Advanced Camera for Surveys (ACS), also on board
{\em HST} \cite{acs1,acs2}. The {\em HST} archive also contains WFPC2 $V$ and $I$ observations of the 
one bona fide old SMC cluster (NGC 121), four of the six established Sagittarius clusters (excluding 
Pal. 12 and NGC 4147; however Pal. 12 does possess recent ACS observations), and the Fornax cluster which 
was not included in the study of Mackey \& Gilmore \shortcite{rrlyr} (cluster 4). These
WFPC2 observation sets are listed in Table \ref{t:hbobs}.

\subsection{Data reduction and photometry}
Each set of observations was reduced via the photometry and calibration pipeline described by 
Mackey \& Gilmore \shortcite{rrlyr}. Briefly, this process contains the following steps. First, 
for each cluster we identified a reference image -- these are listed in Table \ref{t:hbobs}. All 
the other images for a cluster were aligned with its reference image using the {\sc iraf} task
{\sc imalign}. Next we reduced each image using {\sc HSTphot} \cite{dolphin:00a}. This reduction
included preliminary preparation (such as the masking of bad pixels, and the construction of a 
background image) followed by PSF fitting photometry with a minimum threshold for object detection 
of $3\sigma$ above the background. Artificial star tests were also carried out. Photometric 
goodness-of-fit parameters were employed to select only objects with high quality 
measurements. Only measurements for which an object was classified as stellar ({\sc HSTphot} 
types 1, 2, and 3), and for which $\chi \le 2.5$, S/N\ $\ge 3.0$, $-0.3 \le$\ sharpness\ $\le 0.3$,
and errors in the flight magnitude $\sigma_{F} \le 0.1$ were retained. This resulted in at most 
$5$ F555W and $6$ F814W photometric measurements of a star (except in NGC 121, which had $10$ 
F555W and $8$ F814W frames). These measurements were calibrated and converted to Johnson-Cousins
$V$ and $I$ magnitudes according to the recipe of Holtzman et al. \shortcite{holtzman:95}, and 
using the latest update to the Dolphin \shortcite{dolphin:00b} charge transfer efficiency (CTE) and
zero-point calibrations. The fake star measurements were also passed through the full reduction 
procedure, so that a completeness value could be assigned to each real stellar detection.

Unlike for the RR Lyrae study, we did not perform a variability search on these data -- there were 
in general too few observations per star to be able to obtain an accurate and complete selection 
of variable objects across the cluster sample. Instead, our aim was to construct a high quality CMD
of the HB region for each cluster, and use the instability strip edges measured from the
Fornax RR Lyrae study to split the HB stars into three groups -- a $B$ group (stars bluer than
the blue edge of the instability strip), an $R$ group (stars redder than the red edge of the
instability strip), and a $V$ group (stars between the two edges -- that is, hypothetical variable
HB objects). The most common parametrization of HB morphology is the index of Lee, Demarque \& 
Zinn \shortcite{lee:94} (hereafter referred to as the HB type): $(B-R)/(B+V+R)$.

\begin{figure}
\includegraphics[width=0.5\textwidth]{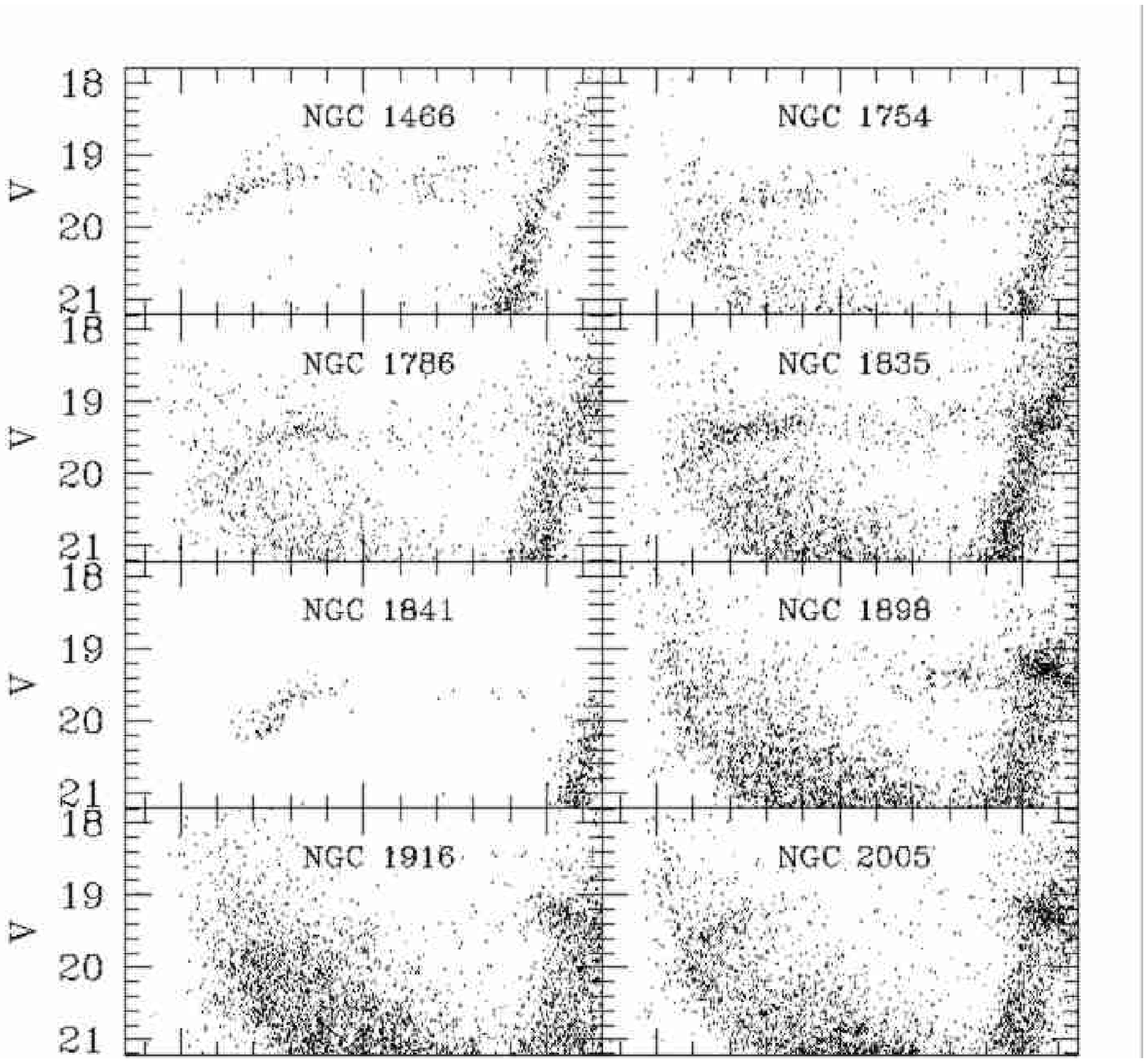} \\
\vspace{-5mm}
\includegraphics[width=0.5\textwidth]{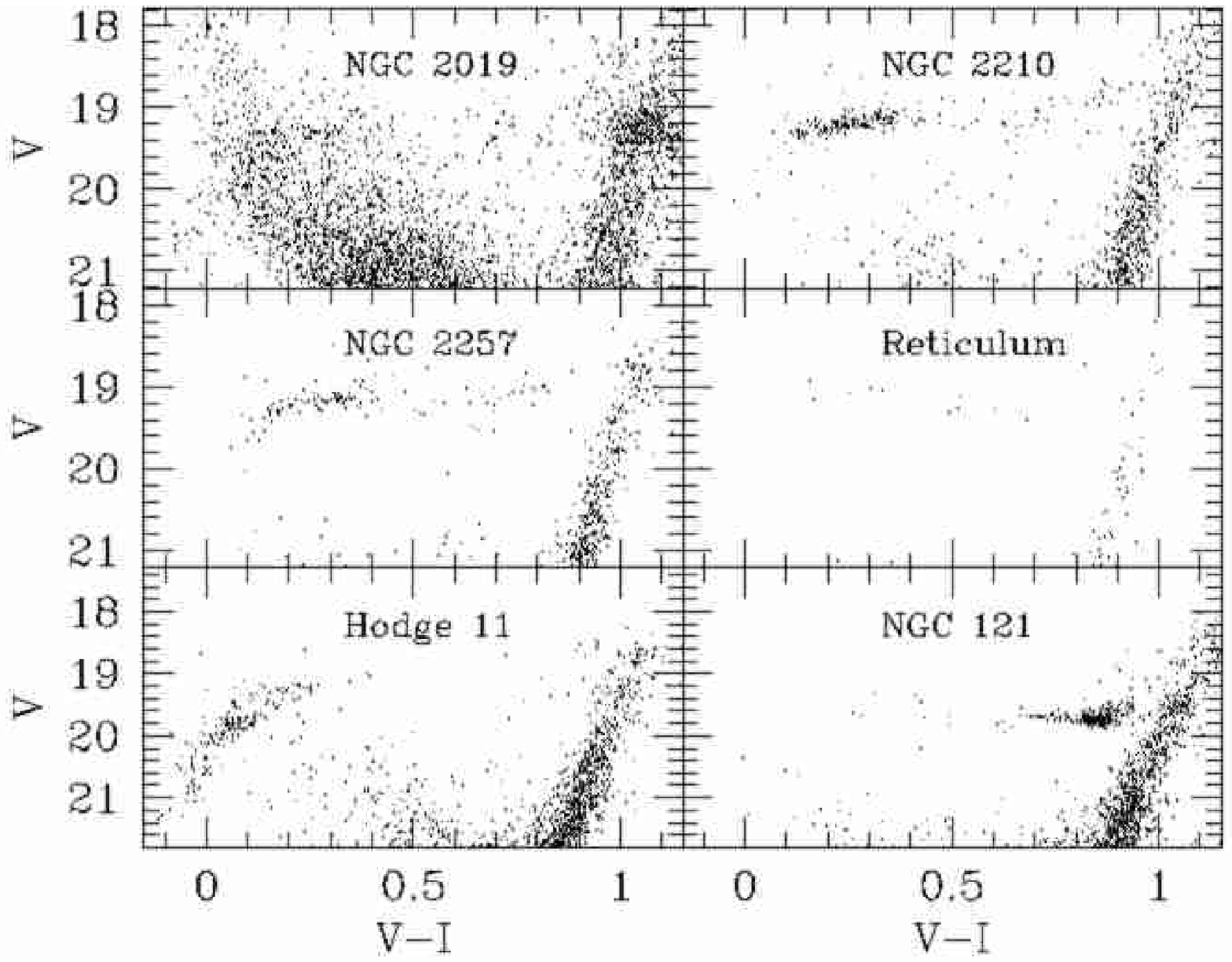}
\caption{Colour-magnitude diagrams showing the horizontal branches of all the old LMC and SMC clusters with $(V,I)$ WFPC2 images. All measured stars are plotted, with no reddening corrections. Note the strong field star contamination in the LMC bar clusters.}
\label{f:hbmcc}
\end{figure}

\begin{figure}
\begin{center}
\includegraphics[width=52mm]{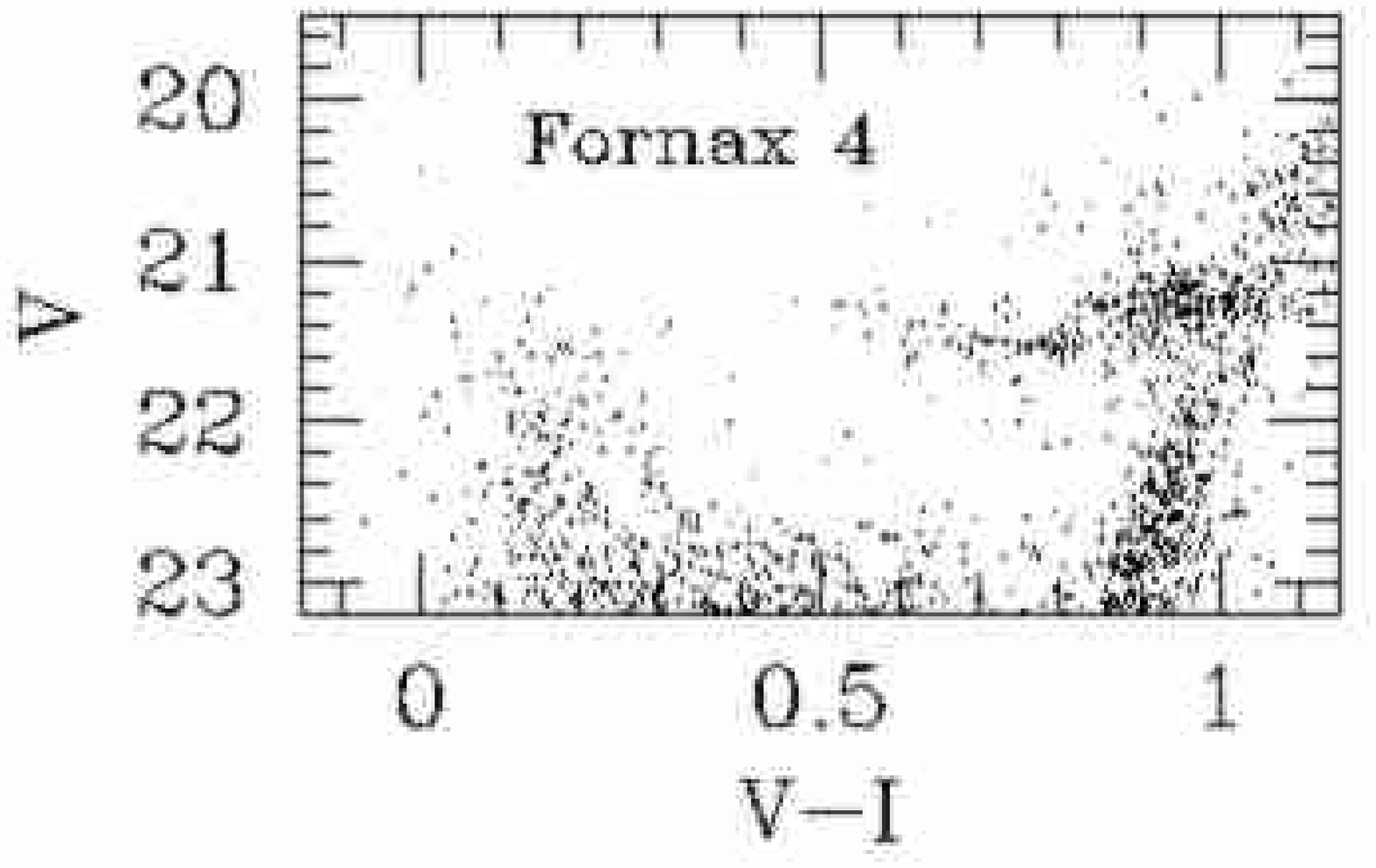} \\
\vspace{-3mm}
\includegraphics[width=52mm]{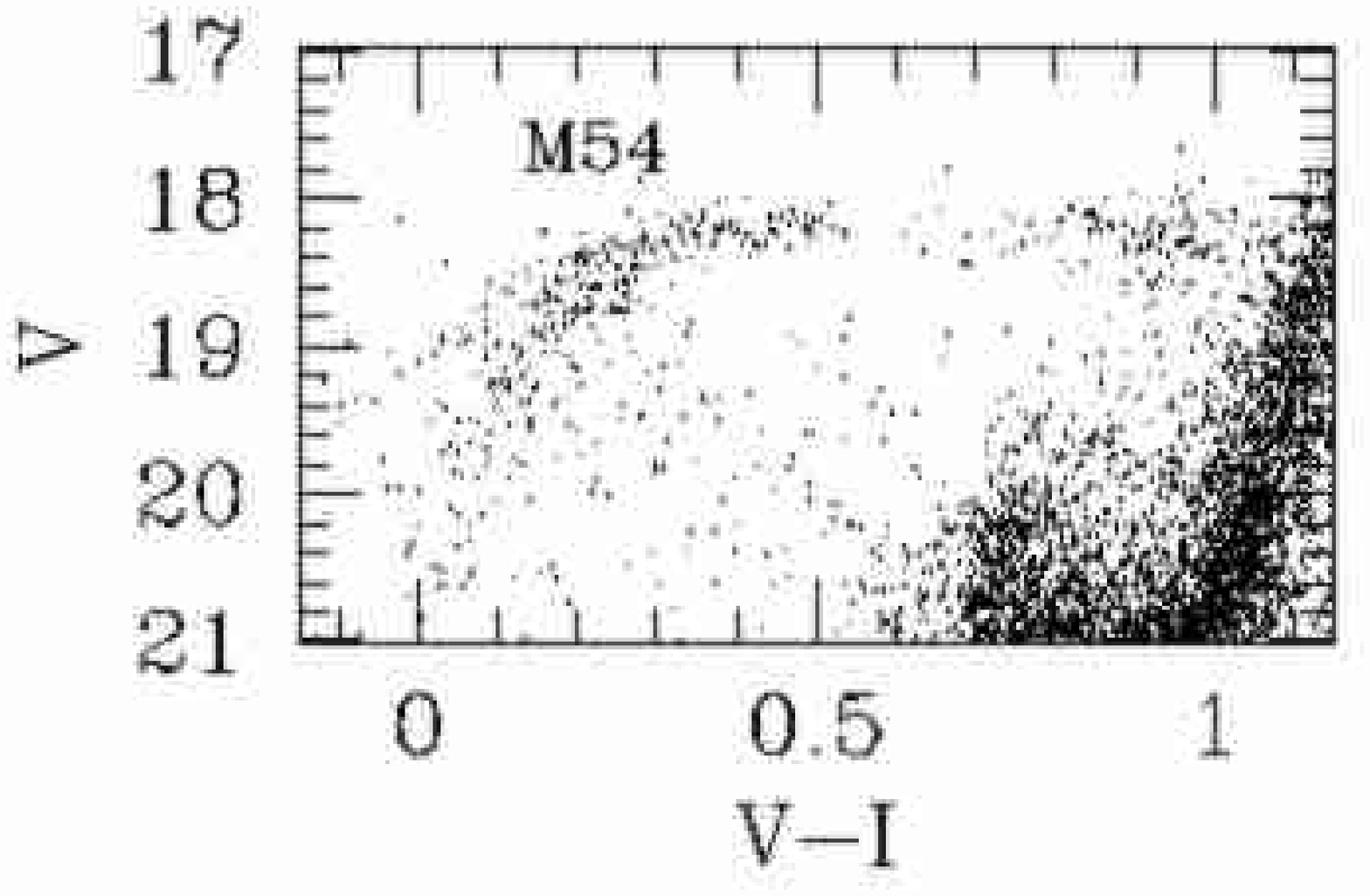}
\end{center}
\caption{Colour-magnitude diagrams showing the horizontal branches of the central clusters of the Fornax and Sagittarius dwarf galaxies -- Fornax 4 (upper), and M54 (lower). The data are from archival $V$ and $I$ WFPC2 images. No corrections have been made for reddening or field star contamination.}
\label{f:hbdsph}
\end{figure}

In order to calculate these parameters, we first combined all measurements in a particular 
passband for a given cluster in a weighted average. This allowed the construction of an accurate 
CMD for that cluster. For $N_V$ measurements $V_i \pm \sigma_{V,i}$ in F555W, and $N_I$ 
measurements $I_i \pm \sigma_{I,i}$ in F814W, we calculated the mean measurements:
\begin{eqnarray}
V = \frac{\sum_{i=0}^{N_V} \frac{V_i}{\sigma_{V,i}^{2}}}{\sum_{i=0}^{N_V} \frac{1}{\sigma_{V,i}^{2}}}\hspace{10mm} \sigma_{V}^{2} =  \frac{1}{{\sum_{i=0}^{N_V} \frac{1}{\sigma_{V,i}^{2}}}} \\
I = \frac{\sum_{i=0}^{N_I} \frac{I_i}{\sigma_{I,i}^{2}}}{\sum_{i=0}^{N_I} \frac{1}{\sigma_{I,i}^{2}}}\hspace{10mm} \sigma_{I}^{2} =  \frac{1}{{\sum_{i=0}^{N_I} \frac{1}{\sigma_{I,i}^{2}}}} .
\end{eqnarray}
Upon the completion of the CMDs, it became immediately apparent that the data for the three 
Sagittarius clusters Terzan 7, Terzan 8, and Arp 2 were not suitable for the desired measurements. 
With no exposure durations shorter than $300$ s for these clusters, all HB stars were saturated on 
every image. The HB regions for all other clusters are presented in Fig. \ref{f:hbmcc} and 
\ref{f:hbdsph}.

In these CMDs all detected stars have been plotted. More complications are evident. For a 
significant proportion of the LMC clusters, as well as Fornax 4 and M54, field star contamination 
is severe. For the LMC bar clusters in particular, the horizontal branches are essentially wiped 
out by field stars. Furthermore, for the LMC cluster NGC 1841, there is another saturation problem.
This time, the blue HB region has been measured; however the saturation limit cuts across
the HB and almost no RR Lyrae stars or red horizontal branch stars are detected. Literature CMDs 
show that a significant number of such objects exist \cite{walker:1841,brocato:96} -- hence
the WFPC2 observations of NGC 1841 are also not suitable for the present study. In addition, as 
noted by Mackey \shortcite{thesis}, the only archival WFPC2 observations of the LMC cluster 
Reticulum do not cover the cluster core. This is not crucial for measurements of HB morphology; 
however Reticulum is an extremely sparse cluster and missing the core means a very low density of 
HB stars -- as is clear from the CMD. Hence, any determination of HB type from the present 
measurements is extremely uncertain, and far better measurements exist in the literature 
\cite{walker:ret,acs1}. Thus, the WFPC2 Reticulum observations were also not analysed 
further.

\subsection{Results}
\label{ss:results}
The results of the Fornax cluster RR Lyrae survey placed the intrinsic red and blue edges of the instability
strip at $(V-I)_{RE} = 0.59 \pm 0.02$ and $(V-I)_{BE} = 0.28 \pm 0.02$, respectively, on a
$(V,\,V-I)$ CMD \cite{rrlyr}. Even if no RR Lyrae stars had been detected in the four Fornax 
clusters from this survey, application of these limits to the counting of HB stars into $R$, $B$, 
and $V$ groups would have resulted in HB types extremely similar to those derived using the
RR Lyrae variables (see Table \ref{t:hbresults}). Hence, application of these limits to the 
counting of HB stars in the present clusters provides a uniform means of calculating HB types 
across the sample. This procedure requires an accurate reddening to be known for any cluster 
under consideration -- however, literature values are available for all but one of the sample 
clusters. We have compiled these in Table \ref{t:hbresults}. For the LMC clusters the main
sources are a series of papers by Walker, summarized in Walker \shortcite{walker:sum}, which 
predominantly deal with the outer LMC objects; and the work of Olsen et al. \shortcite{olsen:98} 
which concern the LMC bar clusters. For the SMC the main work is that of Crowl et al. 
\shortcite{crowl:01}, while for the dSph clusters the values are taken from the compilation in
Mackey \& Gilmore \shortcite{sbp3} except for Fornax cluster 1, 2, 3, and 5, where the results
from Mackey \& Gilmore \shortcite{rrlyr} supersede these. 

As mentioned above, several clusters suffer from field star contamination, and this must be
accounted for in any HB calculations. However, no offset field observations were available. 
It was therefore necessary to devise a means of estimating the field contamination directly from 
the cluster images. First, for contaminated clusters, we only used the very central cluster 
regions to count HB stars -- this minimizes the number of field star interlopers (by taking a 
relatively small area), and maximizes the cluster HB stars because a cluster centre is the region 
of greatest density. Second, we selected a region on each image as far from the cluster centre as 
possible, and performed additional star counts. These ``field'' regions were reasonably large in 
area, to ensure good statistics, and typically at radii $r\sim 80-100\arcsec$ from a cluster 
centre. Counting stars in such a region is only an approximation to the field because regions at 
such radii are almost invariably still within the cluster limits, and hence possibly still contain 
cluster HB stars. However, Mackey \& Gilmore \shortcite{sbp1,sbp2,sbp3} performed background
subtractions during the construction of their surface brightness profiles for these clusters, and
it is clear from their results that for the worst affected clusters (e.g., the LMC bar clusters),
the cluster contribution is very much below the field level (by up to $\sim 3$ mag arcsec$^{-2}$). 
Hence, the approximation to a pure field region appears valid.

With the contribution of field stars to the $R$, $V$, and $B$ HB regions determined for a given
cluster, we subtracted these values from the central star counts to obtain a field corrected HB 
index. Completeness corrected values were used in all star counts, to ensure that completeness 
variations along the HB (due to BHB stars being intrinsically fainter in F814W imaging than RHB 
stars) and as a function of cluster radius were accounted for. Taking a central value of the HB 
index as representative of an entire cluster is contingent upon the fact that the HB index does 
not vary radially within the cluster -- while this is still somewhat of an open matter 
(see e.g., Alcaino et al. \shortcite{alcaino:96a}), we observed no evidence for systematic radial 
HB dependence in the clusters with no field contamination (e.g., NGC 1466). 

\begin{figure}
\includegraphics[width=0.5\textwidth]{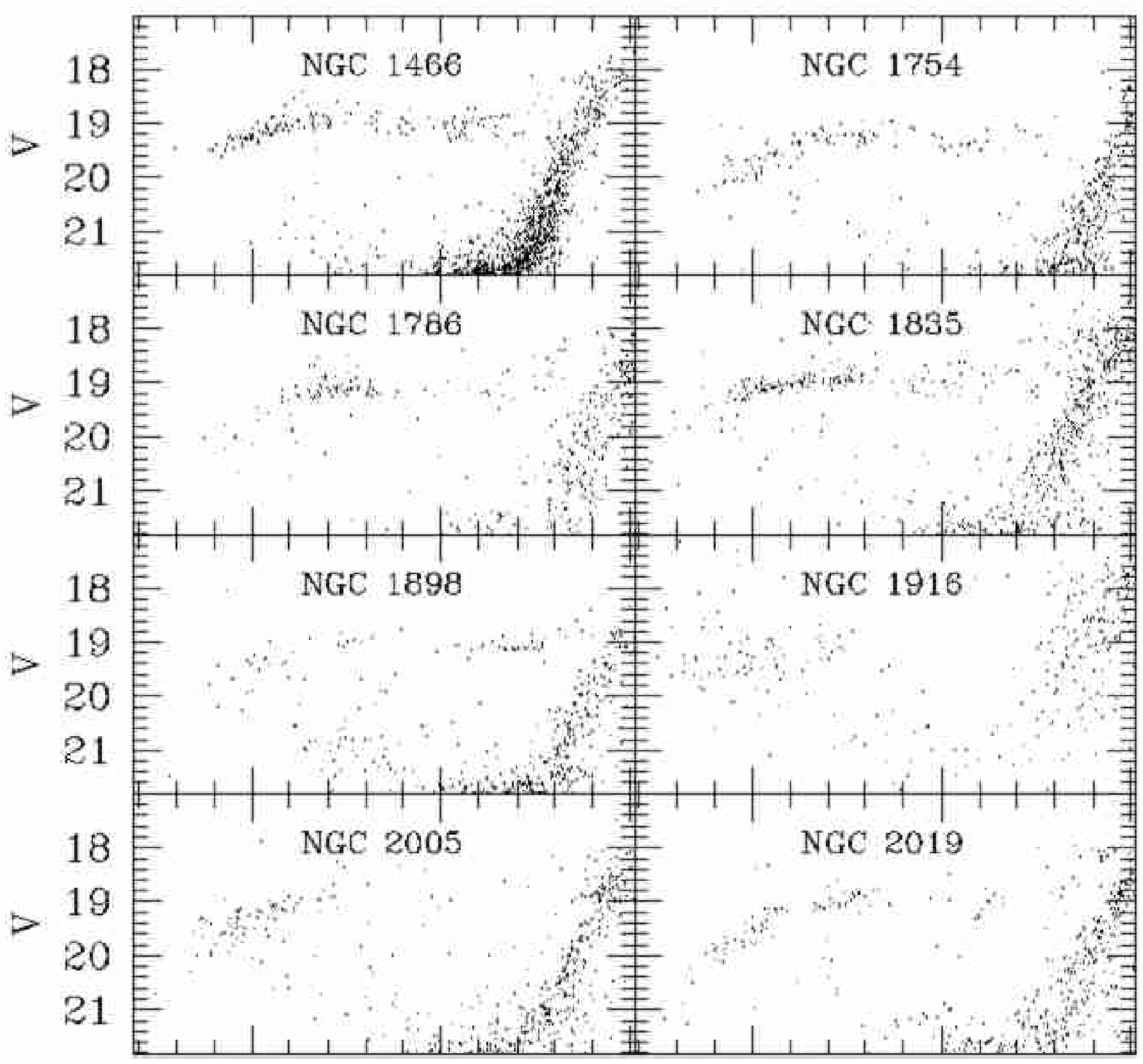} \\
\vspace{-5mm}
\includegraphics[width=0.5\textwidth]{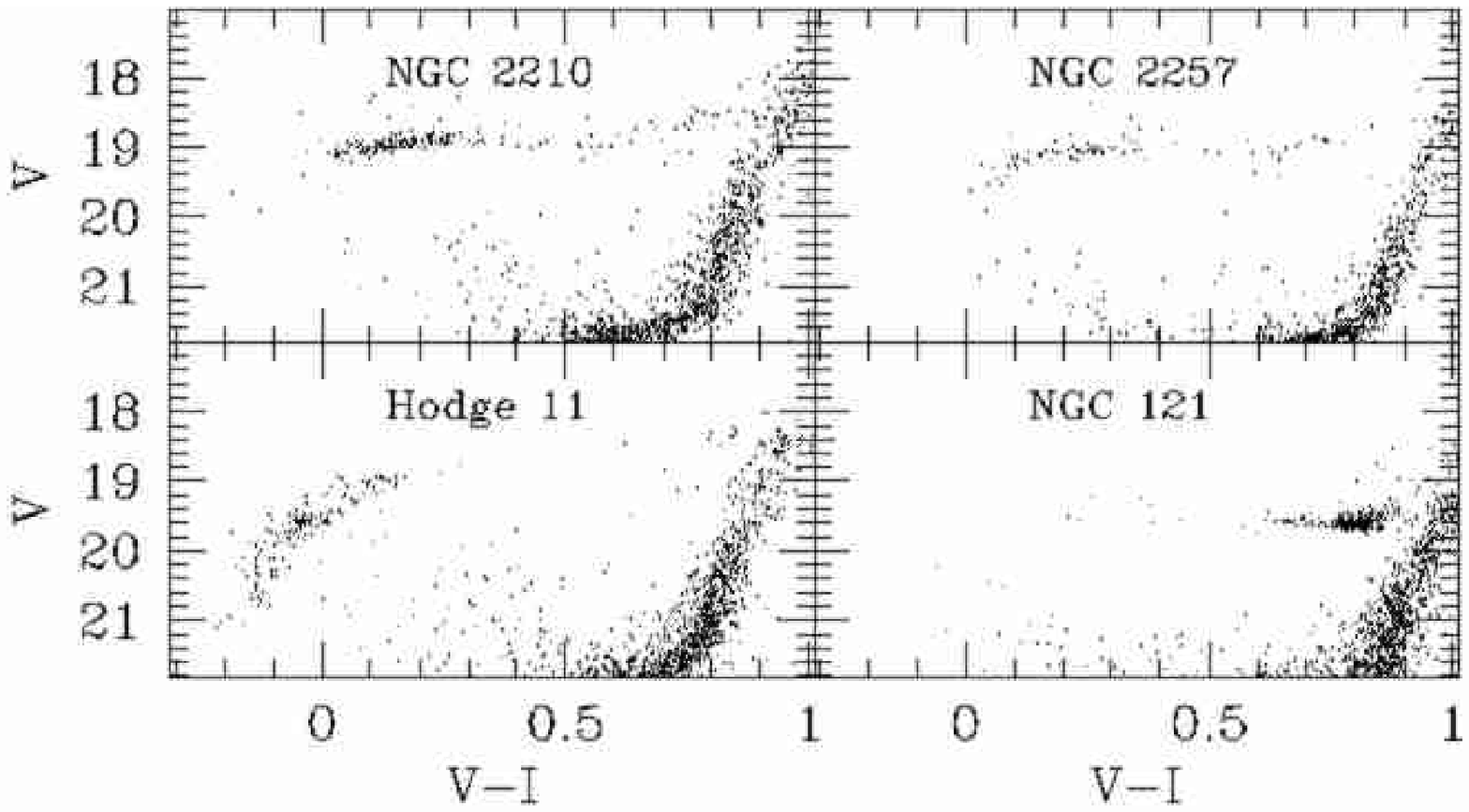}
\caption{Colour-magnitude diagrams showing the horizontal branches for $12$ of the old LMC and SMC clusters from Fig. \ref{f:hbmcc}. The photometry has been corrected for reddening using the literature values from Table \ref{t:hbresults}. For targets badly affected by field star contamination (NGC 1754, 1786, 1835, 1898, 1916, 2005, 2019), only stars within $10\arcsec$ of a given cluster centre are plotted. For the remaining clusters, only stars within $50\arcsec$ of a given centre are plotted.}
\label{f:hbmccrcorr}
\end{figure}

\begin{figure}
\begin{center}
\includegraphics[width=52mm]{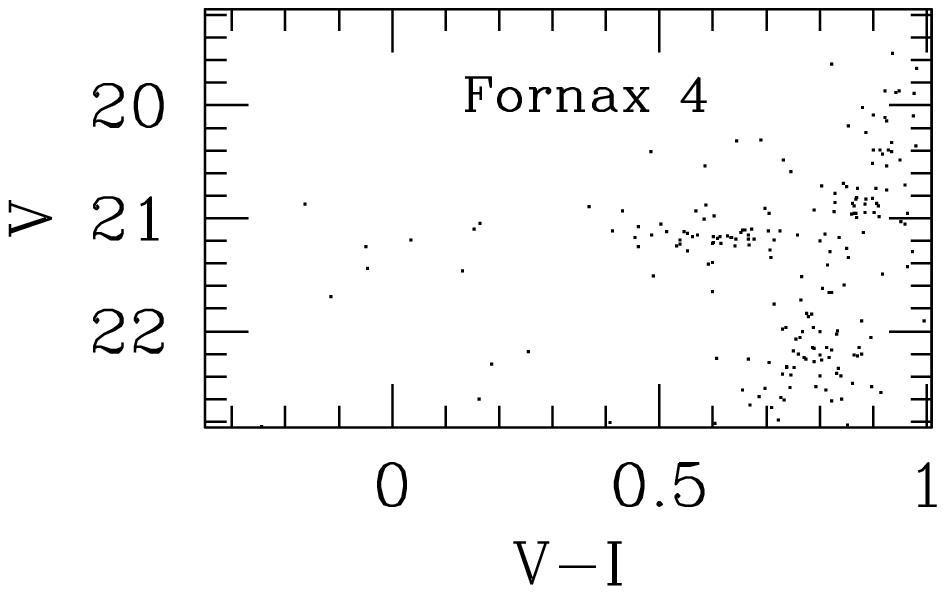} \\
\vspace{-3mm}
\includegraphics[width=52mm]{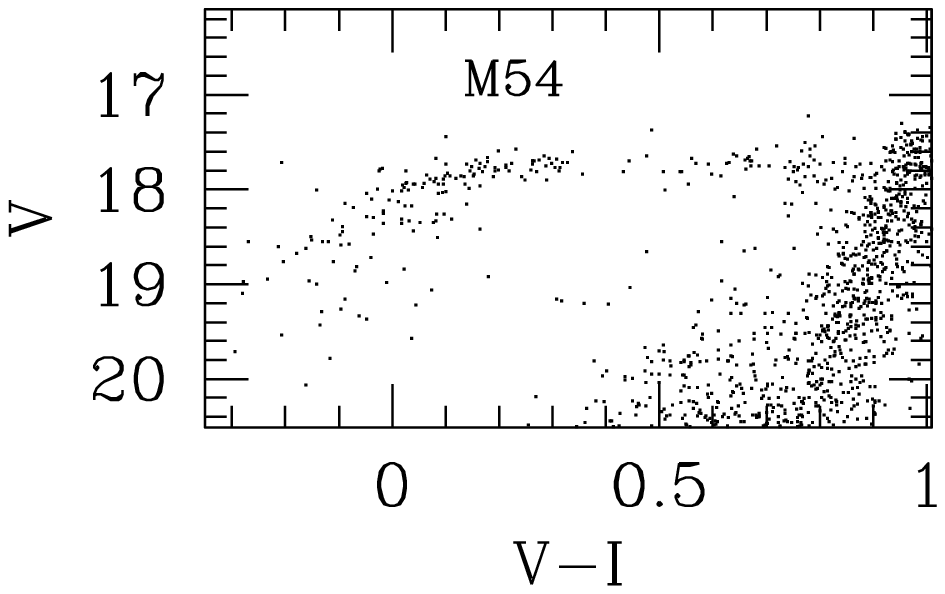}
\end{center}
\caption{Colour-magnitude diagrams showing the horizontal branches of Fornax 4 (upper), and M54 (lower). The photometry has been corrected for reddening using the literature values from Table \ref{t:hbresults}. For Fornax 4, only stars within $12\arcsec$ of the cluster centre are plotted, while for M54, only stars within $25\arcsec$ of the centre are plotted.}
\label{f:hbdsphrcorr}
\end{figure}

The reddening corrected central-region HB CMDs are presented in Figs. \ref{f:hbmccrcorr} and
\ref{f:hbdsphrcorr}. It is clear that by taking the central regions only, field star contamination
is significantly alleviated, although not cured completely. The results of the HB index calculations
are listed in Table \ref{t:hbresults}, along with measurements from the literature where available.
In general the agreement is good, demonstrating the validity of the field star subtractions.
In several cases, there is a significant discrepancy, or the procedure required some form of 
alteration. These clusters are discussed individually below:

\vspace{3mm}{\it NGC 1786:} Although there are previously published CMDs for this cluster, we could
not locate any record of a calculation of its HB index. The value measured here for the first time, 
$0.39 \pm 0.08$, is the reddest of the compact LMC bar clusters, excluding NGC 1898. NGC 1786 
is strongly affected by field star contamination; however the good agreement between the present 
measurements for the other bar clusters (NGC 1754, NGC 1835, NGC 1898, NGC 2005, and NGC 2019), 
and those of Olsen et al. \shortcite{olsen:98} who used a full iterative statistical subtraction 
method, lends confidence in the accuracy of the result.

{\it NGC 1916:} This LMC bar cluster suffers badly from differential reddening in addition
to field star contamination, as is evident in both Fig. \ref{f:hbmcc} and Fig. \ref{f:hbmccrcorr}.
Although Olsen et al. \shortcite{olsen:98} present a CMD for this cluster, no measurements of 
reddening or HB type exist. Examination of Fig. \ref{f:hbmccrcorr} shows an almost exclusively 
blue horizontal branch, similar in appearance to that of NGC 2005 or Hodge 11. For these two 
clusters, the red end of the BHB region signifies the level of the horizontal branch. To estimate 
a mean reddening towards the centre of NGC 1916, we measured this level ($V = 19.6 \pm 0.1$). 
Assuming the relation between HB absolute magnitude and cluster metallicity from the review of 
Chaboyer \shortcite{chaboyer:99} and adopting the spectroscopic metallicity measurement for NGC 1916 
from Olszewski et al. \shortcite{olszewski:91} ($[$Fe$/$H$] = -2.08 \pm 0.2$), the HB has an intrinsic 
brightness $M_V(HB) = 0.45 \pm 0.05$. Assuming an LMC distance modulus of $18.5$, this implies a mean 
extinction $A_V = 0.65 \pm 0.15$ mag to the centre of NGC 1916, and hence a mean colour excess 
$E(V-I) = 0.27 \pm 0.06$. This is significantly larger than for the other bar clusters; however 
the level of differential reddening suffered by NGC 1916 is also much larger than for these 
clusters, and it also has the highest level of field star contamination -- hence the reddening 
estimate does not seem inappropriate. Using this value, a reasonable estimate of the 
HB type for NGC 1916 can be made. As expected, this parametrizes the HB morphology as almost 
completely blue -- just as for the comparable clusters NGC 2005 and Hodge 11.

{\it Hodge 11:} This cluster has previously been assumed to possess an entirely blue 
horizontal branch \cite{walker:h11,mighell:96}. However, the present measurements suggest
$2$ potential variable stars within $50\arcsec$ of the cluster centre, and $4$ likely RHB stars. 
These objects are clearly visible in Fig. \ref{f:hbmccrcorr}, and alter the derived HB index
to $0.96 \pm 0.04$, slightly down from the $1.00$ limit usually quoted in the literature.

{\it Fornax 4:} From Fig. \ref{f:hbdsph}, it can be seen that Fornax 4 suffers field
star contamination in both the blue and red parts of its horizontal branch. Buonanno et al. 
\shortcite{buonanno:99} have published a CMD for this cluster, based on WFPC2 imaging, and suggest 
an entirely red HB index ($-1.00 \pm 0.2$). This is based on their observation that the mean red 
edge to the instability strip is $(V-I)_{RE} = 0.46 \pm 0.06$, and that all but a handful of the 
Fornax 4 HB stars are redder in colour than this limit. However, this red edge is significantly
bluer than that which we are using here ($(V-I)_{RE} = 0.59 \pm 0.02$). Hence, we find
considerably more RR Lyrae candidates in Fornax 4 than Buonanno et al. \shortcite{buonanno:99} counted. 
In addition, there is apparently a weak blue extension to the HB. This is masked by field stars in 
Fig. \ref{f:hbdsph} but is evident in Fig. \ref{f:hbdsphrcorr}. Careful field star counts in 
several outer regions of the Fornax 4 image result in consistent field star corrections
for the BHB region, which are not large enough to account for this extension. Furthermore, 
close examination of the eight BHB stars from Fig. \ref{f:hbdsphrcorr} shows them to be centrally 
concentrated spatially, with six lying within $r = 7\arcsec$. This lends weight to the present 
calculation that the true HB index of Fornax 4 is significantly bluer than previous measurements, 
at $-0.42 \pm 0.10$.

{\it M54:} This is the largest and brightest of all the clusters in the present study,
and with no exposures shorter than $\sim 300$ s it suffers very badly from crowding and saturation 
in the central regions of its images. This, combined with its large tidal radius 
($r_t \sim 450\arcsec$ -- see e.g., Harris \shortcite{harris:96}), renders it impractical to 
carry out the field subtraction as outlined above, even though M54 does suffer somewhat from 
contamination due to the Sagittarius dwarf. There has been some confusion in the past with 
relation to the M54 HB, because of the presence of an ``anonymous'' RHB. As described
by Layden \& Sarajedini \shortcite{layden:00}, it is not clear whether this RHB belongs to M54 or 
the Sagittarius dwarf. Using measurements with a minimum inner radius of $60\arcsec$, 
Layden \& Sarajedini \shortcite{layden:00} calculated the HB index to be $0.75 \pm 0.04$ if the 
RHB belongs to Sagittarius, or $0.44 \pm 0.09$ if it belongs to M54. The present images can extend 
the measurements to a minimum inner radius $r\sim 10\arcsec$. The RHB persists strongly even at 
such small radii, and it therefore seems likely that it belongs to M54. The new BHB and variable 
star counts are in the same proportion as those of Layden \& Sarajedini \shortcite{layden:00}, so 
it seems that the field star contamination within $\sim 20\arcsec$ is not too severe. Hence our
calculated HB index of $0.54 \pm 0.07$, which is in reasonable agreement with the value from 
Layden \& Sarajedini, including the RHB, appears appropriate even without any field star 
subtraction.

\begin{table*}
\begin{minipage}{165mm}
\caption{Measured and literature HB types, plus reddenings and metallicities, for all sample clusters.}
\begin{tabular}{@{}lccccccc}
\hline \hline
Cluster & Reddening & Ref. & Field-star Correction & Calculated & Literature & Ref. & Metallicity \\
Name & $E(V-I)$ & & $(R,V,B)$ (arcsec$^{-2}$) & HB Type & HB Type & & $[$Fe$/$H$]^{a}$ \\
\hline
\vspace{-3mm} & & & & & & & \\
{\bf LMC Clusters} & & & & & & & \vspace{1mm} \\
NGC 1466     & $0.13 \pm 0.03$ & $25,26$ & $(0.00\,,\,0.00\,,\,0.00)$ & $0.42 \pm 0.05$ & $0.40 \pm 0.04$ & $25$ & $-2.17 \pm 0.20$ \\
NGC 1754     & $0.12 \pm 0.03$ & $13$ & $(0.00\,,\,0.00\,,\,0.0176)$ & $0.45 \pm 0.08$ & $0.47 \pm 0.07$ & $13$ & $-1.54 \pm 0.20$ \\
NGC 1786     & $0.11 \pm 0.04$ & $2,21,26$ & $(0.00\,,\,0.00\,,\,0.0155)$ & $0.39 \pm 0.08$	& $-$ & $-$ & $-1.87 \pm 0.20$ \\
NGC 1835     & $0.14 \pm 0.03$ & $13,26,27$ & $(0.00\,,\,0.00\,,\,0.0105)$ & $0.57 \pm 0.07$ & $0.48 \pm 0.05$ & $13$ & $-1.79 \pm 0.20$ \\
NGC 1841     & $0.24 \pm 0.03$ & $1,2,23,26$ & $-$ & $0.71 \pm 0.05$ & $0.72 \pm 0.05$ & $23,25$ & $-2.11 \pm 0.10$ \\
NGC 1898     & $0.09 \pm 0.03$ & $13$ & $(0.00\,,\,0.00\,,\,0.0151)$ & $0.03 \pm 0.09$ & $-0.08 \pm 0.10$ & $13$ & $-1.37 \pm 0.20$ \\
NGC 1916     & $0.27 \pm 0.06$ & $29$ & $(0.0075\,,\,0.00\,,\,0.0397)$ & $0.97 \pm 0.03$ & $-$ & $-$ & $-2.08 \pm 0.20$ \\
NGC 1928     & $0.08 \pm 0.02$ & $9$ & $-$ & $0.94 \pm 0.06$ & $-$ & $-$ & $-1.27 \pm 0.14$ \\
NGC 1939     & $0.16 \pm 0.03$ & $9$ & $-$ & $0.94 \pm 0.06$ & $-$ & $-$ & $-2.10 \pm 0.19$ \\
NGC 2005     & $0.13 \pm 0.03$ & $13$ & $(0.00\,,\,0.00\,,\,0.0204)$ & $0.90 \pm 0.06$ & $0.87 \pm 0.04$ & $13$ & $-1.92 \pm 0.20$ \\
NGC 2019     & $0.08 \pm 0.03$ & $13$ & $(0.0025\,,\,0.00\,,\,0.0204)$ & $0.66 \pm 0.08$ & $0.56 \pm 0.07$ & $13$ & $-1.81 \pm 0.20$ \\
NGC 2210     & $0.10 \pm 0.03$ & $2,19,26$ & $(0.00\,,\,0.00\,,\,0.0025)$ & $0.65 \pm 0.05$ & $0.57 \pm 0.04$ & $25$ & $-1.97 \pm 0.20$ \\
NGC 2257     & $0.05 \pm 0.02$ & $18,22,26$ & $(0.00\,,\,0.00\,,\,0.00)$ & $0.42 \pm 0.05$ & $0.49 \pm 0.04$ & $22,25$ & $-1.63 \pm 0.21$ \\
Hodge 11    & $0.10 \pm 0.02$ & $11,28$ & $(0.00\,,\,0.00\,,\,0.00)$ & $0.96 \pm 0.04$ & $1.00 -\,0.04$ & $25,28$ & $-2.06 \pm 0.20$ \\
Reticulum   & $0.07 \pm 0.02$ & $9$ & $-$ & $0.00 \pm 0.15$ & $-0.04 \pm 0.06$ & $24,25$ & $-1.66 \pm 0.12$ \\
ESO121-SC03 & $0.04 \pm 0.03$ & $10$ & $-$ & $-1.00 +\,0.04$ & $-1.00$ & $15$ & $-0.93 \pm 0.10$\vspace{1mm} \\
\hline
\vspace{-3mm} & & & & & & & \\
{\bf SMC Clusters} & & & & & & & \vspace{1mm} \\
NGC 121      & $0.05 \pm 0.02$ & $5,20$ & $(0.005\,,\,0.00\,,\,0.00)$ & $-0.95 \pm 0.05$ & $-$ & $-$ & $-1.71 \pm 0.10$ \\
NGC 339      & $0.05 \pm 0.02$ & $5$ & $-$ & $-1.00 +\,0.05$ & $-$ & $-$ & $-1.50 \pm 0.14$ \\
NGC 361      & $0.12 \pm 0.03$ & $5$ & $-$ & $-1.00 +\,0.05$ & $-$ & $-$ & $-1.45 \pm 0.11$ \\
NGC 416      & $0.10 \pm 0.03$ & $5$ & $-$ & $-1.00 +\,0.05$ & $-$ & $-$ & $-1.44 \pm 0.12$ \\
Kron 3      & $0.03 \pm 0.02$ & $5$ & $-$ & $-1.00 +\,0.05$ & $-$ & $-$ & $-1.16 \pm 0.09$ \\
Lindsay 1   & $0.03 \pm 0.02$ & $5$ & $-$ & $-1.00 +\,0.05$ & $-$ & $-$ & $-1.35 \pm 0.08$ \\
Lindsay 38  & $0.03 \pm 0.02$ & $5$ & $-$ & $-1.00 +\,0.05$ & $-$ & $-$ & $-1.65 \pm 0.12$\vspace{1mm} \\
\hline
\vspace{-3mm} & & & & & & & \\
\multicolumn{2}{l}{\hspace{-2.3mm}{\bf Fornax dSph Clusters}} & & & & & & \vspace{1mm} \\
Fornax 1    & $0.10 \pm 0.01$ & $8$ & $-$ & $-0.30 \pm 0.04$ & $-0.20 \pm 0.20$ & $3$ & $-2.05 \pm 0.10$ \\
Fornax 2    & $0.07 \pm 0.01$ & $8$ & $-$ & $0.50 \pm 0.06$ & $0.38 \pm 0.07$ & $3$ & $-1.83 \pm 0.07$ \\
Fornax 3    & $0.05 \pm 0.01$ & $8$ & $-$ & $0.44 \pm 0.05$ & $0.50 \pm 0.06$ & $3$ & $-2.04 \pm 0.07$ \\
Fornax 4    & $0.15 \pm 0.06$ & $4$ & $(0.0029\,,\,0.0008\,,\,0.0043)$ & $-0.42 \pm 0.10$ & $-1.00 +\,0.20$ & $4$ & $-1.95 \pm 0.15$ \\
Fornax 5    & $0.04 \pm 0.01$ & $8$ & $-$ & $0.52 \pm 0.04$ & $0.44 \pm 0.09$ & $3$ & $-1.90 \pm 0.06$\vspace{1mm} \\
\hline
\vspace{-3mm} & & & & & & & \\
\multicolumn{2}{l}{\hspace{-2.3mm}{\bf Sagittarius dSph Clusters}} & & & & & & \vspace{1mm} \\
M54         & $0.18 \pm 0.02$ & $7$ & $-$ & $0.54 \pm 0.08$ & $0.75 \pm 0.04$ & $7$ & $-1.79 \pm 0.08$ \\
Terzan 7    & $0.09 \pm 0.04$ & $16$ & $-$ & $-$ & $-1.00 +\,0.04$ & $6$ & $-0.82 \pm 0.15$ \\
Terzan 8    & $0.15 \pm 0.04$ & $12$ & $-$ & $-$ & $1.00 -\,0.04$ & $6$ & $-1.99 \pm 0.08$ \\
Arp 2       & $0.13 \pm 0.03$ & $16$ & $-$ & $-$ & $0.53 \pm 0.17$ & $17$ & $-1.84 \pm 0.09$ \\
Pal. 12      & $0.03 \pm 0.02$ & $14$ & $-$ & $-$ & $-1.00 +\,0.04$ & $14,15$ & $-0.94 \pm 0.10$ \\
NGC 4147     & $0.03 \pm 0.02$ & $6$  & $-$ & $0.66 \pm 0.11$ & $0.55 \pm 0.14$ & $6$ & $-1.83 \pm 0.10$\vspace{1mm} \\
\hline
\label{t:hbresults}
\end{tabular}
\medskip
\begin{minipage}{164mm}
\flushleft \footnotesize
{\bf Reference list:} 1. Alcaino et al. \shortcite{alcaino:96b}; 2. Brocato et al. \shortcite{brocato:96}; 3. Buonanno et al. \shortcite{buonanno:98}; 4. Buonanno et al. \shortcite{buonanno:99}; 5. Crowl et al. \shortcite{crowl:01}; 6. Harris \shortcite{harris:96}; 7. Layden \& Sarajedini \shortcite{layden:00}; 8. Mackey \& Gilmore \shortcite{rrlyr}; 9. Mackey \& Gilmore \shortcite{acs1}; 10. Mateo et al. \shortcite{mateo:86}; 11. Mighell et al. \shortcite{mighell:96}; 12. Montegriffo et al. \shortcite{montegriffo:98}; 13. Olsen et al. \shortcite{olsen:98}; 14. Rosenberg et al. \shortcite{rosenberg:98}; 15. Sarajedini et al. \shortcite{sarajedini:95}; 16. Sarajedini \& Layden \shortcite{sarajedini:97}; 17. Sarajedini et al. \shortcite{sarajedini:97b}; 18. Testa et al. \shortcite{testa:95}; 19. Walker \shortcite{walker:2210}; 20. Walker \& Mack \shortcite{walker:121}; 21. Malker \& Mack \shortcite{walker:1786}; 22. Walker \shortcite{walker:2257}; 23. Walker \shortcite{walker:1841}; 24. Walker \shortcite{walker:ret}; 25. Walker \shortcite{walker:1466}; 26. Walker \shortcite{walker:sum}; 27. Walker \shortcite{walker:1835}; 28. Walker \shortcite{walker:h11}; 29. This work. \\
$^{a}$ Metallicity values are taken from the compilations of: Mackey \& Gilmore \shortcite{sbp1} (all LMC clusters except NGC 1928, 1939, and Reticulum \cite{acs1}; and ESO121-SC03 \cite{mateo:86}); Mackey \& Gilmore \shortcite{sbp2} (all SMC clusters except Lindsay 1 and 38 \cite{crowl:01}); Mackey \& Gilmore \shortcite{sbp3} (all Fornax and Sagittarius clusters except Pal. 12 and NGC 4147 \cite{harris:96}). As noted in the text, alternative abundances exist for the LMC clusters NGC 1754, 1835, 1898, 2005, and 2019; plus Fornax 4, M54, Terzan 7, and Arp 2. \\
\end{minipage}
\normalsize
\end{minipage}
\end{table*}

As noted above, previously derived values of the HB index from the literature are listed in 
Table \ref{t:hbresults} for many clusters. In a number of cases (e.g., NGC 1466, 1841, 2257, Hodge 11,
and Reticulum) these measurements have made use of variability information to count the number of 
RR Lyrae stars, rather than estimating this quantity via the imposition of instability strip limits 
as we have done here. The good agreement between the present calculations and these literature measurements
leads us to have confidence in the validity of our procedure. 

One issue of potential concern is
the possibility that the edges of the instability strip vary with cluster metallicity. 
For example, Walker \shortcite{walker:98} finds a weak correlation between the position of the red edge of
the instability strip on a $(B,\,B-V)$ CMD and cluster metallicity: $\Delta(B-V)_{RE} = 0.04 \pm 0.02$ mag 
dex$^{-1}$ over a range $[$Fe$/$H$] \sim -1$ to $-2$. We have not included such an effect, for a number
of reasons. First, we know of no such calibration for $V-I$ colours, and with no variability 
information for most of our clusters we are not in a position to construct one. Second, the large majority
of clusters in the sample with morphologies that are not clearly fully red fall in the range 
$[$Fe$/$H$] \sim -1.6$ to $-2.1$, so any shift in the red boundary of the instability strip is not
expected to be large. Indeed, our mean edges were derived from Fornax clusters 1, 2, 3, and 5, which
cover the range $[$Fe$/$H$] \sim -1.8$ to $-2.1$. Even considering the two highest metallicity clusters 
with non-red HB morphologies in the sample (NGC 1898 and 1928) the possible shift in our red edge
is less than the uncertainties in the reddenings adopted for these clusters. Ultimately, we see that
the agreement between our present measurements and those from the literature is very close, demonstrating
that any possible shift in the boundaries of the instability strip due to metallicity effects is
small enough not to have introduced a significant systematic error into our results.

\subsection{Additional literature results}
\label{ss:additional}
\subsubsection{Horizontal branch indices}
\label{sss:hbmextra}
For comparisons to be made between the different globular cluster systems, it is important that 
each sample be as complete as possible, so selection effects do not influence the results. 
We therefore supplemented the above measurements with high quality literature results where
appropriate. These data are described below, and included in Table \ref{t:hbresults}.

For the LMC clusters, only those objects older than (or within the upper echelons of) the age gap 
are of interest. As discussed earlier, we assume this group to contain $16$ members. For eleven 
of these, new HB measurements have been made, while for NGC 1841 and Reticulum the WFPC2 
observations were insufficient, and for NGC 1928, 1939, and ESO121-SC03, no WFPC2 observations 
exist. However, four of these five clusters (excluding NGC 1841) have been the subject of a recent 
ACS photometric study \cite{acs1,acs2}. As part of this, HB types were measured for each, using 
exactly the same procedure (and instability strip limits) as in the present work. The CMD for 
ESO121-SC03 shows this cluster to be securely associated with a completely red HB morphology 
$(-1.00)$, confirming the earlier results of Mateo et al. \shortcite{mateo:86} and Sarajedini, 
Lee \& Lee \shortcite{sarajedini:95}. For NGC 1928 and 1939, the HB measurements are somewhat 
uncertain, due to very heavy field star contamination and an uncertain transformation from the ACS
fight magnitude system to Johnson-Cousins $V$ and $I$. Nonetheless, it is qualitatively clear
that both clusters possess predominantly blue HB morphologies, which matches well their
index estimates (both $0.94^{+0.06}_{-0.04}$). For Reticulum, a HB index $0.00 \pm 0.15$ was 
measured, which is entirely consistent with the result of Walker \shortcite{walker:ret} 
($-0.04^{+0.00}_{-0.05}$), made using variability information. The lends credibility both to the 
magnitude transformation used by Mackey \& Gilmore \shortcite{acs1} as well as the HB measurement 
procedure employed here. Finally, returning to NGC 1841, Brocato et al. \shortcite{brocato:96} 
provide a high quality $(V,\,V-I)$ CMD. Imposing the instability strip limits on this CMD 
allows $54$ BHB stars, $8$ variable stars, and $6$ RHB stars to be counted, resulting in a HB 
index of $0.71 \pm 0.05$. This matches very well the result of Walker \shortcite{walker:1841} who
found $0.72 \pm 0.05$, again by using variability information.

For the two dwarf spheroidal systems, the Fornax cluster sample is complete but only one of
the Sagittarius clusters (M54) had images suitable for HB index measurement. Of the remaining five
Sagittarius clusters, Terzan 7 clearly has a completely red HB (as is evident from the CMD of 
Sarajedini \& Layden \shortcite{sarajedini:97}), while Terzan 8 has a completely blue HB (clear from the 
CMD of Montegriffo et al. \shortcite{montegriffo:98}). Exactly matching measurements are listed by 
Harris \shortcite{harris:96}. For Arp 2, a CMD is presented by Sarajedini \& Layden 
\shortcite{sarajedini:97}, and a detailed calculation including field star subtraction made by
Sarajedini, Chaboyer \& Demarque \shortcite{sarajedini:97b} to obtain a HB index of 
$0.53 \pm 0.17$. Pal. 12 is a comparatively young globular cluster and, like Terzan 7, has a completely 
red HB -- as is clear from the CMD of Rosenberg et al. \shortcite{rosenberg:98} and confirmed 
by Sarajedini et al. \shortcite{sarajedini:95}. Finally, NGC 4147 is listed by Harris \shortcite{harris:96}
as having a HB index of $0.55$ -- this measurement originally comes from Lee et al. \shortcite{lee:94}
who count $7$ RHB stars, $41$ BHB stars and $14$ variable (V) stars and list the index error as $\pm 0.14$. 
The CMD of Auri\`{e}re \& Lauzeral \shortcite{auriere:91} confirms the red extension to the HB;
however the recent WFPC2 CMD of Piotto et al. \shortcite{piotto:02} shows only a very few RHB stars
-- we count $43$ BHB stars, $8$ V stars and $2$ RHB stars for a HB index of $0.77 \pm 0.07$ (see Section
\ref{ss:ggcupdated}). The reality of the RHB measured by Lee et al. \shortcite{lee:94} and
Auri\`{e}re \& Lauzeral \shortcite{auriere:91} is therefore unclear -- for present purposes we take the
mean of the two above HB indices to obtain a compromise value of $0.66 \pm 0.11$.

It is more difficult to ensure that the SMC cluster sample is complete, because the SMC system is 
far larger in number than the two dwarf galaxy cluster systems, and does not possess a convenient
age cut-off like the LMC age gap. An examination of the youngest clusters in the Sagittarius dwarf 
and the Galactic globular cluster system (e.g., Pal. 12, Terzan 7, Ruprecht 106, IC 4499) suggests
an age cut-off of $\tau \sim 6$ Gyr is in order. Crowl et al. \shortcite{crowl:01} have compiled 
data for all the massive SMC clusters older than this limit which possess high 
resolution CMDs. These clusters are NGC 121, 339, 361, 416, Kron 3, and Lindsay 1 and 38. NGC 121 
has been measured above, while CMDs for the other clusters may be found in Mighell, Sarajedini
\& French \shortcite{mighell:98}; Rich et al. \shortcite{rich:00}; and Piatti et al. 
\shortcite{piatti:01}. Examination of these CMDs shows them all to have horizontal branches redder 
than that of NGC 121, and hence a HB index of $-1.00 +\, 0.05$ is assumed for each. This is consistent
with the result of Walker \shortcite{walker:smc} who searched for RR Lyrae stars in several known and
candidate old SMC clusters, but only found such stars in NGC 121. We remark that it is likely
that additional SMC clusters older than $\sim 6$ Gyr exist; however, with no available 
information on these, the present sample will have to suffice.

\subsubsection{Cluster metallicities and core radii}
\label{sss:extramet}
In addition to the new HB measurements, we list both a metallicity (Table \ref{t:hbresults}) and core
radius (Table \ref{t:spfreq}) for each cluster. These are important for later discussion (Section
\ref{s:comparison}). We have taken both sets of quantities from Mackey \& Gilmore 
\shortcite{sbp1,sbp2,sbp3}, who compiled indicative metallicities from the literature and
performed core radius measurements from high quality surface brightness profiles. 

Several clusters were not covered by these studies. On the core radius front, profiles
were not constructed for Pal. 12, NGC 4147, ESO121-SC03, Reticulum, NGC 1928 and 1939, and Lindsay 1 and 38. 
For the first four of these clusters we adopt the measurements of Rosenberg et al. 
\shortcite{rosenberg:98}, Harris \shortcite{harris:96}, Mateo et al. \shortcite{mateo:86}, and Suntzeff et al.
\shortcite{suntzeff:92}, respectively; however, we could not locate structural measurements for NGC 1928 and 
1939, or for Lindsay 1 and 38. Recently, Mackey \& Gilmore \shortcite{acs1} published the results of a 
photometric study of NGC 1928 and 1939 using ACS on {\em HST}. Close examination of the images used in this 
work reveals NGC 1928 and 1939 to both be very compact objects. A direct comparison with WFPC2 images of
similar compact clusters (e.g., NGC 1835, 1916, 2005, 2019, etc) shows that the core radii of NGC 1928 and 
1939 must be less than $\sim 1$ pc and $\sim 1.5$ pc, respectively. We have no estimates available for
the core radii of Lindsay 1 or 38.

Metallicity estimates are also required for these eight clusters. For NGC 1928, 1939 and Reticulum, we
have adopted the photometric measurements of Mackey \& Gilmore \shortcite{acs1}, while for Lindsay 1 and
38 we take the values from the compilation of Crowl et al. \shortcite{crowl:01}. Mateo et al. 
\shortcite{mateo:86} provide an estimate for ESO121-SC03, and the database of Harris \shortcite{harris:96}
lists values for Pal. 12 and NGC 4147, to complete the compliation.

It is important to note that several of the clusters in the sample have significant discrepancies
in the metallicity measurements which appear for them in the literature. These discrepancies must be 
taken into account in the later discussion (Section \ref{ss:hbcompare}). Photometric metallicities were
obtained for five of the LMC bar clusters by Olsen et al. \shortcite{olsen:98}, who measured 
$[$Fe$/$H$] = -1.42, -1.62, -1.18, -1.35, -1.23 \pm 0.15$ for NGC 1754, 1835, 1898, 2005, and 2019, 
respectively. These are significantly higher than the spectroscopic metallicities listed in Table
\ref{t:hbresults} (originally from Olszewski et al. \shortcite{olszewski:91} for these five clusters).
Several of the clusters in the Fornax and Sagittarius dwarf galaxies show an opposite discrepancy,
whereby the cluster CMDs indicate metal poor abundances which are in contradiction with much 
more metal rich spectroscopic measurements (the photometric abundances are listed in Table 
\ref{t:hbresults}). Such a contradiction is well known for several Galactic 
clusters -- in particular Rup 106 -- and it may be linked to the fact that these clusters 
appear to be $[\alpha/$Fe$]$-deficient relative to other globular clusters (Buonanno et al. 
\shortcite{buonanno:99}; see also Sarajedini \& Layden \shortcite{sarajedini:97} for a detailed 
discussion). For Fornax cluster 4, photometric measurements indicate
$[$Fe$/$H$] \sim -1.9$ (e.g., Beauchamp et al. \shortcite{beauchamp:95}, Buonanno et al. 
\shortcite{buonanno:99}) while spectroscopy suggests $[$Fe$/$H$] \sim -1.4$ (e.g., Dubath, Meylan \& Mayor
\shortcite{dubath:92}, Strader et al. \shortcite{strader:03}). The Sagittarius globular clusters Terzan 7,
Arp 2 and M54 show a similar result. For Terzan 7, the CMD study of Sarajedini \& Layden 
\shortcite{sarajedini:97} measured $[$Fe$/$H$] =-0.82 \pm 0.15$, which is consistent with the study of
Buonanno et al. \shortcite{buonanno:95a}. However, Da Costa \& Armandroff \shortcite{dacosta:95} obtained 
$[$Fe$/$H$] =-0.36 \pm 0.09$ from spectroscopy of the Ca {\sc ii} triplet. Similarly, for Arp 2 
Sarajedini \& Layden \shortcite{sarajedini:97} measured $[$Fe$/$H$] =-1.84 \pm 0.09$ (in good agreement
with Buonanno et al. \shortcite{buonanno:95b}) while Da Costa \& Armandroff \shortcite{dacosta:95} 
observed $[$Fe$/$H$] =-1.70 \pm 0.11$. For M54, again there appears a difference between the 
photometric measurement of Sarajedini \& Layden \shortcite{sarajedini:95} ($[$Fe$/$H$] =-1.79 \pm 0.08$) 
and the spectroscopy of Da Costa \& Armandroff \shortcite{dacosta:95} ($[$Fe$/$H$] =-1.55 \pm 0.10$).
However, like for the massive globular cluster $\omega$ Cen, Sarajedini \& Layden \shortcite{sarajedini:95} 
have suggested that M54 may possess an internal metallicity dispersion, of $\sigma([$Fe$/$H$]) = 0.16$ dex. 
Da Costa \& Armandroff conclude that if this is indeed the case then the two measurements are consistent.

\subsection{Implications for RR Lyrae star populations}
\label{ss:rrlyr}
The new measurements of HB morphology described above provide some information about the likely
RR Lyrae star population of each cluster in the sample, even though we made no variability 
detections. It is worthwhile briefly considering this information, because a significant number
of these clusters have not been surveyed at high resolution (i.e., including their central regions)
for RR Lyrae stars. 

In a given cluster, the {\em specific frequency} of RR Lyrae stars is defined to be the size of the
population normalized to an integrated cluster luminosity of $M_V = -7.5$:
\begin{eqnarray}
S_{\rm{RR}} = N_{\rm{RR}}\,\,10^{0.4(7.5 + M_V)} \,.
\end{eqnarray}
Mackey \& Gilmore \shortcite{rrlyr} calculated specific frequencies for the four Fornax dSph clusters
in their RR Lyrae survey. They found specific frequencies of $S_{\rm{RR}} = 118.2$, $70.5$, $90.2$, and $68.4$ 
for the four clusters, respectively. As discussed by the authors, these numbers are lower limits because
in no cases did the WFPC2 observations cover an entire cluster. Hence, the samples of RR Lyrae stars are 
certainly incomplete, but in calculating $S_{\rm{RR}}$, the $N_{\rm{RR}}$ values have been normalized to the 
{\em full cluster} luminosities of Mackey \& Gilmore \shortcite{sbp3}. Accounting for this effect is not 
trivial, because the spatial distribution of RR Lyrae stars in any given cluster is unknown, and because 
the WFPC2 geometry is complicated. Given that the centre of each cluster was measured, it is expected 
that the $N_{\rm{RR}}$ values should be $80-90\%$ complete. Hence, the $S_{\rm{RR}}$ quantities may be 
$10-20\%$ greater than the quoted limits.

It is also possible to investigate the other external clusters. By assuming that all stars on the 
horizontal branch lying between the adopted red and blue edges of the instability strip are RR Lyrae stars, 
the approximate number of these objects in each cluster may be calculated, and the specific frequency 
estimated. The results of this process are listed in Table \ref{t:spfreq}. In general, the completeness 
corrected numbers have been adopted, and these have been adjusted to account for field star contamination 
where applicable. The specific frequencies have been normalized to the calculated asymptotic luminosities 
($L_\infty$) from the compilations of Mackey \& Gilmore \shortcite{sbp1,sbp2,sbp3}, and hence are lower 
limits (perhaps $10-20\%$ smaller than the true values), just as described above. Given the
demonstrated consistency between our method of determining HB indices via the imposition of instability 
strip edge limits, and more precise measurements involving variability detections, we expect the values
of $S_{\rm{RR}}$ calculated using the above technique to be relatively accurate indicators of the true RR Lyrae
populations in the sample clusters.

\begin{table}
\caption{Implied RR Lyrae populations and specific frequencies, plus core radii, for all sample clusters.}
\begin{center}
\begin{tabular}{@{}lcccc}
\hline \hline
Cluster & $N_{\rm{RR}}\,^{a}$ & $M_V\,^{b}$ & $S_{\rm{RR}}$ & Core radius \\
Name & & & & $r_c$ (pc)$^{c}$ \\
\hline
\vspace{-3mm} & & & & \\
{\bf LMC Clusters} & & & & \vspace{1mm} \\
NGC 1466     & $55.2$ & $-7.30$ & $66.4$ & $2.61 \pm 0.09$ \\
NGC 1754     & $47.4$ & $-7.33$ & $55.4$ & $0.88 \pm 0.12$ \\
NGC 1786     & $74.4$ & $-7.78$ & $57.5$ & $1.33 \pm 0.09$ \\
NGC 1835     & $101.6$ & $-8.33$ & $47.3$ & $1.16 \pm 0.06$ \\
NGC 1841     & $22^{*}$ & $-6.85$ & $40.0$ & $7.77 \pm 0.17$ \\
NGC 1898     & $55.1$ & $-8.60$ & $20.0$ & $2.04 \pm 0.13$ \\
NGC 1916     & $29.1$ & $-8.33$ & $13.5$ & $0.81 \pm 0.04$ \\
NGC 1928     & $-$ & $-$ & $< 15$ & $< 1.0$ \\
NGC 1939     & $-$ & $-$ & $< 15$ & $< 1.5$ \\
NGC 2005     & $36.0$ & $-7.50$ & $36.0$ & $0.88 \pm 0.08$ \\
NGC 2019     & $54.2$ & $-7.90$ & $37.5$ & $0.88 \pm 0.06$ \\
NGC 2210     & $28.9$ & $-7.53$ & $28.1$ & $1.99 \pm 0.06$ \\
NGC 2257     & $31.1$ & $-7.34$ & $35.9$ & $6.50 \pm 0.18$ \\
Hodge 11    & $2.0$ & $-7.93$ & $1.3$ & $2.95 \pm 0.16$ \\
Reticulum   & $32^{*}$ & $\sim -6.7$ & $\sim 67$ & $14.6 \pm 4.50$ \\
ESO121-SC03 & $0.0$ & $-5.10$ & $0.0$ & $8.30 \pm 1.20$ \vspace{1mm} \\
\hline
\vspace{-3mm} & & & & \\
{\bf SMC Clusters} & & & & \vspace{1mm} \\
NGC 121     & $8.1$ & $-7.94$ & $5.4$ & $2.81 \pm 0.10$ \\
NGC 339     & $0.0$ & $-7.15$ & $0.0$ & $7.38 \pm 0.30$ \\
NGC 361     & $0.0$ & $-7.40$ & $0.0$ & $5.31 \pm 0.28$ \\
NGC 416     & $0.0$ & $-7.70$ & $0.0$ & $2.84 \pm 0.10$ \\
Kron 3     & $0.0$ & $-8.00$ & $0.0$ & $6.04 \pm 0.18$ \\
Lindsay 1  & $0.0$ & $-$ & $0.0$ & $-$ \\
Lindsay 38 & $0.0$ & $-$ & $0.0$ & $-$ \vspace{1mm} \\
\hline
\vspace{-3mm} & & & & \\
\multicolumn{2}{l}{\hspace{-2.3mm}{\bf Fornax dSph Clusters}} & & & \vspace{1mm} \\
Fornax 1 & $16.0$ & $-5.33$ & $118.2$ & $10.03 \pm 0.29$ \\
Fornax 2 & $46.6$ & $-7.05$ & $70.5$ & $5.81 \pm 0.19$ \\
Fornax 3 & $118.9$ & $-7.80$ & $90.2$ & $1.60 \pm 0.07$ \\
Fornax 4 & $45.4$ & $-6.88$ & $80.4$ & $1.75 \pm 0.18$ \\
Fornax 5 & $45.2$ & $-7.05$ & $68.4$ & $1.38 \pm 0.11$ \vspace{1mm} \\
\hline
\vspace{-3mm} & & & & \\
\multicolumn{2}{l}{\hspace{-2.3mm}{\bf Sagittarius dSph Clusters}} & & & \vspace{1mm} \\
M54      & $41.6^{d}$ & $-8.55^{d}$ & $15.8$ & $0.91 \pm 0.04$ \\
Terzan 7 & $0.0$ & $-4.35$ & $0.0$ & $1.63 \pm 0.12$ \\
Terzan 8 & $0.0$ & $-5.35$ & $0.0$ & $9.50 \pm 0.72$ \\
Arp 2    & $4^{*}$ & $-5.60$ & $23.0$ & $13.67 \pm 1.85$ \\
Pal. 12   & $0.0$ & $-4.48$ & $0.0$ & $3.50 \pm 0.20$ \\
NGC 4147 & $\sim 11^{*}$ & $-6.16$ & $\sim 38$ & $0.56 \pm 0.20$ \vspace{1mm} \\
\hline
\label{t:spfreq}
\end{tabular}
\end{center}
\medskip
\vspace{-7mm}
\flushleft \footnotesize
$^{a}$ The listed $N_{\rm{RR}}$ are completeness corrected (and hence non-integer) values. Quantities marked with a $(^*)$ are taken from the literature, as discussed in the text. \\
$^{b}$ The listed $M_V$ are derived from the integrated luminosities calculated by Mackey \& Gilmore \shortcite{sbp1,sbp2,sbp3}, except for Pal. 12 and NGC 4147 \cite{harris:96}, ESO121-SC03 (derived from the measurments of Mateo et al. \shortcite{mateo:86} using an LMC distance modulus of $18.5$), and Reticulum (derived from the dynamical mass estimate of Suntzeff et al. \shortcite{suntzeff:92} using a mass-to-light ratio calculated from the stellar population models of Fioc \& Rocca-Volmerange \shortcite{fioc:97}). \\
$^{c}$ The listed $r_c$ are taken from Mackey \& Gilmore \shortcite{sbp1,sbp2,sbp3}, except for Pal. 12 \cite{rosenberg:98}, NGC 4147 \cite{harris:96}, ESO121-SC03 \cite{mateo:86}, and Reticulum \cite{suntzeff:92}. \\
$^{d}$ Measured to $r = 60\arcsec$ for M54, as discussed in the text.
\normalsize
\end{table}

For the clusters which did not have new HB measurements, we made some attempt to estimate $S_{\rm{RR}}$
using the literature data from which the HB types were obtained. It is easy to make these estimates for
most such clusters, since their HB types are either $\pm 1.00$ -- so $S_{\rm{RR}} = 0$. 
For NGC 1841, we used the measurements of Walker \shortcite{walker:1841}, who discovered $22$ RR Lyrae 
stars, while for Reticulum we adopted the results of Walker \shortcite{walker:ret} ($32$ RR Lyrae stars). 
In both of these clusters (especially NGC 1841) central crowding would have been an issue for the 
terrestrial observations involved, and hence the estimates of $S_{\rm{RR}}$ for these clusters are lower limits.
For Arp 2, we took the measurements of Sarajedini et al. \shortcite{sarajedini:97b} -- these 
authors detected $4$ RR Lyraes, but again this is certainly a lower limit. Finally, for NGC 4147 we estimated
a population of $\sim 11$ RR Lyraes, since Lee et al. \shortcite{lee:94} counted $14$ stars on the HB
instability region, while we count $8$ from the CMD of Piotto et al. \shortcite{piotto:02} (see Section
\ref{ss:ggcupdated}). This results in $S_{\rm{RR}} \sim 38$, which matches reasonably well the value listed
by Harris ($S_{\rm{RR}} = 30.8$). We were unable to calculate $S_{\rm{RR}}$ for NGC 1928 and 1939 because no integrated 
luminosities for these clusters have been published. In any case, given their extremely blue HB morphologies, 
we expect low specific frequencies for these clusters (i.e., $S_{\rm{RR}} \leq 15$, similar to NGC 1916 or Hodge 
11).

Examining the results in Table \ref{t:spfreq} for all the LMC targets, it seems that there are a 
large number of RR Lyrae stars remaining to be discovered in these clusters. Measurements have
been made for several (e.g., Walker 
\shortcite{walker:2210,walker:2257,walker:1841,walker:ret,walker:1466,walker:1835,walker:1786});
however, these have all been ground-based observations and therefore generally do not extend 
to the central regions of these clusters because of crowding. Thus, while there is some overlap
between these surveys and the present measurements, a large percentage of the stars identified
here are likely to be previously unknown RR Lyraes, and suitable targets for future surveys.
A great deal about a cluster can be learned from its population of variable stars, so it would be a 
worthwhile investment of telescope time to obtain high resolution time series observations of the 
clusters listed in Table \ref{t:spfreq} with the aim of measuring the characteristics of each set 
of RR Lyrae stars.

Compared with the Galactic globular clusters, the LMC clusters generally seem to have high values
of $S_{\rm{RR}}$. Of the $132$ Galactic globular clusters with measured $S_{\rm{RR}}$ in the catalogue of Harris
\shortcite{harris:96}, only $15\%$ ($20$ clusters) have $S_{\rm{RR}} > 30$, and only seven of these ($5\%$) 
have $S_{\rm{RR}} > 50$. For the $14$ LMC clusters, at least eight ($\sim 60\%$) have $S_{\rm{RR}} > 30$, and of 
these, four have $S_{\rm{RR}} > 50$. All five Fornax clusters have even larger specific frequencies 
($S_{\rm{RR}} \geq 70$). To the four clusters studied by Mackey \& Gilmore \shortcite{rrlyr}, an estimate has 
now been added for Fornax 4. This measurement is highly consistent with those for the other Fornax 
clusters, with $S_{\rm{RR}} \sim 80.4$. Only two Galactic globulars are listed with $S_{\rm{RR}} > 100$. 
The largest value, that for Palomar 13, is $S_{\rm{RR}} = 127.5$. Given the incompleteness in the Fornax RR 
Lyrae survey discussed above, it is very likely that Fornax 1 has $S_{\rm{RR}}$ significantly greater than this. 
It is certainly intriguing that only a tiny fraction of Galactic globular clusters have very high $S_{\rm{RR}}$, 
whereas {\em all} of the Fornax clusters do, as do many of the LMC clusters -- this is a fact worthy of 
further attention. In contrast, the SMC clusters have very low values for $S_{\rm{RR}}$. Walker 
\shortcite{walker:smc} showed that SMC clusters younger than NGC 121 ($\sim 11$ Gyr) apparently do not contain 
RR Lyrae stars -- hence the low $S_{\rm{RR}}$ values. The dearth of RR Lyrae stars in SMC clusters can possibly 
be attributed to the relative youth of these objectss, which consequently have completely red HB morphologies.
The Sagittarius clusters are mixed. Three have $S_{\rm{RR}} = 0$, but for different reasons -- 
Terzan 7 and Pal. 12 are youthful (see e.g., Sarajedini \& Layden \shortcite{sarajedini:97} and 
Rosenberg et al. \shortcite{rosenberg:98}) and have red horizontal branches, while the older Terzan 8 has a 
completely blue HB. The remaining three clusters -- M54\footnote{$S_{\rm{RR}}$ for this cluster was calculated 
from the present measurements, but only out to a radius $60\arcsec$. The integrated luminosity to this 
radius was calculated using Eq. $12$ from Mackey \& Gilmore \shortcite{sbp1} and the structural 
parameters from Mackey \& Gilmore \shortcite{sbp3}, with $r=60\arcsec$. This was necessary because 
$r_t \sim 450\arcsec$ for M54, so the approximation of using $L_\infty$ to calculate $S_{\rm{RR}}$ is 
completely invalid in this case. The value so calculated ($S_{\rm{RR}} = 15.8$) is probably good to $\pm 5$.}, 
NGC 4147, and Arp 2 -- have relatively modest values of $S_{\rm{RR}}$. M54 has a large number of variables 
(see also Layden \& Sarajedini \shortcite{layden:00}) but is an extremely luminous cluster. Arp 2 
and NGC 4147 are much less luminous and also have a considerably smaller number of variables. These clusters 
appear to match the Galactic globulars much more closely than the LMC and Fornax objects.

\section{Properties of the Galactic globular clusters}
\label{s:ggc}
As discussed earlier, Zinn \shortcite{zinn:93a} proposed a classification scheme for the $\sim 150$
Galactic globular clusters, according to metallicity and HB type. In order to facilitate comparisons with 
the external cluster systems studied in the previous Section, it is useful to repeat this sub-division 
using updated data (Zinn's original study used only the $87$ clusters with suitable measurements at the
time). To do this requires accurate data on cluster metallicities and HB types, which we address below.
Given that we are also interested in examining cluster structures, we further include data on cluster core 
radii in our discussion.

\subsection{Updated measurements and classifications}
\label{ss:ggcupdated}
\begin{figure}
\includegraphics[width=0.5\textwidth]{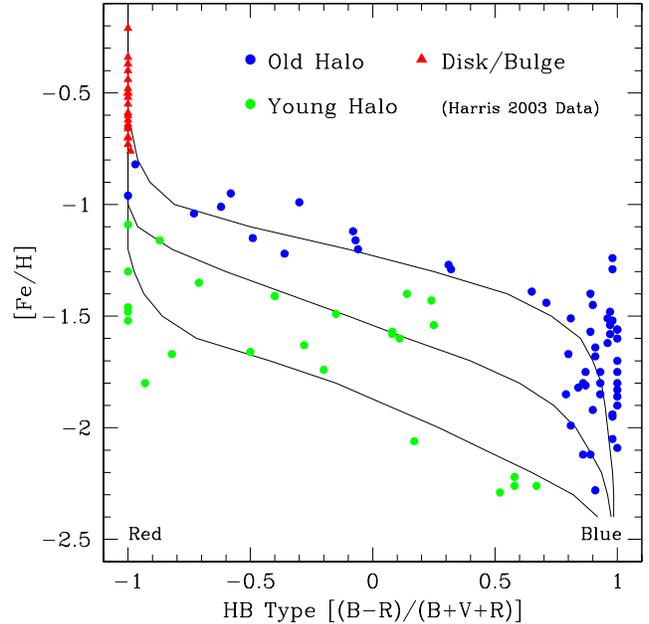}
\caption{HB-type vs. metallicity diagram for the $108$ Galactic globular clusters with suitable measurements in the Harris \shortcite{harris:96} catalogue. The clusters have been divided into three sub-systems, as labelled, according to criteria very similar to those of Zinn \shortcite{zinn:93a}. The over-plotted isochrones are from Rey et al. \shortcite{rey:01} and constitute the latest versions of the Lee et al. \shortcite{lee:94} synthetic HB models. The two lower isochrones are, respectively, $1.1$ Gyr and $2.2$ Gyr younger than the top isochrone.}
\label{f:hbmetorig}
\end{figure}

Perhaps the most useful resource concerning Galactic globular clusters is the database compiled and 
maintained by Harris \shortcite{harris:96}, which contains information on many key parameters 
(including distances, metallicities, HB types, and structural measurements) for all of the $144$ known
Galactic globular clusters\footnote{Actually, the database contains entries for $150$ clusters; however, 
six of these are assumed to be members of the Sagittarius dwarf in the present work (M54, Terzan 7, 
Terzan 8, Arp 2, Pal. 12, and NGC 4147).}. In the most recent version of the Harris database (February 2003, 
at the time of writing), $108$ clusters have sufficient information listed to be classified into the 
Zinn \shortcite{zinn:93a} sub-groups ($27$ bulge/disk; $56$ old halo; and $25$ young halo). Plotting HB-type 
vs. metallicity (Fig. \ref{f:hbmetorig}) illustrates the three sub-systems. Over-plotted on this diagram
are the latest isochrones from the synthetic HB models of Rey et al. \shortcite{rey:01}, which show how age
is expected to govern HB morphology. 

We obtained the sub-groups using a method very similar to that of Zinn \shortcite{zinn:93a}. A histogram 
of metallicity measurements shows a bimodal distribution, with $[$Fe$/$H$] = -0.8$ a suitable division line. 
Zinn \shortcite{zinn:93a} split the halo clusters (i.e., those with $[$Fe$/$H$] < -0.8$) by plotting HB 
type against $[$Fe$/$H$]$ for those objects with $R_{\rm{gc}} < 6$ kpc (the inner halo clusters) and drawing 
a fiducial line through the correlation so defined. He then defined a quantity $\Delta$HB-type, which is 
the difference in HB index between a certain cluster and the fiducial line at given 
metallicity\footnote{That is, $\Delta$HB-type\ $= \left[\frac{(B-R)}{(B+V+R)}\right]_{\rm{cluster}} - \hspace{2mm}\left[\frac{(B-R)}{(B+V+R)}\right]_{\rm{fiducial}}$ at constant $[$Fe$/$H$]$.}. 
His histogram of $\Delta$HB-type showed a suitable dividing line to be
at $\Delta$HB-type\ $= -0.4$, thus defining the old and young halo samples. For the present sample however, we
preferred not to use any spatial information in defining the two halo sub-systems. Rather than use
the fiducial line of Zinn, we noted that the upper isochrone of Rey et al. \shortcite{rey:01} traces well 
the correlation between the clusters with the bluest HB types at given metallicity (as can be seen in
Fig. \ref{f:hbmetorig}). Using this isochrone as our fiducial, a histogram of $\Delta$HB-type shows that
$\Delta$HB-type\ $=-0.3$ provides cleanly separated halo sub-samples. This corresponds to an age difference
of around $-0.6$ Gyr in the centre of Fig. \ref{f:hbmetorig}, according to the isochrones. At either end,
where the isochrones are degenerate, the implied age difference is likely somewhat larger.

The Harris database clearly contains some missing information ($36$ clusters could not be classified 
above because of a missing metallicity measurement, missing HB index measurement, or both). It is possible 
to fill in some of this data from the literature. First, only two clusters do not 
have listed metallicities. The first, BH 176, is described by Ortolani, Bica \& Barbuy 
\shortcite{ortolani:95b} to have near solar metallicity, and, in reasonable agreement, has been 
spectroscopically measured by Bica et al. \shortcite{bica:98} to have metallicity $[$Fe$/$H$] > 0.1$.
The second cluster is 2MASS-GC02, and no metallicity estimates are available for this heavily
obscured Galactic bulge cluster.

The situation with missing HB index measurements is more severe. Thirty-five clusters do not have HB 
types listed in the Harris catalogue. However, the situation is not as bad as it might initially seem.
Of these thirty-five clusters, nine\footnote{NGC 6256, 6380, 6388, 6441, Terzan 3, Terzan 12, 
Pal. 11, IC 1276, and Ton. 2.} have metallicities $[$Fe$/$H$] > -0.8$, and so, for present purposes,
may be assumed to be members of the bulge/disk population, with HB index $-1.00$. A further twelve
clusters\footnote{NGC 5634, 5946, 6273, 6284, 6325, 6355, 6401, 6453, 6517, 6540, 6569, and 6642.} appear 
in the compilation of WFPC2 CMDs for $74$ Galactic globular clusters, published by Piotto et al. 
\shortcite{piotto:02}. This $(B,\,V)$ photometry is freely available, so it is possible to calculate 
HB types for eleven of these clusters (the CMD of NGC 6540 is too blurred by differential reddening). We 
did this in a manner very similar to that employed for the external clusters from Section \ref{s:hbm}. 
The red and blue edges of the instability strip on the $(B,\,B-V)$ plane have been more extensively 
studied than on the $(V,\,V-I)$ plane. Smith \shortcite{smith:95} lists in his Table 3.1 an assortment 
of measurements for several globular clusters -- the convergence of this data is at 
$(B-V)_{RE} = 0.41 \pm 0.02$ and $(B-V)_{BE} = 0.17 \pm 0.02$. These boundaries are fully consistent
with more recent measurements such as those by Corwin, Carney \& Nifong \shortcite{corwin:99} who found
$(B-V)_{RE} = 0.38$ and $(B-V)_{BE} = 0.15$ for NGC 5466 ($[$Fe$/$H$] = -2.22$), and by Corwin \& Carney
\shortcite{corwin:01} who found $(B-V)_{RE} = 0.402$ and $(B-V)_{BE} = 0.184$ for M3 ($[$Fe$/$H$] = -1.46$).
We used our adopted limits to estimate HB types 
for the eleven clusters from Piotto et al. \shortcite{piotto:02}. As with the external clusters we examined 
in this way, it is necessary to know a reddening value for each target to make such measurements; however 
these quantities are available in the Harris catalogue. To monitor the accuracy of this procedure we also 
selected a number of ``control'' clusters from the Piotto et al. compilation, with each already possessing 
secure measurements of HB type in the literature. We ensured that these covered as wide a range in metallicity 
and extinction as possible.

The results of the measurement process are listed in Table \ref{t:galhbtypes}. We calculated HB indices
for six control clusters, eleven new clusters, and the Sagittarius cluster NGC 4147 (see Section 
\ref{sss:hbmextra}). In the main, the eleven new clusters measured here all lie on lines of sight 
towards the Galactic bulge, and are therefore quite heavily obscured ($0.5 \le E(B-V) \le 1.0$ for many). 
Our calculated HB types are thus best regarded as estimates; however, due to the resolution of WFPC2, they 
are certainly far more accurate than any measurements available in the literature, and suitable for 
distinguishing between membership in the old and young halo sub-systems. The results for the control clusters
show that our procedure allows HB types to be derived which are entirely consistent with those in
the literature (in general to $\pm 0.04$ or better). As with the external clusters, variation in the 
position of the edges of the instability strip might be expected with changing metallicity; however, we 
again find no evidence that this has introduced any serious systematic uncertainty into the measurements. 
There is perhaps a hint that HB indices measured for those clusters with $[$Fe$/$H$] > -1.2$ are too red by 
$0.04-0.09$ (see the results for NGC 362 and 1851). Only three of the eleven new clusters have such high 
metallicities, and in none of these cases would a change in HB index of even $0.1$ to the red change their 
classification. We are therefore confident in our assignation of the eleven new clusters to their respective
sub-systems according to the method described above.

\begin{table*}
\begin{minipage}{150mm}
\begin{center}
\caption{HB indices for $18$ globular clusters from the $(B,\,B-V)$ WFPC2 CMDs of Piotto et al. \shortcite{piotto:02}.}
\begin{tabular}{@{}lccccccccc}
\hline \hline
Cluster & \hspace{3mm} & Reddening & Metallicity & $N_{RHB}$ & $N_{VHB}$ & $N_{BHB}$ & Calculated & Literature & Type \\
Name & & $E(B-V)$ & $[$Fe$/$H$]$ & $(R)$ & $(V)$ & $(B)$ & HB Index & HB Index & \\
\hline
\vspace{-3mm} & & & & & & & & & \\
NGC 362  & & $0.05$ & $-1.16$ & $214 \pm 20$ & $31 \pm 4$ & $13 \pm 6$   & $-0.78 \pm 0.05$ & $-0.87$ & Control \\
NGC 1851 & & $0.02$ & $-1.22$ & $178 \pm 15$ & $18 \pm 4$ & $87 \pm 14$  & $-0.32 \pm 0.07$ & $-0.36$ & Control \\
NGC 3201 & & $0.23$ & $-1.58$ & $8 \pm 2$    & $6 \pm 1$  & $10 \pm 2$   & $0.08 \pm 0.12$  & $0.08$  & Control \\
NGC 5024 & & $0.02$ & $-1.99$ & $13 \pm 3$   & $12 \pm 2$ & $201 \pm 10$ & $0.83 \pm 0.03$  & $0.81$  & Control \\
NGC 7078 & & $0.10$ & $-2.26$ & $31 \pm 5$   & $61 \pm 7$ & $293 \pm 20$ & $0.68 \pm 0.03$  & $0.67$  & Control \\
NGC 7089 & & $0.06$ & $-1.62$ & $3 \pm 3$    & $8 \pm 1$  & $163 \pm 10$ & $0.92 \pm 0.03$  & $0.96$  & Control\vspace{1mm} \\
\hline
\vspace{-3mm} & & & & & & & & & \\
NGC 5634 & & $0.05$ & $-1.88$ & $2 \pm 2$    & $9 \pm 2$   & $132 \pm 7$  & $0.91 \pm 0.03$  & $-$ & OH \\
NGC 5946 & & $0.54$ & $-1.38$ & $14 \pm 6$   & $12 \pm 4$  & $101 \pm 10$ & $0.69 \pm 0.09$  & $-$ & OH \\
NGC 6273 & & $0.41$ & $-1.68$ & $7 \pm 3$    & $1 \pm 1$   & $393 \pm 15$ & $0.96 \pm 0.02$  & $-$ & OH \\
NGC 6284 & & $0.28$ & $-1.32$ & $8 \pm 4$    & $1 \pm 1$   & $133 \pm 11$ & $0.88 \pm 0.05$  & $-$ & OH \\
NGC 6325 & & $0.89$ & $-1.17$ & $1 \pm 1$    & $7 \pm 3$   & $49 \pm 12$  & $0.84 \pm 0.06$  & $-$ & OH \\
NGC 6355 & & $0.75$ & $-1.50$ & $8 \pm 2$    & $12 \pm 3$  & $53 \pm 10$  & $0.62 \pm 0.07$  & $-$ & OH \\
NGC 6401 & & $0.72$ & $-0.98$ & $15 \pm 3$   & $10 \pm 4$  & $37 \pm 10$  & $0.35 \pm 0.12$  & $-$ & OH \\
NGC 6453 & & $0.66$ & $-1.53$ & $10 \pm 5$   & $5 \pm 2$   & $142 \pm 15$ & $0.84 \pm 0.06$  & $-$ & OH \\
NGC 6517 & & $1.08$ & $-1.37$ & $12 \pm 4$   & $44 \pm 12$ & $121 \pm 20$ & $0.62 \pm 0.07$  & $-$ & OH \\
NGC 6569 & & $0.55$ & $-0.86$ & $165 \pm 25$ & $8 \pm 3$   & $13 \pm 3$   & $-0.82 \pm 0.04$ & $-$ & OH \\
NGC 6642 & & $0.41$ & $-1.35$ & $23 \pm 3$   & $12 \pm 4$  & $21 \pm 7$   & $-0.04 \pm 0.14$ & $-$ & YH\vspace{1mm} \\
\hline
\vspace{-3mm} & & & & & & & & & \\
NGC 4147 & & $0.02$ & $-1.83$ & $2 \pm 2$ & $8 \pm 2$ & $43 \pm 4$ & $0.77 \pm 0.07$ & $0.55$ & Sgr\vspace{1mm} \\
\hline
\label{t:galhbtypes}
\end{tabular}
\end{center}
\end{minipage}
\end{table*}

We searched the literature for observations of the remaining $15$ clusters from the Harris catalogue
without HB index measurements, obtaining the following results:

$\mit{\omega}$ {\it Cen:} (NGC 5139) This is the most massive of the Galactic globular clusters
and is unique for many reasons -- one being the wide spread in metallicity of its member stars
($-1.8 \le [$Fe$/$H$] \le -0.6$) (see e.g., Platais et al. \shortcite{platais:03} and references therein).
The HB morphology is correspondingly complicated, with a large number of variable stars and
an extremely extended blue tail. It has been suggested that $\omega$ Cen is the surviving nucleus of
a dwarf galaxy accreted and stripped by the Galaxy (see e.g., Bekki \& Freeman \shortcite{bekki:03}, and
references therein). If this hypothesis is correct, $\omega$ Cen is not a globular cluster in the normal 
sense, and it is not appropriate to classify it according to the present criteria. We omit this object 
from the remainder of our discussion.

{\it NGC 6540:} Bica, Ortolani \& Barbuy \shortcite{bica:94} show a sparse CMD for this reddened cluster, 
which lies along a line of sight towards the Galactic centre. The HB morphology is quite blue, but with 
some red stars, and quite similar to that of M5, NGC 288, and NGC 6752 -- all old halo members. Bica et al. 
\shortcite{bica:94} calculate that $\frac{B}{(B+R)} = 0.65 \pm 0.15$. If the number of HB variable stars 
is assumed to be small (or zero) then this may be rearranged to show that 
$\frac{(B-R)}{(B+V+R)} = 0.30 \pm 0.30$, which places NGC 6540 also in the old halo group. This type of 
HB morphology is compatible with the very blurred CMD of Piotto et al. \shortcite{piotto:02}, which clearly 
does not show a HB concentrated to either the red or blue.

{\it NGC 6558:} This cluster also lies towards the Galactic bulge, and again suffers from
reasonably severe obscuration. Rich et al. \shortcite{rich:98} present a CMD, showing NGC 6558 to have a 
predominantly blue HB similar to that for HP 1, NGC 6522, and NGC 6540, which are all old halo clusters. 
From their CMD it is possible to estimate that the HB type $\sim 0.7$ $(3R,\,4V,\,27B)$, which makes
NGC 6558 also an old halo member.

{\it Terzan 1:} A {\em HST} WFPC2 CMD for this cluster has been published by Ortolani et al.
\shortcite{ortolani:99a}, which shows Terzan 1 to be a relatively metal poor cluster 
($[$Fe$/$H$] \sim -1.3$ according to Harris \shortcite{harris:96}) with a completely
red HB morphology (HB type $= -1.00$). This renders Terzan 1 unusual in that it is a second parameter
(young halo) cluster physically located in the Galactic bulge.

{\it Terzan 4:} This is another heavily reddened globular cluster located in the Galactic bulge. 
Ortolani et al. \shortcite{ortolani:01} observed Terzan 4 with {\em HST} NICMOS and provide a CMD 
which shows it to be a metal poor cluster with an exclusively blue HB morphology. Thus, Terzan 4 is an
old halo cluster located towards the bulge.

{\it Terzan 9:} Ortolani, Bica \& Barbuy \shortcite{ortolani:99b} present $V$ and $I$ photometry for 
this obscured cluster projected on the Galactic bulge. From their CMD it is possible to estimate that 
the HB type $\sim 0.25$ $(9R,\,6V,\,17B)$, meaning that Terzan 9 is most likely a young halo cluster.

{\it Pal. 2:} In contrast, this cluster lies away from the bulge, towards the Galactic 
anti-centre. Because it is still in the plane of the disk however, it suffers heavy extinction. Harris
et al. \shortcite{harris:97} observed Pal. 2, finding it to be an outer halo cluster with a well populated
RHB and an extension to the blue, similar to the horizontal branches of NGC 1261, NGC 1851, and 
NGC 6229. Two of these clusters are young halo members (NGC 1261 and NGC 6229), while the other is 
(marginally) an old halo member. From the CMD of Harris et al. \shortcite{harris:97}, it is possible to 
estimate that the HB type $\sim -0.1$ $(38R,\,27V,\,28B)$ for Pal. 2, which places it (just) as a 
young halo member, consistent with the other similar clusters.

{\it Djorg 1:} No CMDs for this bulge cluster sufficient for obtaining a HB type are available. Ortolani,
Bica \& Barbuy \shortcite{ortolani:95a} suggest (from a very sparse CMD) that Djorg 1 is a relatively 
metal rich cluster ($[$Fe$/$H$] \sim -0.4$) with a red HB morphology -- this would make it a bulge/disk 
member. The spectroscopy of Bica et al. \shortcite{bica:98} also suggests that Djorg 1 is metal rich. 
However, C\^ot\'e \shortcite{cote:99} shows Djorg 1 to have a very high radial velocity which would place 
it in the outer halo at apo-Galacticon, while the infra-red CMD of Davidge \shortcite{davidge:00} suggests 
an extremely metal poor cluster. Given these contradictions, we are at present unable to classify this
cluster.

{\it E3:} This cluster is sparsely populated, and relatively metal rich ($[$Fe$/$H$] \sim -0.8$) -- on 
the border of being assigned as a bulge/disk member (although it has a large Galactocentric distance 
($7.6$ kpc) for such a cluster). Three CMDs are available in the literature -- those of 
McClure et al. \shortcite{mcclure:85}, Gratton \& Ortolani \shortcite{gratton:87}, and Mochejska,
Kaluzny \& Thompson \shortcite{mochejska:00}. However, none of these show any HB stars, so it is 
impossible to classify E3 for the purposes of the present discussion.

{\it AM 4:} Inman \& Carney \shortcite{inman:87} have studied this halo cluster, and show it to be 
probably the most under-populated of all known Galactic globular clusters. In fact, it is so sparse that, 
like E3, no HB stars appear on its CMD. Hence, we also cannot classify this cluster.

{\it HP 1:} This is another obscured cluster projected towards the Galactic bulge. Ortolani, Bica \&
Barbuy \shortcite{ortolani:97} provide $V$ and $I$ photometry, and from this it is possible to estimate a
relatively blue HB type $\sim 0.75$ $(2R,\,7V,\,33B)$, placing HP 1 in the old halo ensemble.

{\it 1636-283:} (ESO452-SC11) This is a reasonably obscured, poorly populated cluster, which has been 
studied by Bica, Ortolani \& Barbuy \shortcite{bica:99a}. From their CMDs, a red HB index $\sim -0.4$ 
$(6R,\,5V,\,1B)$ may be estimated. This renders 1636-283 a young halo member.

{\it ESO280-SC06:} This is a newly discovered, very sparse, halo globular cluster. Ortolani et al.
\shortcite{ortolani:00} present a CMD; however, similarly to E3 and AM 4, there are very few HB stars 
indeed -- not sufficient for determining a HB type. 

{\it 2MASS-GC01:} No suitable CMDs exist to determine the HB type of this newly discovered
bulge cluster.

{\it 2MASS-GC02:} Similarly, no suitable CMDs are available for observing the HB type of 
this newly discovered cluster.

\begin{figure}
\includegraphics[width=0.5\textwidth]{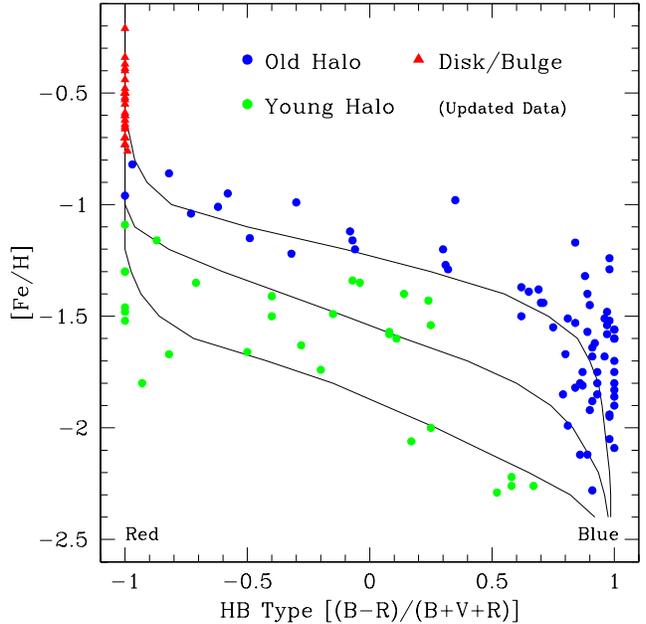}
\caption{HB-type vs. metallicity diagram just as Fig. \ref{f:hbmetorig}, but now including an additional $29$ clusters with updated data, as described in the text. These take the total sample to $137$ -- all but $7$ of the known Galactic globular clusters. The isochrones are as in Fig. \ref{f:hbmetorig}.}
\label{f:hbmetupdated}
\end{figure}

Having exhausted the available literature, the classification is now as complete as possible. 
The bulge/disk sample contains $37$ clusters; the old halo sample contains $70$ clusters; and the
young halo sample contains $30$ clusters. Only seven clusters suffer from incomplete data or
could not be classified ($\omega$ Cen, E3, AM 4, Djorg 1, ESO280-SC06, 2MASS-GC01, and
2MASS-GC02). The HB-type vs. metallicity diagram is thus considerably filled out, as shown in
Fig. \ref{f:hbmetupdated}. 

\begin{figure*}
\begin{minipage}{175mm}
\begin{center}
\includegraphics[width=58.4mm]{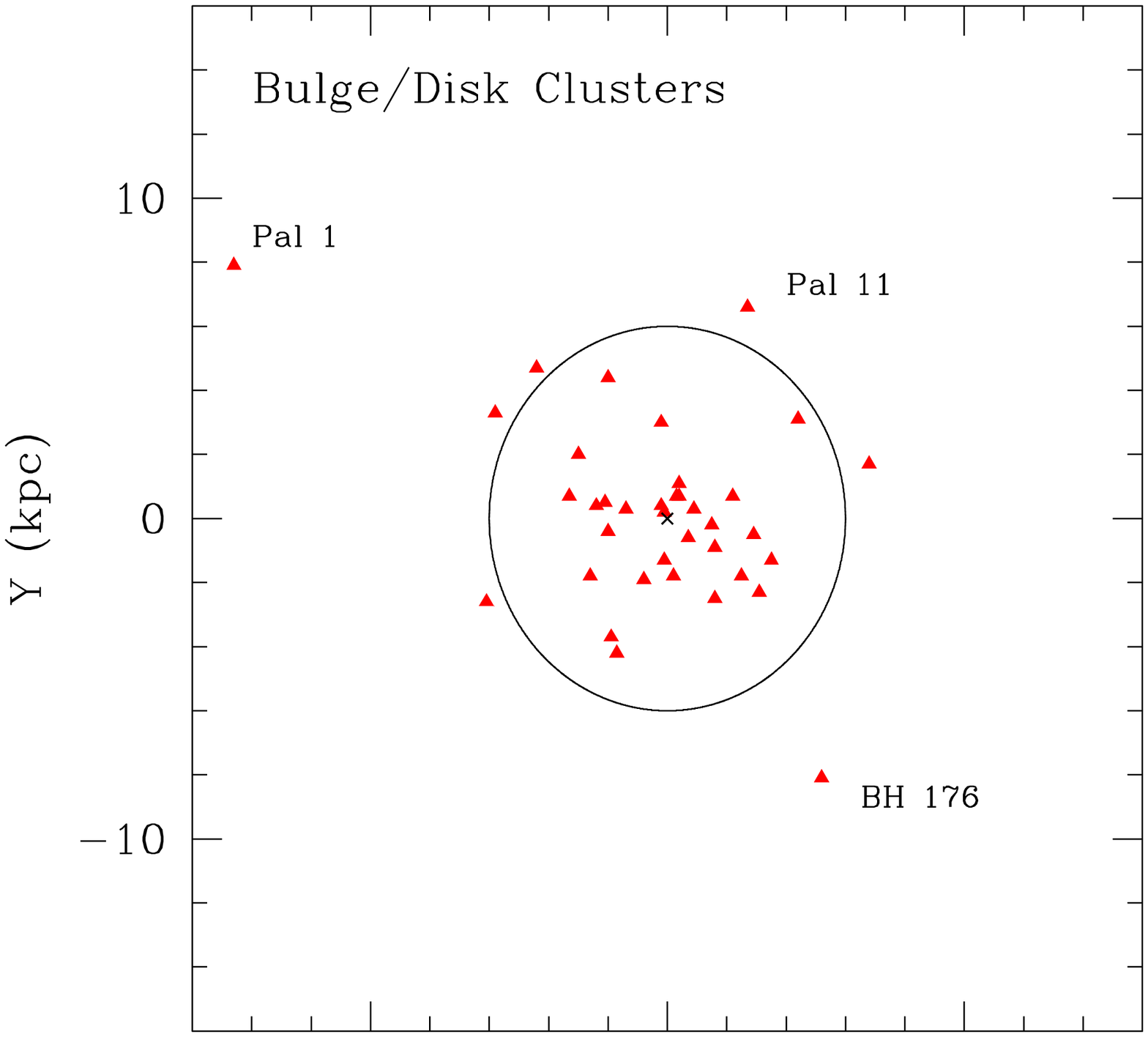}
\hspace{-2mm}
\includegraphics[width=58.4mm]{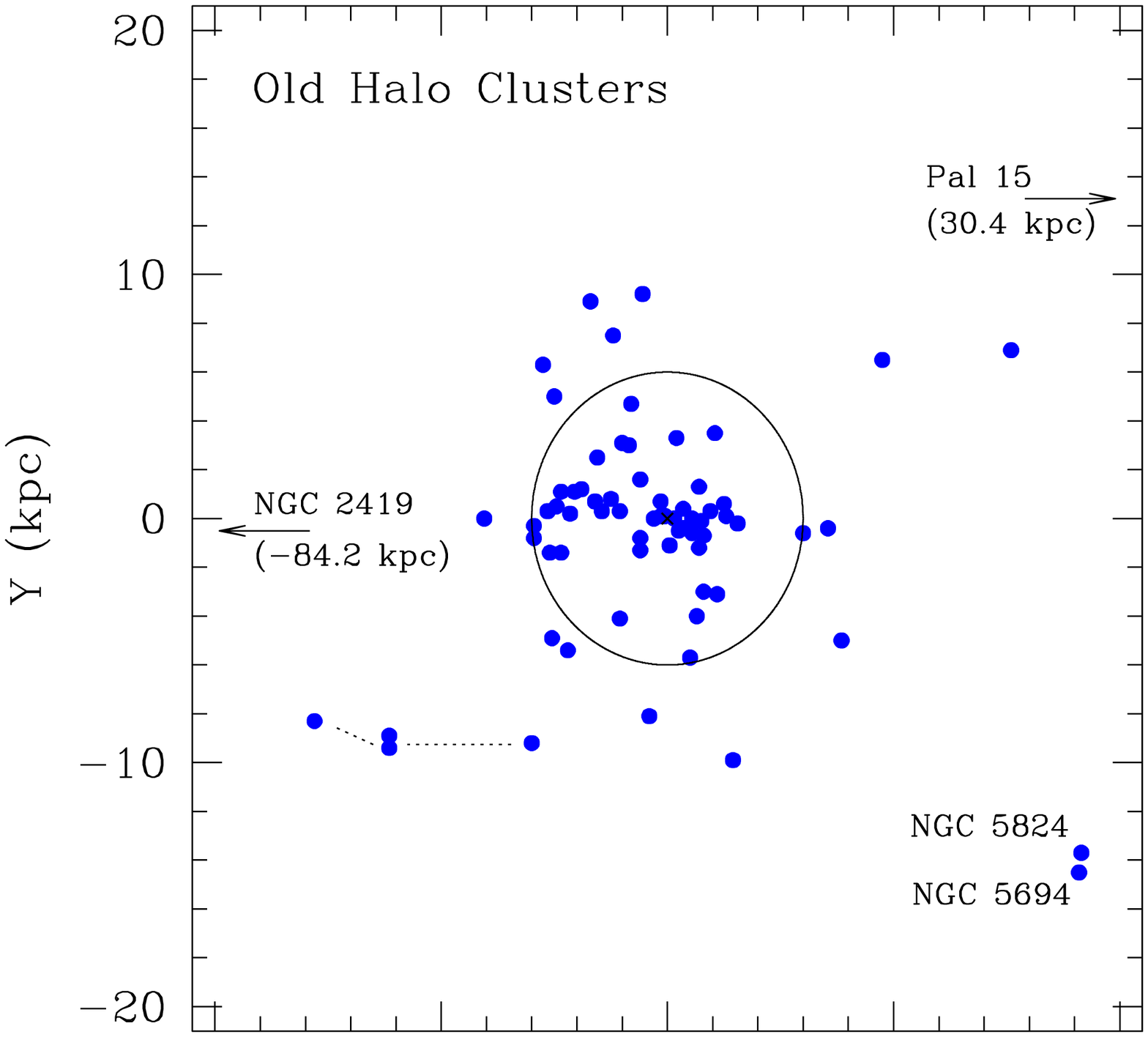}
\hspace{-2mm}
\includegraphics[width=58.4mm]{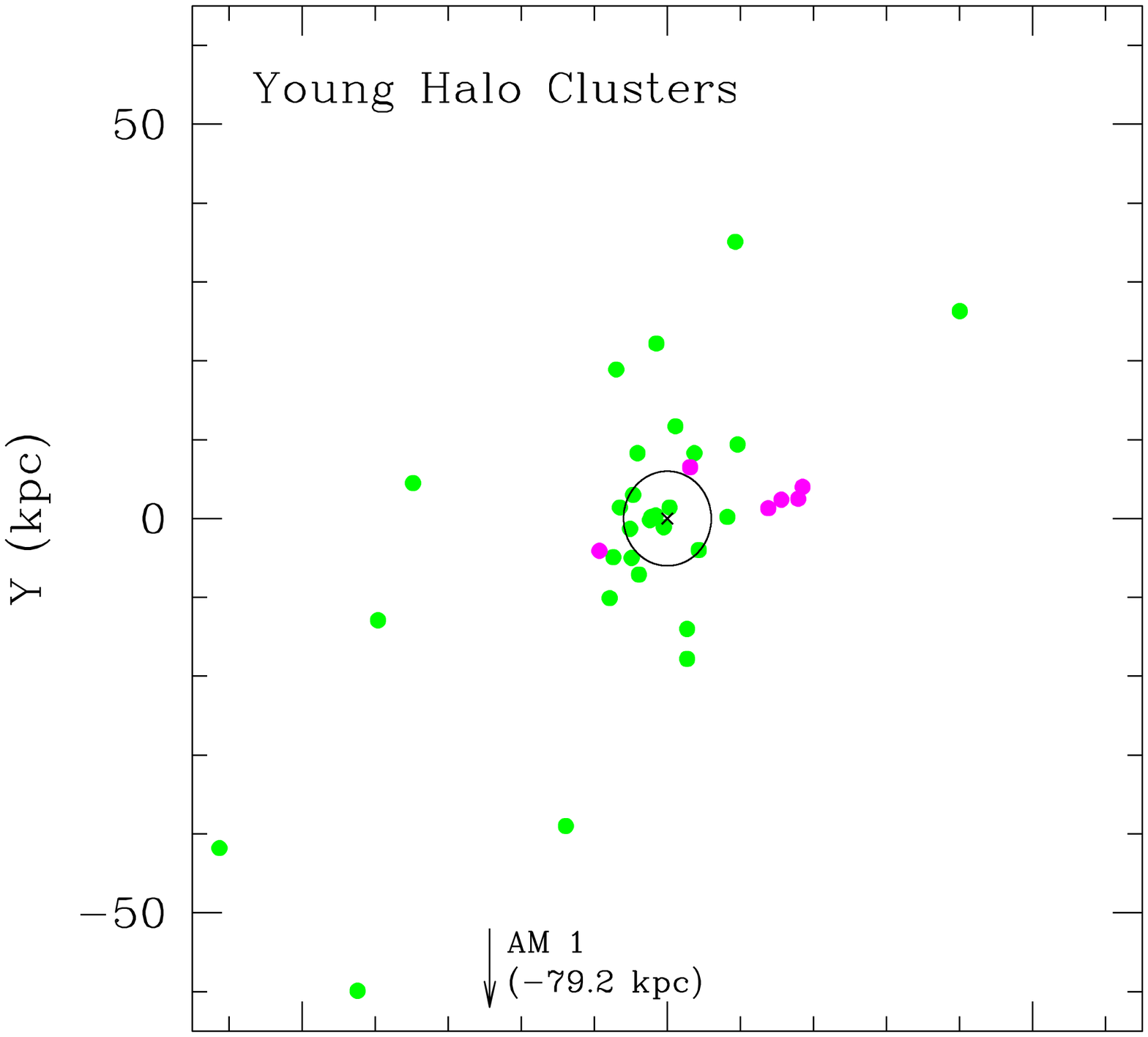} \\
\vspace{-9.8mm}
\includegraphics[width=58.4mm]{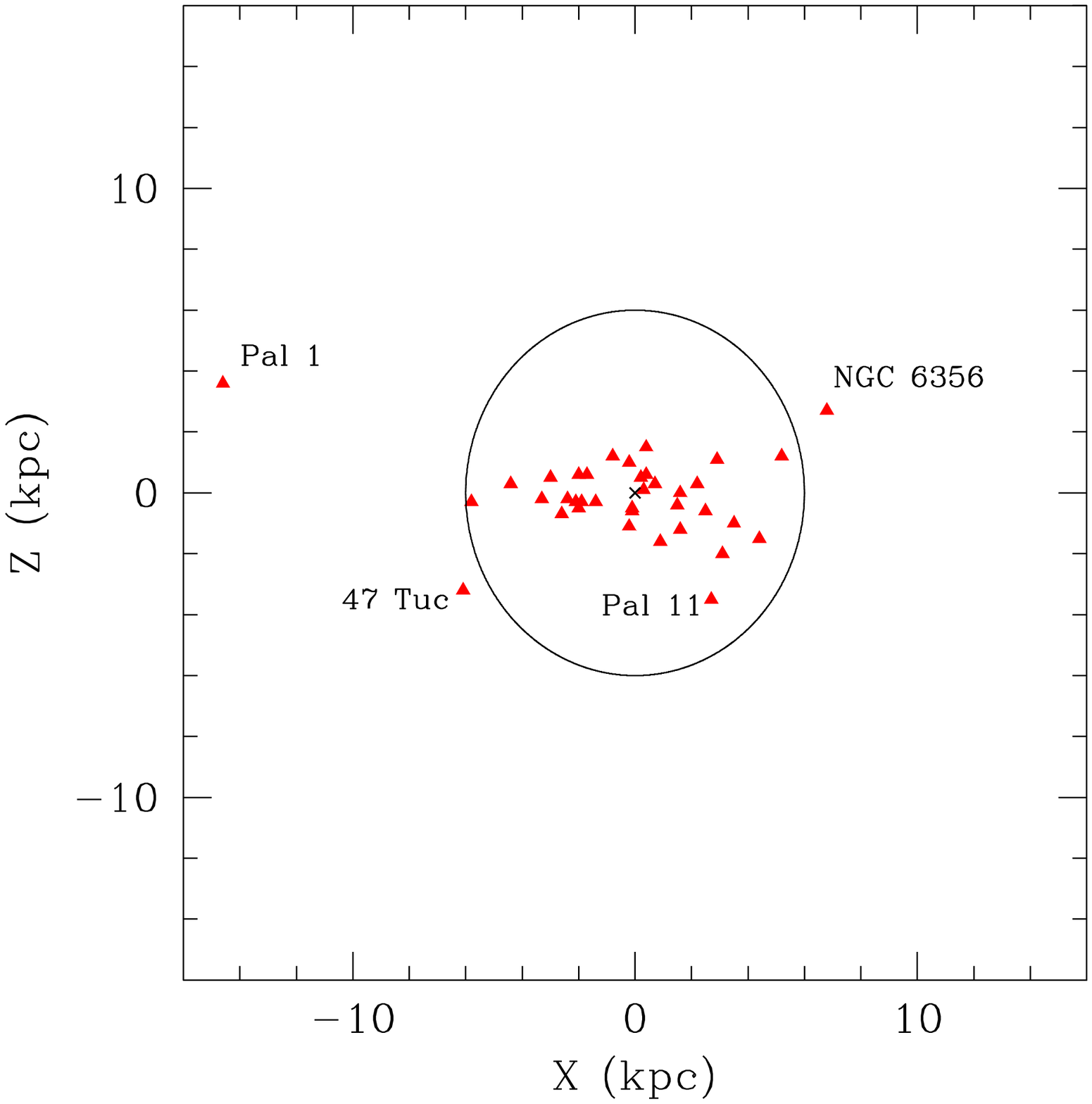}
\hspace{-2mm}
\includegraphics[width=58.4mm]{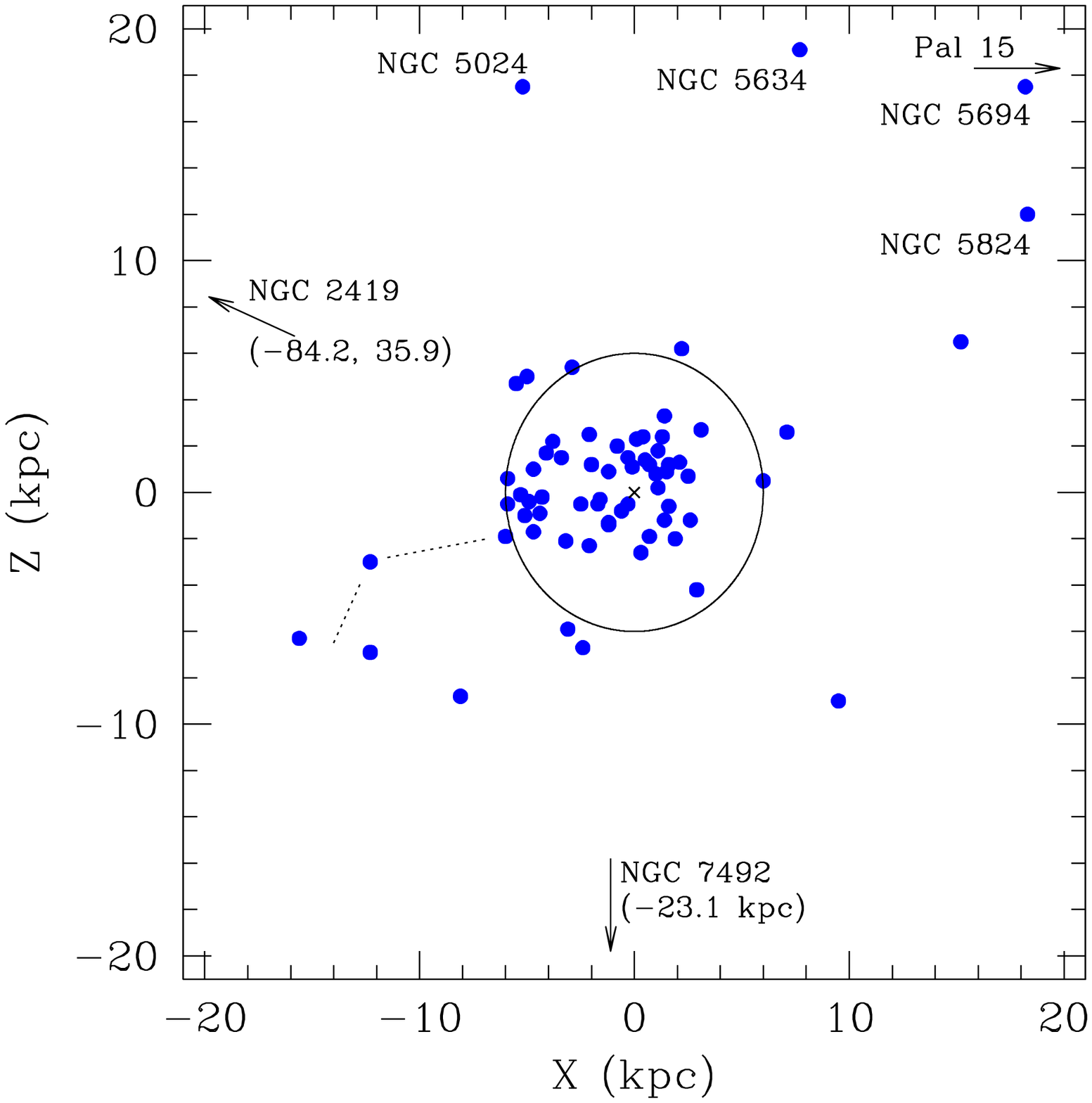}
\hspace{-2mm}
\includegraphics[width=58.4mm]{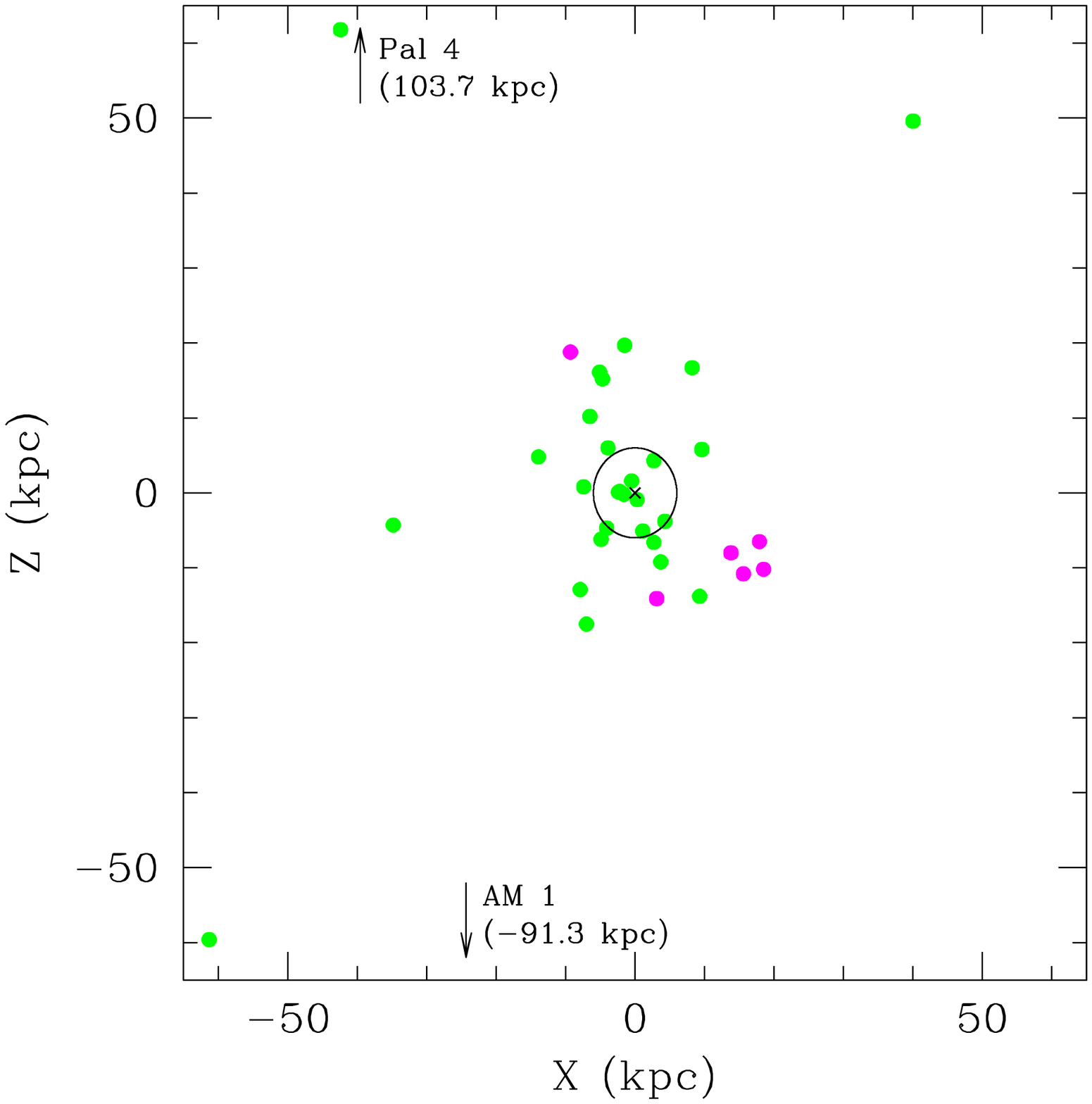}
\caption{Spatial positions of globular clusters in each of the Galactic sub-systems, as labelled, in Galactocentric cartesian coordinates. In this system the sun is at $(X,Y,Z) = (-8,0,0)$. In each diagram a circle of radius $6$ kpc is marked. This helps give an indication of the relative volumes occupied by the three systems. Clusters discussed at various points in the text are labelled, as are those objects which fall outside the range of a given plot. The six Sagittarius clusters are marked in magenta in the young halo diagram, while the four clusters linked with the Canis Major dwarf -- NGC 1851, 1904, 2298, and 2808 (Martin et al. 2004) -- are joined with a dotted line in the old halo plot.}
\label{f:clusterpos}
\end{center}
\end{minipage}
\end{figure*}

Finally, we are also interested in cluster structures. Nine of the $144$ clusters do not have core radius 
measurements listed in the Harris catalogue. Of these, we located suitable data in the literature for five. 
Four of these measurements (those for Terzan 4, Terzan 10, 1636-283 (also known as ESO452-SC11), and BH 176) 
come from the compilation of Webbink \shortcite{webbink:85}. These data are listed as angular sizes, so 
we converted them to linear sizes (in pc) using the distances from the Harris catalogue. The fifth 
measurement is for IC 1257, and comes from C\^ot\'e \shortcite{cote:99}, who lists this cluster as 
having $r_c \sim 0.05\arcmin$. Of the remaining four clusters, Lyng\aa\ 7 has no available structural 
measurements, while the other three are very newly discovered (and hence have no structural measurements 
as yet): ESO280-SC06 \cite{ortolani:00} and two from the 2MASS survey -- 2MASS-GC01 and 2MASS-GC02 
\cite{hurt:00}. 

In total therefore, we have core radius measurements for $140$ of the $144$ Galactic globular clusters
listed by Harris, and classifications on the HB-type vs. metallicity diagram for $137$ of the $144$ clusters.
The union of the two sets consists of $136$ clusters (the excluded objects are Lyng\aa\ 7, which has 
HB type and metallicity measurements but no core radius measurement, and $\omega$ Cen, E3, AM 4, and Djorg. 1,
which have core radius measurements, but insufficient data to be classified). 

\subsection{Global characteristics of the sub-systems}
\label{ss:characteristics}
The present data set is larger than any which could be located in the literature. In addition, new data 
concerning the ages and kinematics of reasonably large samples of globular clusters have recently been
published (see e.g., Rosenberg et al. \shortcite{rosenberg:99}; Dinescu et al. \shortcite{dinescu:99};
Salaris \& Weiss \shortcite{salaris:02}). It is therefore useful to briefly consider the global 
characteristics of each of the Galactic sub-systems.

We first examine the spatial distribution of the three samples. Previous authors (see e.g., Zinn 
\shortcite{zinn:93a,zinn:96}) have found that the metal-rich clusters are concentrated towards the Galactic 
bulge and disk, while the old halo clusters are most prevalent in the inner halo of the Galaxy (all halo
clusters in Zinn's \shortcite{zinn:93a} sample with $R_{\rm{gc}} < 6.0$ kpc are old halo clusters, as are 
$55$ per cent of the clusters with $6.0 < R_{\rm{gc}} < 40.0$ kpc) and the young halo clusters predominantly 
occupy the remote halo (no young halo clusters in Zinn's sample have $R_{\rm{gc}} < 6.0$, while $5$ of the 
$6$ clusters with $R_{\rm{gc}} > 40.0$ are young halo objects). Equivalent spatial distributions are
observed by Van den Bergh \shortcite{vdb:93} for similar sub-groups.

Similar results hold for the present (larger) data set. We have plotted spatial positions for our
three sub-systems in Fig. \ref{f:clusterpos}, using the values of $X$, $Y$, and $Z$ listed in the 
Harris catalogue. Marked on each plot is a circle of Galactocentric radius 
$R_{\rm{gc}} = 6$ kpc, for use as a reference distance. It is clear that the metal-rich clusters are 
strongly concentrated towards the Galactic centre. All but a couple lie very close to or within the $6$
kpc radius. In fact, $60$ per cent of the metal-rich clusters have $R_{\rm{gc}} < 3.0$ kpc (as listed by 
Harris), while a further $20$ per cent have $3.0 < R_{\rm{gc}} < 5.0$. These objects also form a very 
flattened system. All $37$ objects in the sample have $\vert Z \vert < 3.6$ (where $Z$ is the distance 
above or below the Galactic plane), and in fact all but $4$ have $\vert Z \vert < 2.0$. The innermost 
metal-rich clusters form the major component of the cluster group associated with the Galactic bulge, while 
those with $R_{\rm{gc}} > 4.0$ kpc form a system which can be associated with the Galactic thick disk. The 
approximate scale heights match (Gilmore \& Reid \shortcite{gilmore:83} found a thick disk scale height of 
$\sim 1.5$ kpc), and a sub-sample of the present group of clusters has been shown (e.g., Armandroff 
\shortcite{armandroff:89}) to have disk-like kinematics -- a high rotational velocity ($v_{rot} \sim 190$ 
km s$^{-1}$) and a relatively small line-of-sight velocity dispersion ($\sigma_{los} \sim 60$ km s$^{-1}$).

As with Zinn's \shortcite{zinn:93a} old halo group, the majority of our old halo clusters ($65$ per cent)
reside within $6$ kpc of the Galactic centre. It is interesting to note that the distribution of old halo 
clusters continues to very small Galactocentric radii, so that the old halo physically overlaps with the 
bulge/disk region. Fig. \ref{f:clusterpos} shows some evidence that the inner region of the old halo is
somewhat flattened towards the Galactic plane, a result also noted by Zinn \shortcite{zinn:93a}. 
The remaining old halo clusters all have $R_{\rm{gc}} < 30$ kpc, except for Pal. 15 at $37.9$ kpc and NGC 2419
at $91.5$ kpc. As is clear from Fig. \ref{f:clusterpos}, the outer old halo objects have a rather
asymmetric distribution in terms of their $Z$ coordinates -- $6$ clusters lie high above the Galactic plane
with $Z > 10$ kpc, but only $1$ lies far below the plane with $Z < -10$ kpc. 

Our young halo clusters are somewhat different to Zinn's \shortcite{zinn:93a} sample in that they extend 
into the bulge region, with five clusters ($15$ per cent) having $R_{\rm{gc}} < 6.0$ kpc (in fact all five lie 
within $2.5$ kpc)\footnote{The identification of these five clusters (Pal. 6, NGC 6642, 1636-283, Terzan 1, 
and Terzan 9) as young halo objects has used observations from after 1996}. Apart from this, our classification
is consistent with that of Zinn. The remainder of young halo objects have $R_{\rm{gc}} > 7.0$ kpc, with 
$20$ lying between this radius and $40$ kpc, and another $6$ beyond $40$ kpc (the outermost being at AM 1 at 
$123$ kpc). It is interesting to note that like the outer old halo objects, the outer young halo clusters also 
show an asymmetry in their distribution -- this time in the projection onto the $XY$ plane. This distribution 
appears elongated from the lower left corner to the upper right corner of the diagram. No clusters fall in 
the upper left or lower right corners. The inner objects show a spherical distribution on this plane, 
however, as do all the young halo clusters in the $XZ$ projection. It is also interesting to note that the
six Sagittarius clusters, which have been placed on the young halo plot (since they are mostly outer halo 
objects), appear consistent with the distribution of young halo clusters.

The global kinematics of the old and young halo groups (unsurprisingly) display halo-like properties -- the 
summary by Zinn \shortcite{zinn:93a} shows that his old halo (which is essentially a subset of the present
old halo system) has some mean rotation ($v_{rot} \sim 70$ km s$^{-1}$) and an intermediate line-of-sight 
velocity dispersion ($\sigma_{los} \sim 90$ km s$^{-1}$), while his young halo (again, a subset of the
present young halo system) has rotation consistent with zero (although there is the possibility of a small 
{\em retrograde} mean rotation) but a large dispersion ($\sigma_{los} \sim 150$ km s$^{-1}$). Van den Bergh
\shortcite{vdb:93} examined three similar sub-groups and found that the second parameter (outer halo) objects
typically exhibit retrograde and plunging orbits. We will consider more specific kinematic measurements below.

\begin{figure}
\includegraphics[width=0.5\textwidth]{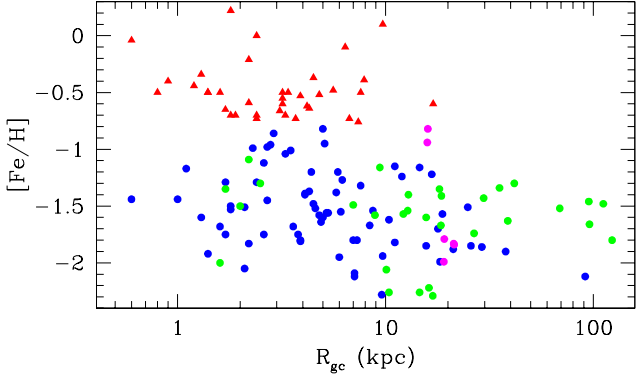}
\caption{Metallicity versus Galactocentric radius for the three Galactic globular cluster sub-systems, plus the Sagittarius clusters. Measurements are taken from the Harris (1996) database, except for the Sagittarius clusters, which have measurements compiled in Tables \ref{t:hbresults} and \ref{t:spfreq}. The bulge/disk clusters are solid red triangles, while the old halo clusters are blue dots, the young halo clusters green dots, and the Sagittarius clusters magenta dots.}
\label{f:metrad}
\end{figure}

Fig. \ref{f:metrad} shows cluster metallicity as a function of Galactocentric radius. As demonstrated
above, the metal-rich clusters are confined to relatively small radii. The old halo objects exhibit a
considerable amount of scatter, but apparently show a weak anti-correlation between metal abundance and 
Galactocentric radius. This matches a result of Zinn \shortcite{zinn:93a}. Crudely binning these clusters, 
we find $\langle [$Fe$/$H$] \rangle = -1.42$ for the $42$ objects with $R_{\rm{gc}} < 6$ kpc; 
$\langle [$Fe$/$H$] \rangle = -1.67$ for those with $6 \le R_{\rm{gc}} < 15$ kpc ($17$ clusters); and
$\langle [$Fe$/$H$] \rangle = -1.77$ for the $11$ clusters with $R_{\rm{gc}} \ge 15$ kpc. In contrast,
the (smaller) young halo sample shows no evidence of such a trend, with $\langle [$Fe$/$H$] \rangle = -1.61$ 
for the $14$ clusters lying within $R_{\rm{gc}} = 15$ kpc, and $\langle [$Fe$/$H$] \rangle = -1.62$ for the
remaining $16$ objects outside this radius. It is interesting to note that van den Bergh \shortcite{vdb:95}
found cluster abundances to correlate in general more strongly with their peri-Galactic radii than their 
present Galactocentric distances. Four of the Sagittarius clusters fall close together on Fig. \ref{f:metrad}, 
while the remaining two (Terzan 7 and Pal. 12) have a somewhat high metallicity for their spatial 
position. As noted by van den Bergh \& Mackey \shortcite{vdb:04}, Pal. 1 has a similarly high metallicity
for its Galactocentric radius.

We also briefly consider the populations of RR Lyrae stars in each of the three cluster sub-systems,
as indicated by the specific frequency, $S_{\rm{RR}}$. In Section \ref{ss:rrlyr} we demonstrated that,
compared with the Galactic globular clusters, the globular clusters in nearby dwarf galaxies generally
have large values of $S_{\rm{RR}}$. Examining the three present sub-systems, the vast majority of
Galactic clusters with large $S_{\rm{RR}}$ fall in the young halo sample. The $27$ clusters in this ensemble
with measured values of $S_{\rm{RR}}$ in the Harris catalogue have $\langle S_{\rm{RR}} \rangle = 35.3$,
compared with $\langle S_{\rm{RR}} \rangle = 8.1$ for the $64$ old halo clusters with measured values,
and $\langle S_{\rm{RR}} \rangle < 0.5$ for the bulge/disk clusters. Five of the $27$ measured young halo
clusters ($18.5$ per cent) have $S_{\rm{RR}} > 50$, while only two of the $64$ measured old halo clusters
do ($3$ per cent). Clearly the young halo clusters strongly resemble the external clusters with respect
to their normalized RR Lyrae populations.

\begin{figure}
\includegraphics[width=0.5\textwidth]{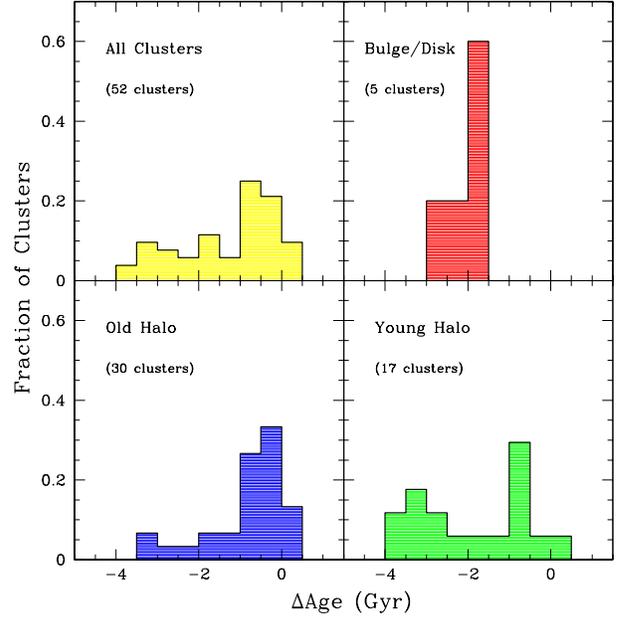}
\caption{Histograms of ages for $52$ Galactic globular clusters, as labelled. The ages are taken from Salaris \& Weiss (2002), calculated relative to the age of M92 as described in the text.}
\label{f:agehist}
\end{figure}

Several recent papers have addressed the question of globular cluster relative ages, and it is
useful to examine these results in the context of the sub-systems described here. The largest sample of
relative ages has been derived by Salaris \& Weiss \shortcite{salaris:02}, for $55$ clusters. Of these, $5$ 
are in our bulge/disk group, $17$ in our young halo, and $30$ in our old halo (the remaining three are the 
Sagittarius clusters Arp 2, Terzan 7, and Pal. 12). Histograms of cluster ages are plotted in Fig.
\ref{f:agehist}. In this Figure we consider the quantity $\Delta$Age, defined as the age relative to
M92 (NGC 6341), which is usually considered to be one of the oldest clusters in the Galactic halo. Salaris
\& Weiss \shortcite{salaris:02} calculate two ages per cluster -- one for each of the metallicity scales
they consider. Here, we assign a cluster's age by calculating the mean of these two results -- in the majority
of cases the difference between them is small. By this method, we assign an age of $12.55$ Gyr to M92.

Examining Fig. \ref{f:agehist} it is clear that the three cluster sub-systems have quite distinct age
distributions. The bulge/disk clusters are confined in a narrow peak at $\Delta$Age\ $\sim -2$ Gyr.
The oldest cluster in this group has $\Delta$Age\ $= -1.85$ Gyr, while the youngest has 
$\Delta$Age\ $= -2.75$ Gyr. The old halo group is also strongly peaked, but at ages essentially coeval
with M92. However, there is also clearly a tail to much younger ages in this sample. If we consider only
clusters in the peaked region of the distribution (i.e., the $24$ clusters with $\Delta$Age\ $\ge -1.5$ Gyr)
we find $\langle \Delta$\ Age$\rangle = -0.4$ Gyr, with a narrow dispersion of only $0.45$ Gyr. The
remaining $6$ clusters, with $\Delta$Age\ $< -1.5$, have $\langle \Delta$\ Age$\rangle = -2.5$ Gyr. It is
interesting to note that two of the three youngest clusters in the group, NGC 1851 and 2808 (with ages
of $\Delta$Age\ $= -3.4$ Gyr and $\Delta$Age\ $= -2.8$ Gyr, respectively) have been suggested as
members of the recently discovered disrupted dwarf galaxy in Canis Major \cite{martin:04}. The other two
prospective members of this galaxy (NGC 1904 and 2298) also lie in the old halo group, but fall in the
older peak with ages of $\Delta$Age\ $= -0.5$ Gyr and $\Delta$Age\ $ = +0.2$ Gyr, respectively.
The young halo clusters show an apparently bimodal distribution in cluster age, with one peak at 
$\Delta$Age\ $\sim -0.75$ Gyr and the other at $\Delta$Age\ $\sim -3.25$ Gyr. The youngest object
in this group is the Eridanus cluster with $\Delta$Age\ $= -3.9$ Gyr. The mean age of the full young halo
sample is $\Delta$Age\ $= -1.9$ Gyr, but with a large dispersion of $1.35$ Gyr. It is interesting to
note that the mean age of the young tail of the old halo sample matches well the ages of the bulge/disk 
clusters, but also matches closely the ages of typical young halo clusters. 

\begin{figure}
\begin{center}
\includegraphics[width=0.5\textwidth]{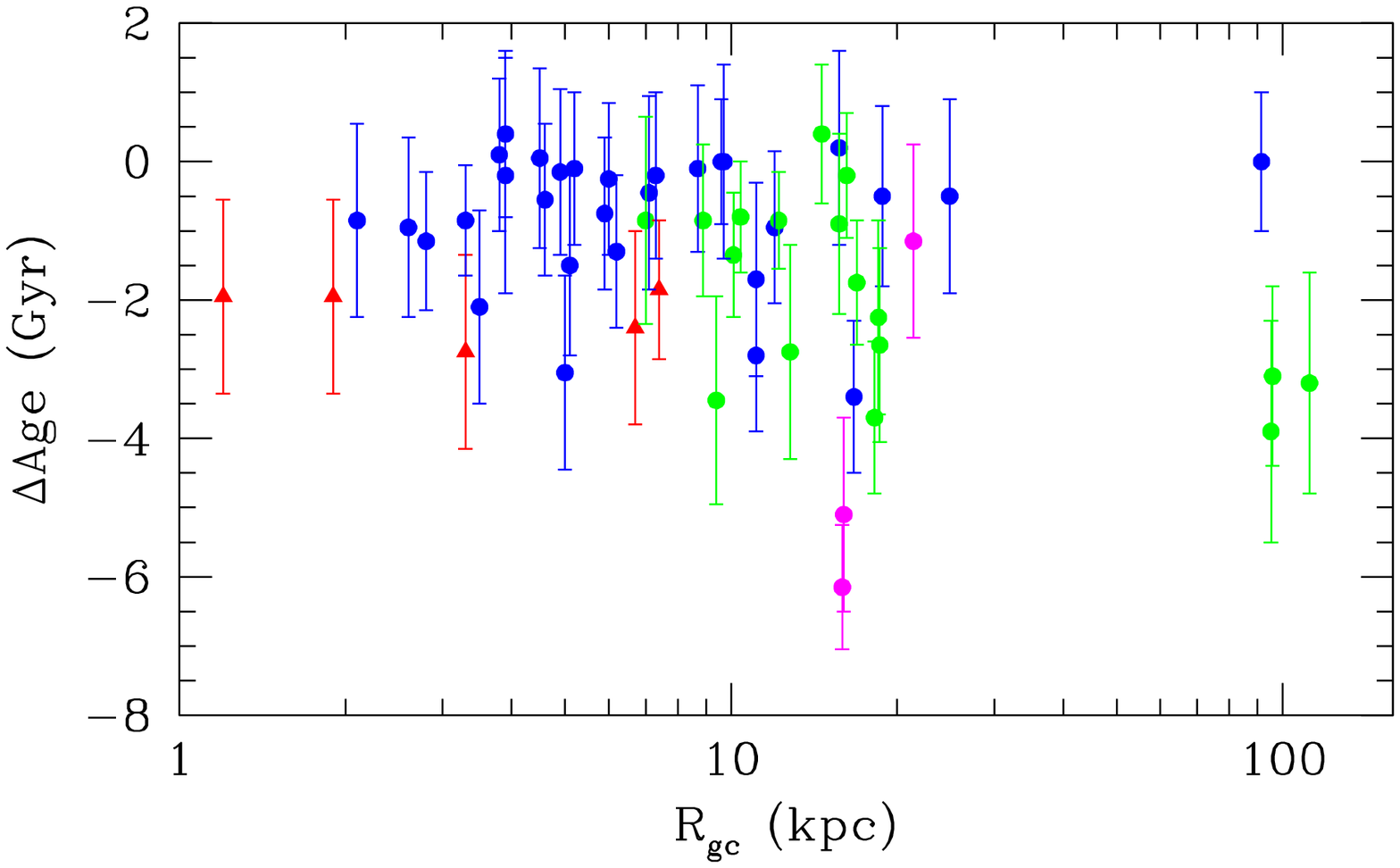} \\
\vspace{-3mm}
\includegraphics[width=0.5\textwidth]{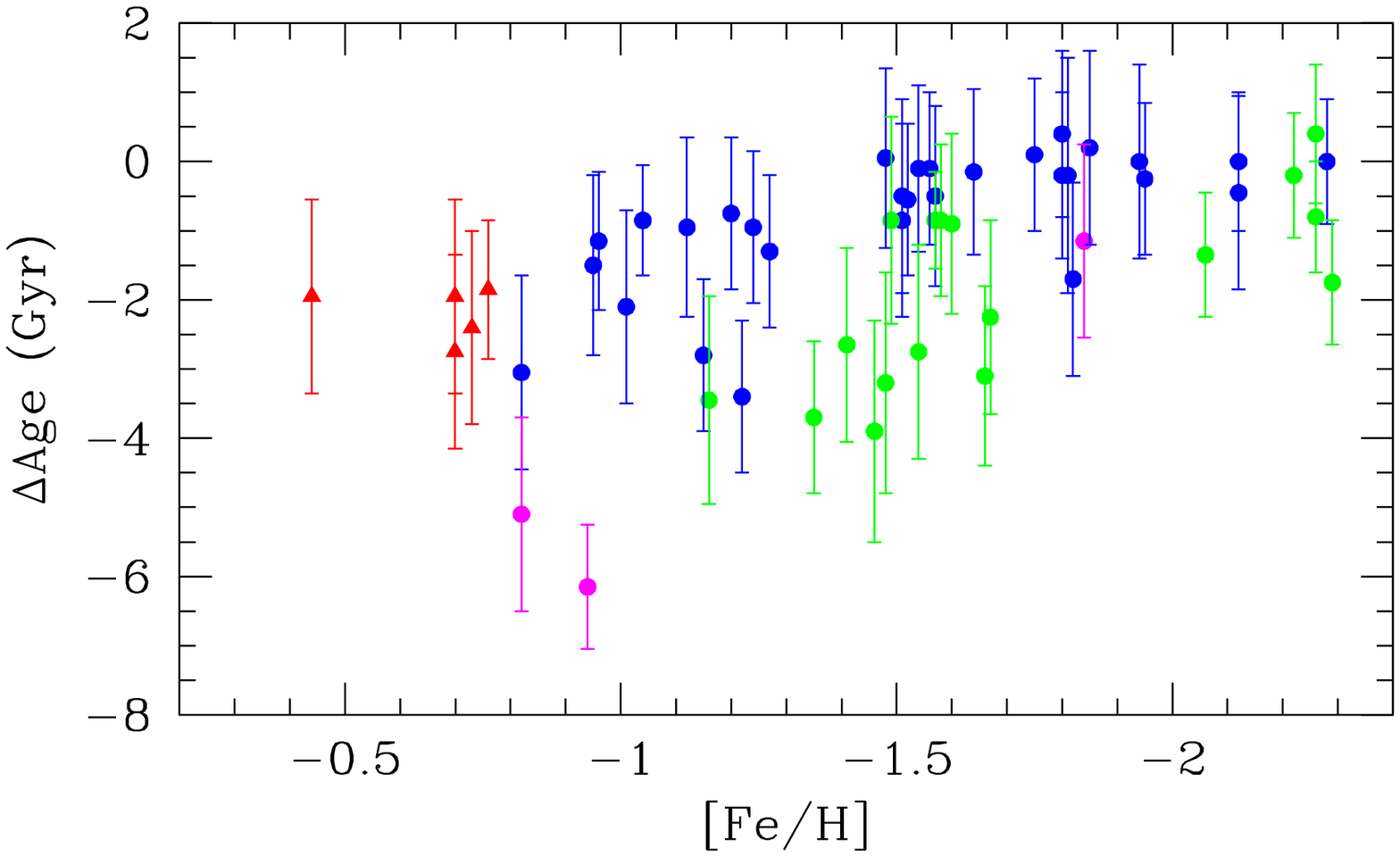}
\caption{Age versus Galactocentric radius (upper panel) and metallicity (lower panel) for $52$ Galactic globular clusters and $3$ Sagittarius clusters. Data are taken from Salaris \& Weiss (2002) (ages) and Harris (1996) (radii and metallicities). The bulge/disk clusters are solid red triangles, while the old halo clusters are blue dots, the young halo clusters green dots, and the Sagittarius clusters magenta dots.}
\label{f:ageradmet}
\end{center}
\end{figure}

These results agree well with both the relative ages from Rosenberg et al. \shortcite{rosenberg:99} 
(a smaller sample of $35$ clusters, fully contained as a subset of the Salaris \& Weiss sample), and the 
recent analysis of Gratton et al. \shortcite{gratton:03}. The latter authors measured very accurate ages 
for the prototypical old halo clusters NGC 6397 and 6752, and the bulge/disk cluster 47 Tuc (NGC 104). 
They found the old halo clusters to be coeval (at $13.4$ Gyr old), and 47 Tuc to be $\sim 2.6$ Gyr
younger. 

It is instructive to further examine the cluster ages in the context of spatial distribution and cluster
abundance. Fig. \ref{f:ageradmet} shows plots of $\Delta$Age versus $R_{\rm{gc}}$ and $[$Fe$/$H$]$.
Considering the upper panel first, it is clear that none of the three sub-systems show a clear correlation
between age and Galactocentric radius. There is however a hint that the innermost old halo clusters
are slightly younger that the outer members of this group, matching well the bulge/disk clusters in age.
The young halo clusters show a large dispersion as expected. It is interesting to note that the three
Sagittarius clusters in the Salaris \& Weiss sample appear to correspond well with the distribution of
young halo objects. The three very outermost young halo clusters in the group (Eridanus, Pal. 3, and Pal. 4)
are all significantly younger ($\Delta$Age\ $\sim -3.5$ Gyr) than the old halo cluster (NGC 2419) in
this region.

Considering next the variation of cluster age with metal abundance, several trends are apparent. First, the 
majority of the old halo clusters follow quite a tight correlation. The oldest clusters are metal-poor, but
moving to younger and younger clusters, the metal abundance increases. The metal-rich bulge/disk clusters
appear to fall naturally at the young end of this sequence. Second, the young halo clusters, although
having a larger dispersion, appear to follow a {\it steeper} trend. The Sagittarius clusters appear to
take this correlation to the youngest ages. It is interesting to note that several of the youngest old 
halo clusters fall more naturally into the young halo trend rather than the quite tight old halo correlation.

It is also important to examine the three sub-systems in the context of recent kinematic results.
In particular, Dinescu, Girard \& van Altena \shortcite{dinescu:99} present measurements of full spatial 
motions and orbital parameters for $38$ globular clusters. To this sample can be added the orbits of
NGC 7006 \cite{dinescu:01}, Pal. 13 \cite{siegel:01}, and Pal. 12 \cite{dinescu:00}. Dinescu et al.
\shortcite{dinescu:99} group their sample into disk, ``BHB'', ``RHB'', and metal poor (MP) sub-systems
(according to the Zinn \shortcite{zinn:96} classification), and explore the orbital properties of these
groups. Our classification is different in a number of cases, so it is worthwhile to examine
the global kinematic characteristics of the present three sub-systems.

\begin{figure}
\includegraphics[width=0.5\textwidth]{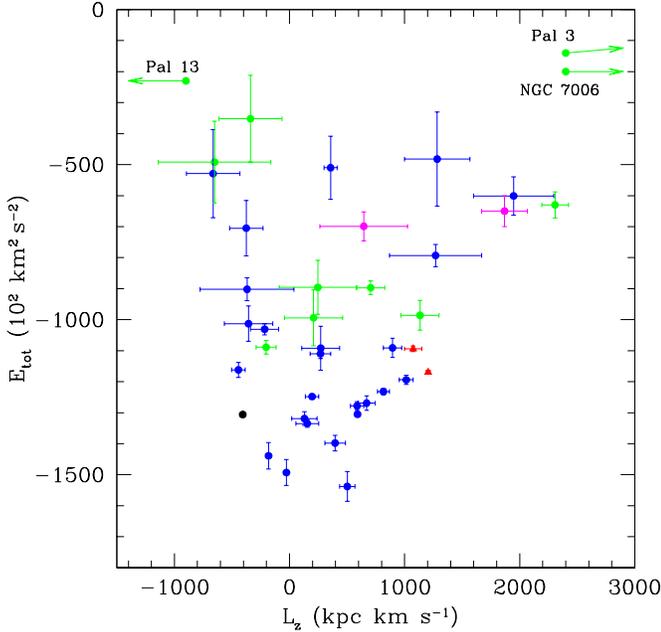}
\caption{Total cluster energy ($E_{\rm{tot}}$) versus orbital angular momentum ($L_{\rm{z}}$) for $41$ globular clusters. Data are taken from the references listed in the text. The bulge/disk clusters are solid red triangles, while the old halo clusters are blue dots, the young halo clusters green dots, and the Sagittarius clusters magenta dots. $\omega$ Cen is plotted as a black dot. Three young halo clusters fall outside the plot boundaries -- these are (as labelled): Pal. 13 at $(E_{\rm{tot}},L_{\rm{z}}) = (-3016 \pm 729, -230 \pm 80)$; NGC 7006 at $(5420 \pm 1732, -200 \pm 90)$; and Pal. 3 at $(9609 \pm 3833, 307 \pm 111)$.}
\label{f:energymom}
\end{figure}

\begin{figure}
\includegraphics[width=0.5\textwidth]{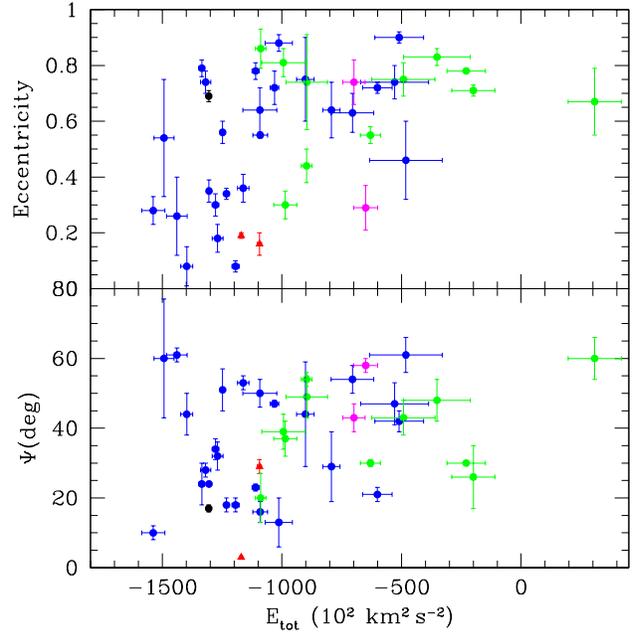}
\caption{Orbital eccentricity ($e$) (upper panel) and inclination ($\Psi$) (lower panel) as a function of total cluster energy for $41$ globular clusters. As before, data are taken from the references listed in the text. The bulge/disk clusters are solid red triangles, while the old halo clusters are blue dots, the young halo clusters green dots, and the Sagittarius clusters magenta dots. $\omega$ Cen is plotted as a black dot.}
\label{f:energyorbit}
\end{figure}

Fig. \ref{f:energymom} is a plot of total energy ($E_{\rm{tot}}$) as a function of orbital angular momentum
for the $41$ clusters from the papers listed above, while Fig. \ref{f:energyorbit} shows orbital 
eccentricity ($e$) and inclination ($\Psi$) to the Galactic plane as a function of energy for these same 
clusters. As with their other characteristics, the three sub-systems appear quite to have reasonably distinct 
kinematic properties. The two bulge/disk clusters in the sample have intermediate orbital energies,
significant prograde rotation, and low eccentricities and inclinations. These are all indicative of
disk-like kinematics. In contrast, the young halo clusters ($11$ objects) form a hot population, with 
energetic orbits and a very wide range in orbital angular momentum, indicating both prograde and retrograde 
rotation. These clusters generally have large eccentricities ($\langle e \rangle = 0.68$) and relatively 
highly inclined orbits ($\langle \Psi \rangle = 40\degr$). The two observed Sagittarius clusters 
(Pal. 12 and NGC 4147) have properties fully consistent with this group.

The old halo clusters cover the full range of orbital parameters. This group contains clusters with highly 
energetic orbits and those with very low energy orbits; clusters with large prograde rotation and those with 
significant retrograde rotation; and clusters spanning the full range of observed eccentricities and orbital 
inclinations. Closer inspection reveals some sub-groups in this system. From Fig. \ref{f:energymom} it
is clear that no young halo clusters have $E_{\rm{tot}} < -1.05 \times 10^5$ km$^2$ s$^{-2}$. There are 
$\sim 7$ old halo clusters in the sample with orbital energies significantly larger 
than this limit. These objects also match the young halo clusters well in terms of their dispersion in 
$L_{\rm{z}}$ (several even have retrograde rotation), and they have high eccentricities 
($\langle e \rangle = 0.70$) and intermediate inclinations ($\langle \Psi \rangle = 43\degr$). It is 
interesting to note that three of these seven clusters are NGC 1851, 1904, and 2298, which are suggested
former members of the CMa dwarf galaxy \cite{martin:04}. 

In addition, Dinescu et al. \shortcite{dinescu:99} have demonstrated that several old halo clusters
apparently possess disk-like kinematics -- specifically, NGC 6254, 6626, and 6752. These objects lie 
close to the two disk clusters in Figs. \ref{f:energymom} and \ref{f:energyorbit}. Dinescu et al.
also demonstrate an apparent correlation between $L_{\rm{z}}$ and $\Psi$ for a number
of old halo clusters, so that the higher the orbital angular momentum, the smaller the inclination.
This trend may well indicate a transition from halo-like to disk-like kinematics among the old halo
population. Dinsecu et al. \shortcite{dinescu:03} studied the kinematics of seven globular clusters lying
in the inner Galaxy, with $R_{\rm{gc}} < 3$ kpc (but did not derive orbital parameters for these objects 
due to the uncertain Galactic models in this region). Four of these clusters (NGC 6304, 6316, 6528, and 
6553) are in our bulge/disk sub-system, while the remaining three (NGC 6522, 6266, and 6723) are in our 
old halo group. Dinescu et al. find that NGC 6522 and 6723 have kinematics consistent with halo membership. 
This demonstrates that the old halo clusters extend far into the inner Galaxy, physically overlapping
with the bulge region. Further, Dinescu et al. find NGC 6266 to have disk-like kinematics, just like
NGC 6254, 6626, and 6752 described above. Of their four metal-rich clusters, they find NGC 6304 and
6553 to belong to a rotationally supported (disk-like) system, while NGC 6316 is apparently a bulge
member. Finally, they suggest that the kinematics of NGC 6528 argue in favour of this cluster belonging
to the Galactic bar. 

Taken together, this information is fully compatible with the presently preferred scenario for the 
formation of the Galaxy and its globular cluster system. Zinn \shortcite{zinn:93a,zinn:96} (see also the 
references therein) proposed that the bulge/disk clusters and the majority of the old halo clusters 
are intrinsic to the Milky Way and likely represent an evolutionary sequence in the dissipative collapse 
which formed the Galaxy (similar to the paradigm originally proposed by Eggen, Lynden-Bell \& Sandage 
\shortcite{eggen:62}). In this scenario globular clusters formed first in the halo, and then in the 
bulge and disk as the collapse continued. This is supported by the observed trends in age, metallicity, 
and kinematics for these clusters. Many of the youngest old-halo clusters seem to match the ages of the 
oldest bulge/disk clusters, and lie at small Galactocentric radii. The youngest old halo clusters also
have abundances matching those of the bulge/disk objects, while the oldest old halo clusters are considerably
more metal-poor. An apparent transition from halo-like kinematics to disk-like kinematics is also
seen in the old halo clusters (although it should be noted that the two proposed transition
clusters with measured ages from Salaris \& Weiss \shortcite{salaris:02} -- NGC 6254 and 6752 -- are not
younger than the oldest old halo clusters, with $\Delta$Age\ $= -0.55$ and $-0.1$, respectively).

By contrast, the young halo clusters and at least a few of the old halo clusters appear to have been 
captured by the Milky Way via the accretion and subsequent destruction of satellite galaxies (similar 
to the paradigm originally proposed by Searle \& Zinn \shortcite{searle:78}). Indeed, at least one such 
event -- the accretion of the Sagittarius dSph and its $\sim 6$ globular clusters -- is observed in progress 
in the present epoch, and newly discovered structures in the Galactic halo are hinting at more. 
The young halo clusters show a large scatter in their properties, with no correlation between position
and metallicity or age. They possess a different relationship between age and metallicity from the majority
of the old halo and bulge/disk clusters, and exhibit highly energetic and eccentric orbits which
carry them to the very distant reaches of the Galactic halo. Their HB morphologies are different from those
of the old halo clusters (by definition), and they possess much larger normalized RR Lyrae populations. 
Taken together these properties cannot easily be reconciled with the dissipative collapse scenario 
outlined above, and are explained most simply if the young halo clusters, along with some proportion
of the old halo clusters, were formed in now-defunct dwarf galaxies and merged into the Galactic halo at
some later date.

\section{Comparing the Galactic and external globular cluster systems}
\label{s:comparison}
Zinn \shortcite{zinn:93a} points out that for the accretion scenario outlined above for the origin of 
the young halo clusters to be viable, {\it ``the globular clusters in at least some dwarf galaxies must
resemble the younger halo clusters in terms of HB type at a given $\mathit{[}$Fe$\mathit{/}$H$\mathit{]}$''}. 
The advent of 
instruments and techniques which are capable of measuring the full orbital motions of globular clusters, 
together with constantly improving dating methods, will surely allow the accretion scenario to be explored fully
with detailed ages and kinematic data in the reasonably near future; however, with the new HB data presented 
in the present work for both the Galactic and external globular cluster samples it is possible to test the 
merger scenario now. The four external systems considered in the present work are all very close to the Milky 
Way even by Local-Group standards -- indeed three of the galaxies (Sagittarius and the LMC and SMC) lie 
considerably within the Galactic halo (which extends well past $60$ kpc). It therefore seems reasonable to 
suppose that within the foreseeable future (perhaps another Hubble time or so) the orbits of these 
galaxies will decay through dynamical friction with the Galactic dark matter halo, and they too will be 
accreted and destroyed. What sort of clusters would the Galaxy obtain if this were to happen? Do these 
objects match the young halo clusters, which are supposed to have been accreted in this manner at earlier 
epochs? We address these questions in the discussion below.

\subsection{Horizontal branch morphologies}
\label{ss:hbcompare}

\begin{figure*}
\begin{minipage}{175mm}
\begin{center}
\includegraphics[width=86mm]{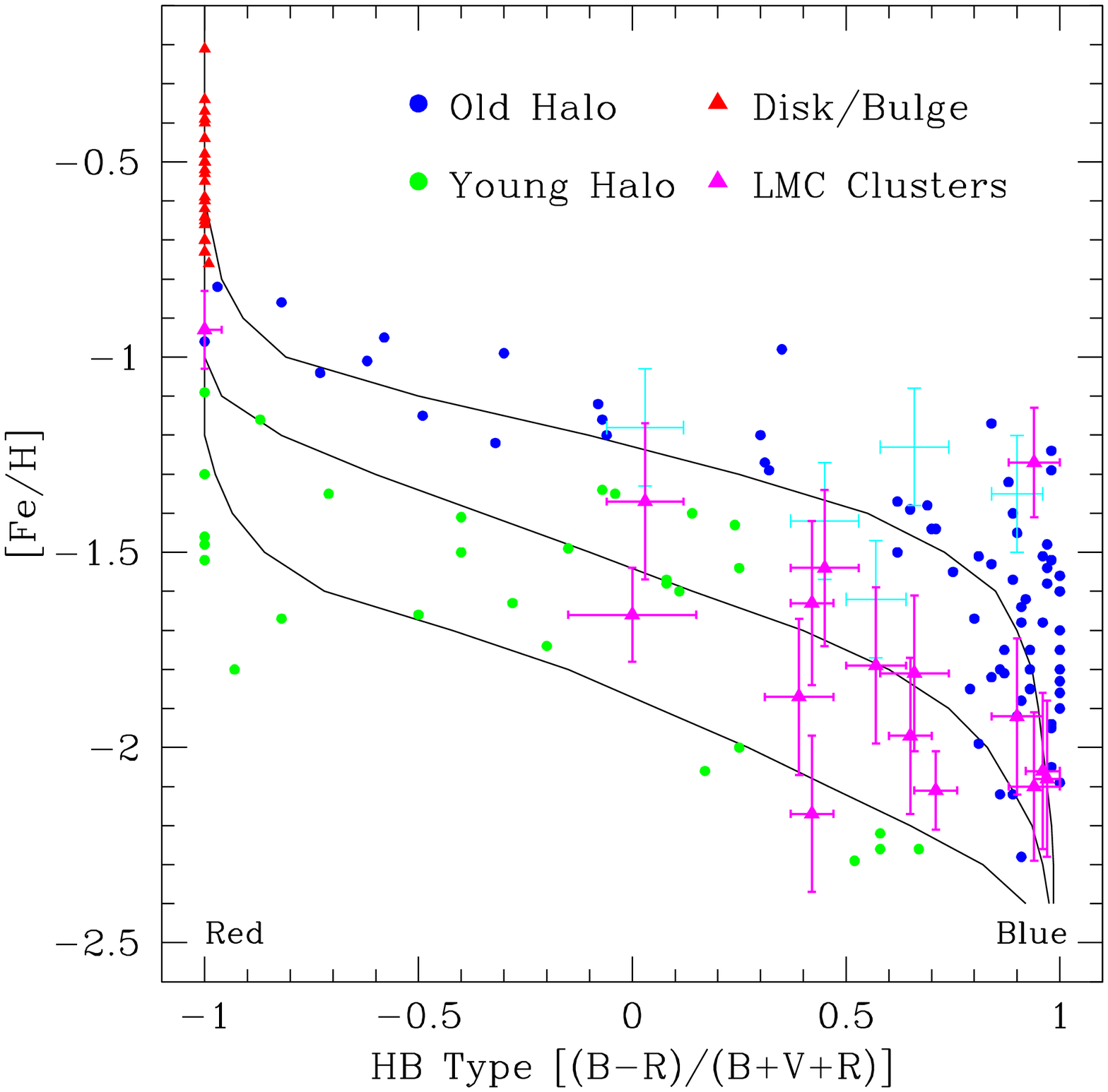}
\hspace{0mm}
\includegraphics[width=86mm]{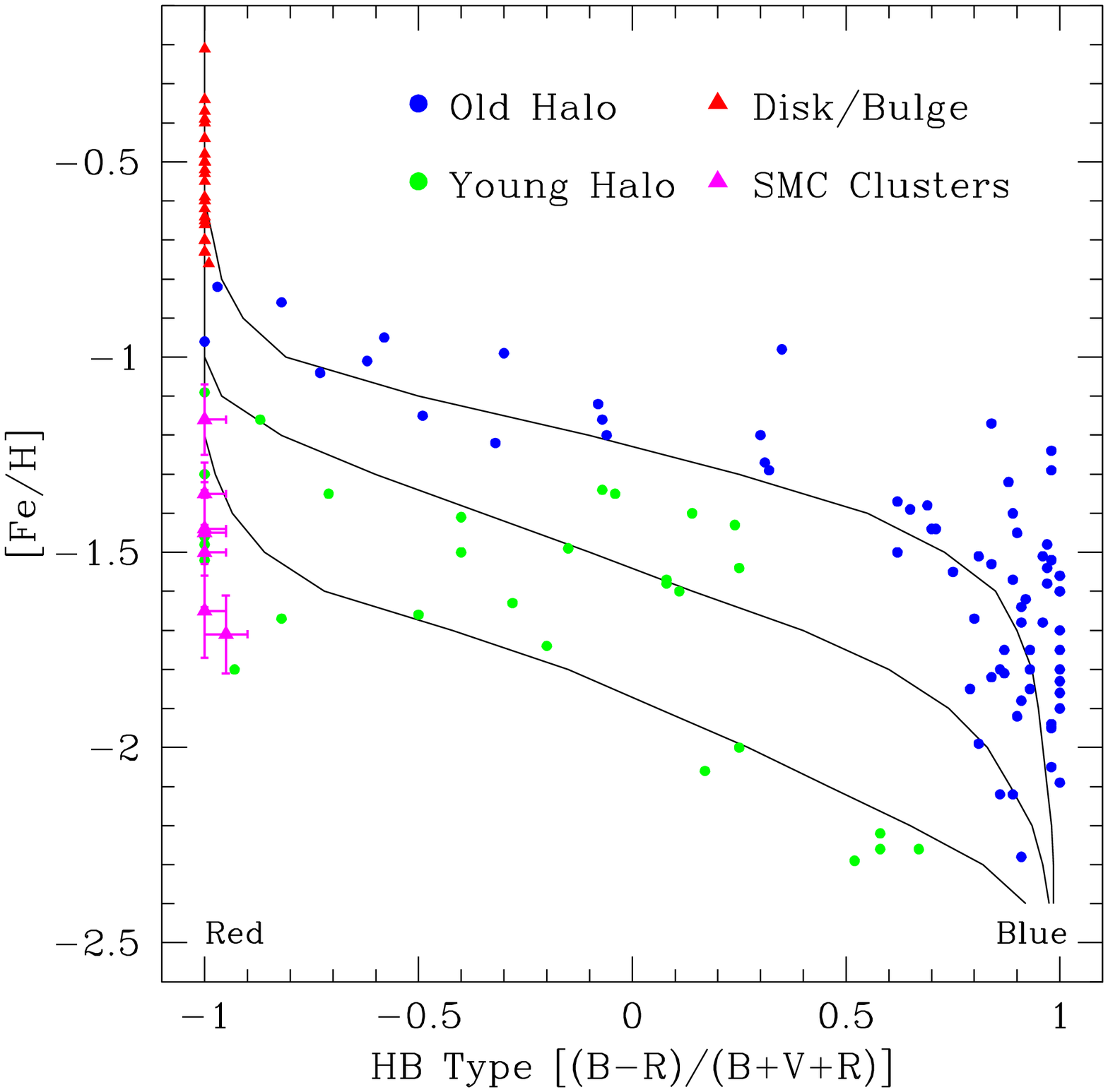} \\
\vspace{1mm}
\includegraphics[width=86mm]{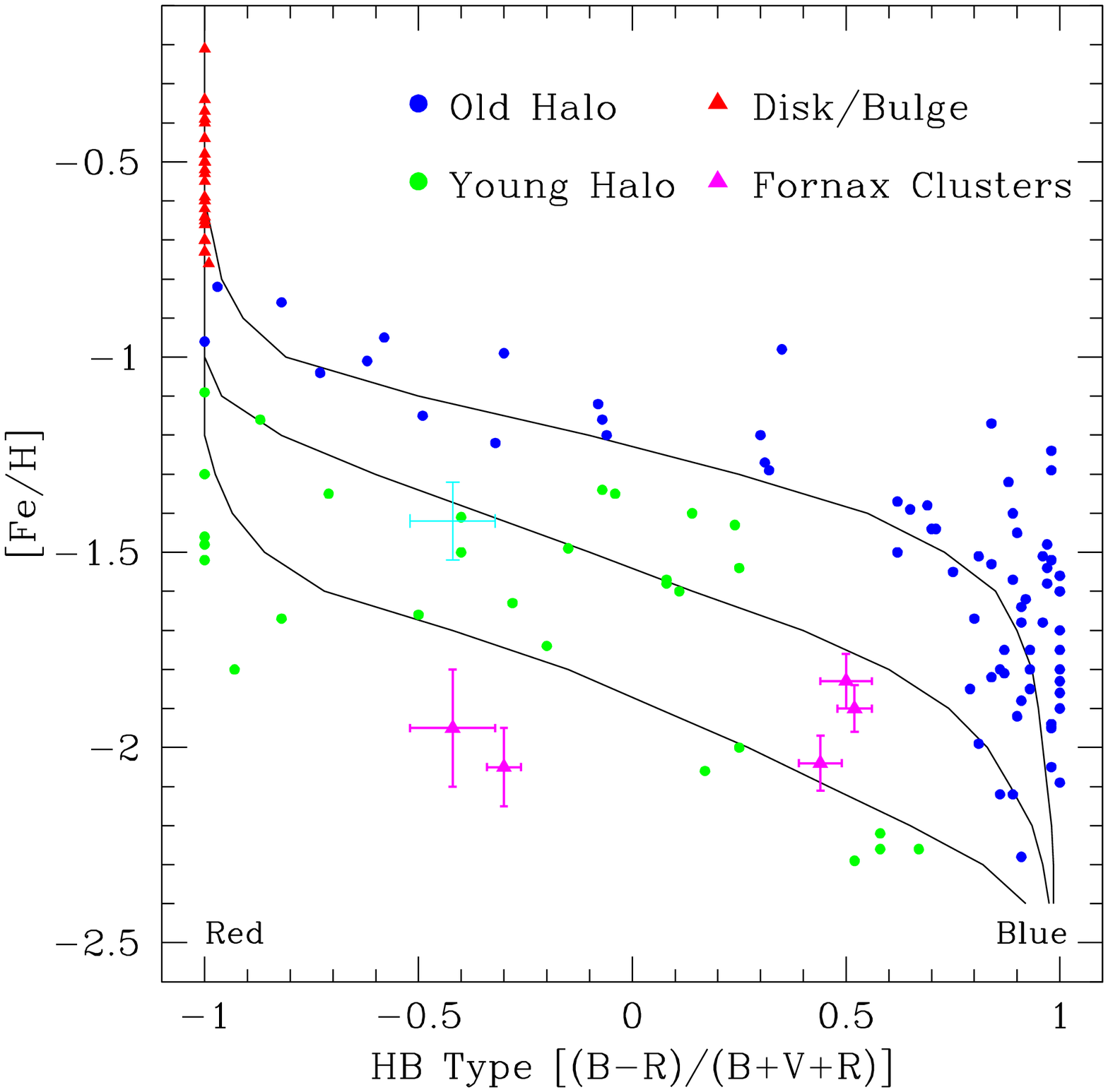}
\hspace{0mm}
\includegraphics[width=86mm]{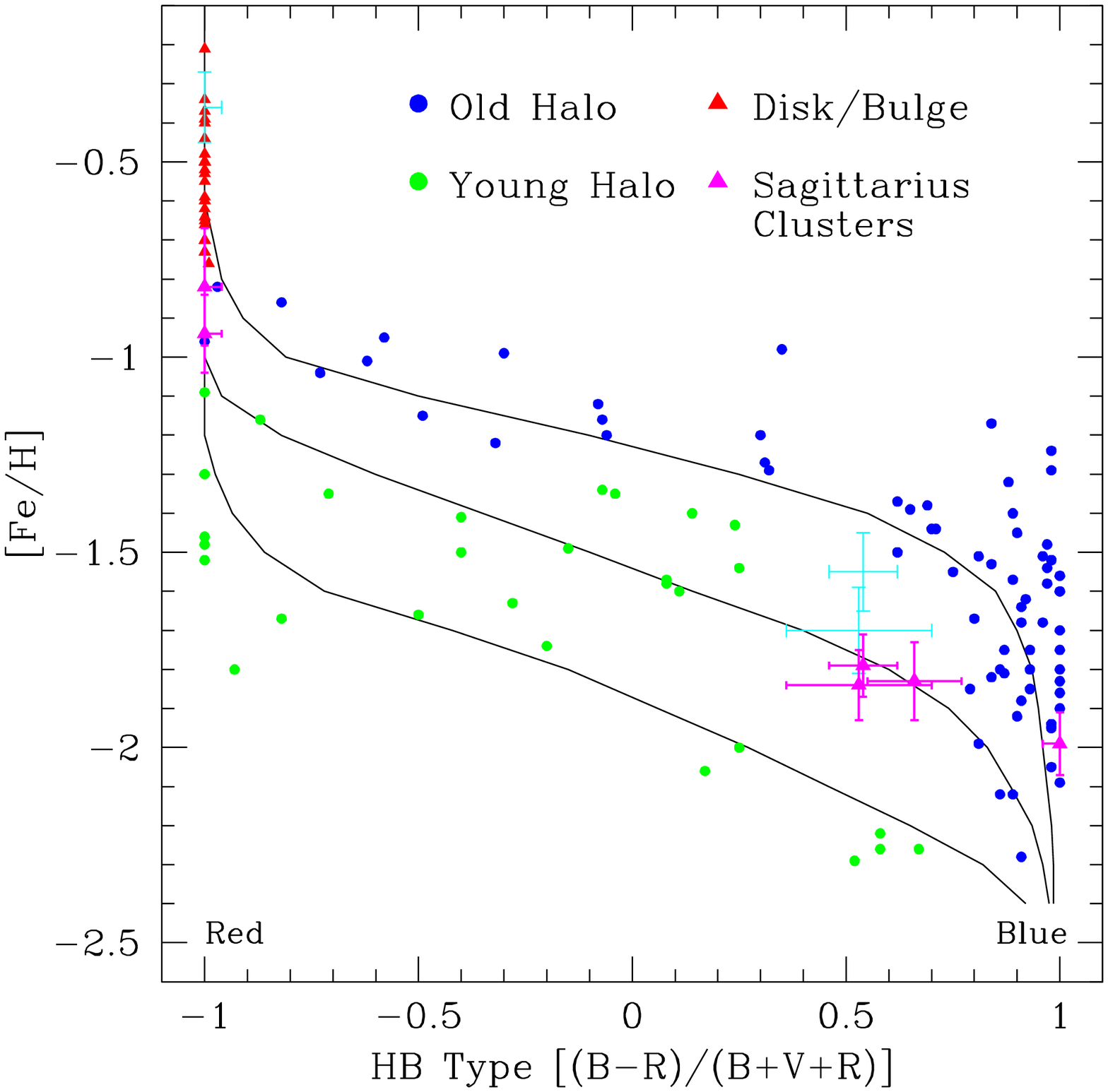}
\caption{HB-type versus metallicity diagrams for the LMC (upper left), SMC (upper right), Fornax dSph (lower left), and Sagittarius dSph (lower right) cluster systems. Points for the three Galactic globular cluster sub-systems are also plotted, as indicated in each diagram. Cyan points represent alternative (higher) metallicity measurements for the clusters directly below them (as described in the text). The overplotted isochrones are as in Fig. \ref{f:hbmetupdated} -- with the two lower isochrones, respectively, 1.1 Gyr and 2.2 Gyr younger than the top isochrone.}
\label{f:hbmetexternal}
\end{center}
\end{minipage}
\end{figure*}

\begin{figure*}
\begin{minipage}{175mm}
\begin{center}
\includegraphics[width=86mm]{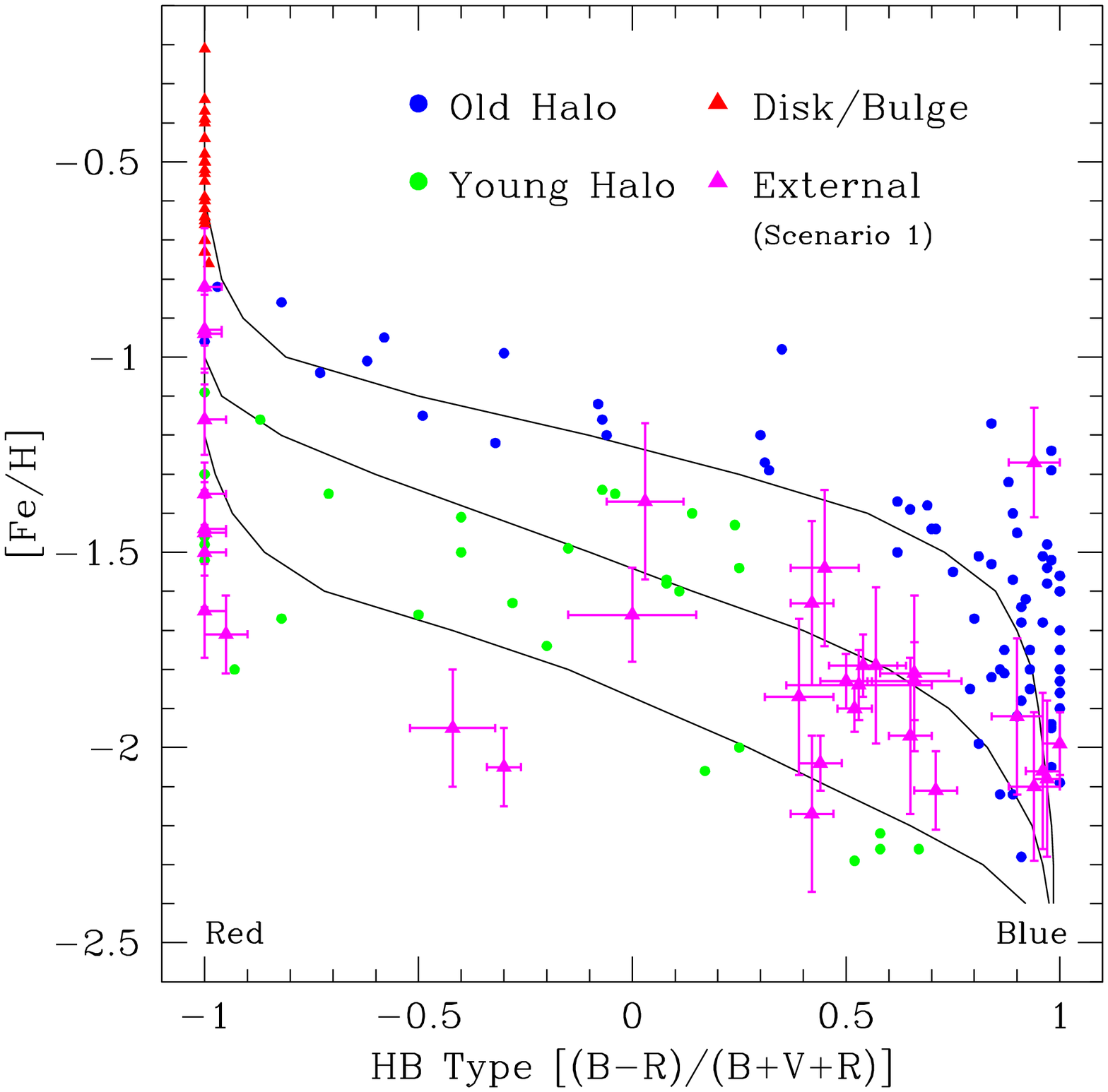}
\hspace{0mm}
\includegraphics[width=86mm]{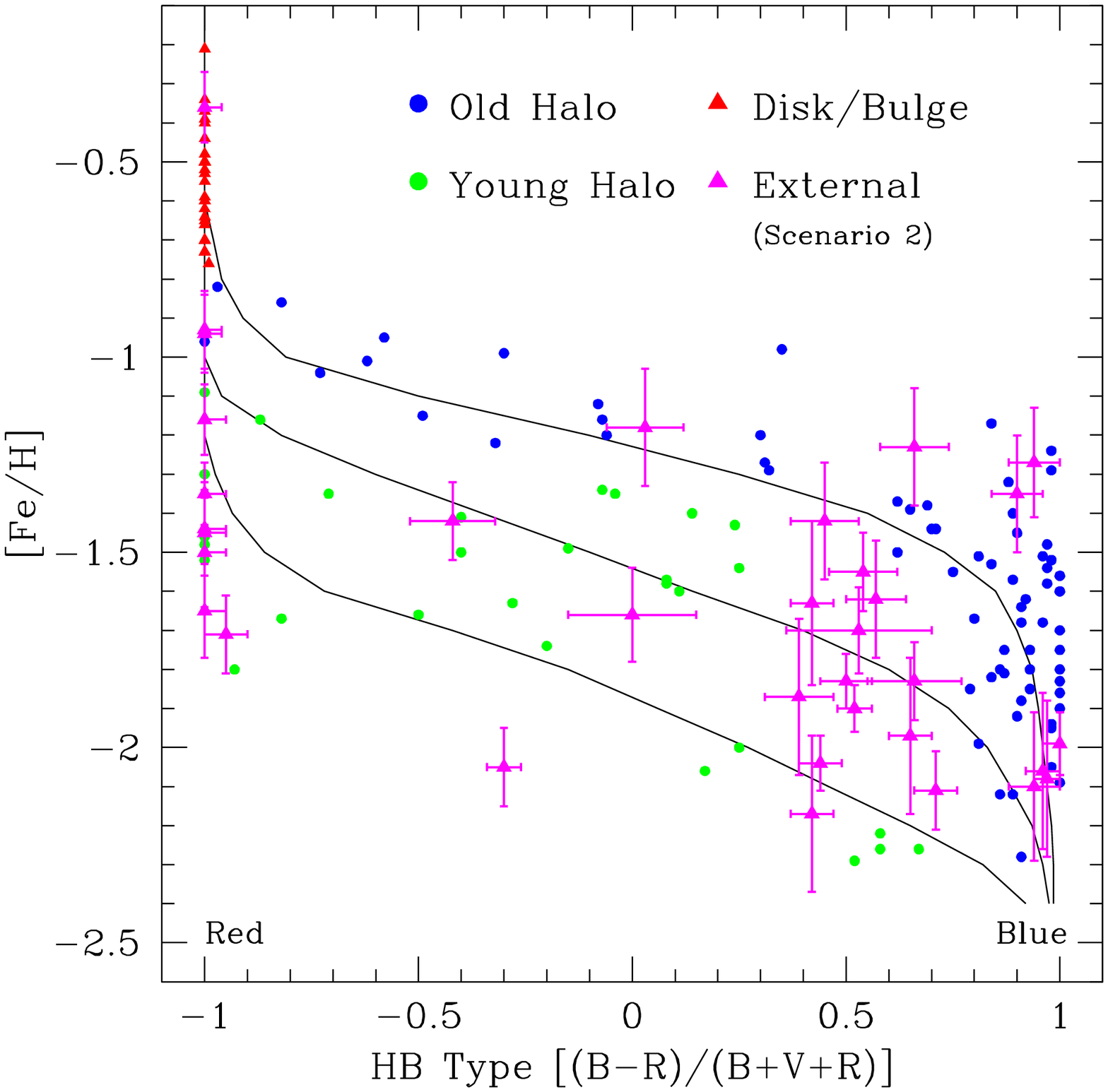}
\caption{HB-type versus metallicity diagrams for the four external cluster systems combined into a single 
external system. {\bf Left:} (Scenario $\# 1$) In this diagram, the more metal-poor points are plotted 
for the clusters with discrepant measurements. {\bf Right:} (Scenario $\# 2$) In this plot, the more 
metal-rich points are plotted for the clusters with discrepant measurements. These two diagrams illustrate 
the estimated maximum variation in the combined Satellite system due to such discrepancies. The overplotted 
isochrones are as in Fig. \ref{f:hbmetupdated}.}
\label{f:hbmetall}
\end{center}
\end{minipage}
\end{figure*}

The simplest and most revealing means of comparing the external globular cluster systems to the
Galactic globular cluster sub-systems is by plotting the external clusters on the Galactic HB-type vs. 
metallicity diagram, Fig. \ref{f:hbmetupdated}. We present such plots for each of the four external 
sub-systems in Fig. \ref{f:hbmetexternal}. For each diagram, we have used the HB measurements and
adopted metallicities from Table \ref{t:hbresults}. The external cluster points are plotted in magenta. 
For some of these points there is an additional point plotted in cyan. These are 
cases where the metallicity is uncertain, as discussed in Section \ref{sss:extramet}. For example, 
for Fornax 4 spectroscopic measurements suggest $[$Fe$/$H$] \sim -1.4$, but photometric measurements 
suggest $[$Fe$/$H$] \sim -1.9$. This is also the case for M54, Arp 2 and Terzan 7. Cyan points are 
plotted for five LMC clusters (NGC 1754, NGC 1835, NGC 1898, NGC 2005, and NGC 2019) which show an
opposite discrepancy (the photometric metallicities of Olsen et al. \shortcite{olsen:98} are
significantly higher than the spectroscopic metallicities of Olszewski et al. \shortcite{olszewski:91}).

In terms of the HB-type vs. $[$Fe$/$H$]$ diagrams, the simplest of the four external systems is the SMC. 
These clusters are a classic example of the second parameter effect. At the indicated metallicities,
old halo clusters generally have entirely blue HB morphologies, whereas the SMC clusters have entirely
red HB morphologies. There are no mysteries here however -- the discrepancy can be explained solely
in terms of age as the second parameter, because the SMC clusters are exclusively significantly younger 
than the old halo clusters (see e.g., Crowl et al. \shortcite{crowl:01}). All the clusters except NGC 121 
lie fully to the left of the diagram, where the isochrones are degenerate. This is consistent with them 
being at least $\sim 3$ Gyr younger than the old halo clusters. NGC 121 is the oldest of the SMC clusters, 
and lies slightly off the region of degeneracy, towards the right. If the isochrone ages of Rey et al.
\shortcite{rey:01} are correct, then extrapolating from the lowest plotted isochrone, NGC 121 must be
$\sim 2-3$ Gyr younger than the old halo clusters. This is exactly the result of Shara et al. 
\shortcite{shara:98} who find NGC 121 to be younger than the oldest Galactic globular clusters by $2-3$ 
Gyr, using a WFPC2 CMD.

The three other external systems are of considerably more interest. The Sagittarius dSph system contains
a wide variety of clusters, of types matching all three Galactic sub-systems. Terzan 8, 
which has been measured to be coeval with the oldest Galactic globular clusters (e.g., Layden \&
Sarajedini \shortcite{layden:00}; Monetgriffo et al. \shortcite{montegriffo:98}), is an old-halo type 
object, while NGC 4147, Arp 2, and Pal. 12 are young-halo type clusters. Pal. 12 is very young 
($\tau \sim 6-8$ Gyr \cite{salaris:02,rosenberg:99}), and falls on the degenerate isochrones to 
the left of the diagram, as expected. Arp 2 is $1-2$ Gyr younger 
than Terzan 8 \cite{layden:00,salaris:02}, and this is consistent with the plotted isochrones if age 
is taken as the sole second parameter. The case of M54 is more complicated. If the lower metallicity 
measurement for this cluster is correct, M54 is a young-halo type object; however, adopting the more 
metal-rich measurement pushes this cluster (just) into the old-halo region of the plot. M54 is coeval 
with Terzan 8 \cite{layden:00}, but given the uncertainty of its measured abundance, and the possibility 
of an internal metallicity spread \cite{dacosta:95}, its position in Fig. \ref{f:hbmetexternal} is not 
necessarily contradictory with age being the sole second parameter. The final Sagittarius cluster considered 
here, Terzan 7, is a very interesting object. Formally, its photometric abundance would classify it (just) 
as an old halo cluster; however, it has a much younger age than any old halo object ($\tau \sim 7$ Gyr 
\cite{layden:00,salaris:02}). Its (higher) spectroscopic metallicity would classify it as a bulge/disk 
object; however its Galactocentric radius is too large (at $R_{\rm{gc}} = 16$ kpc) to be consistent with 
membership in this sub-system. Once Sagittarius has fully merged with the Galaxy, Terzan 7 will likely 
appear as a rather metal-rich cluster with halo-like dynamics. In this sense, it has very few known 
counterparts in the Galactic globular cluster system. As noted by van den Bergh \& Mackey \shortcite{vdb:04}, 
one cluster with similar properties is Pal. 1, which has $R_{\rm{gc}} = 17$ kpc and $[$Fe$/$H$] = -0.6$.

The Fornax dSph contains only young-halo type clusters. These objects are extremely interesting in
the context of the question of whether age is the sole second parameter. Buonanno et al. 
\shortcite{buonanno:98} have measured clusters 1, 2, 3, and 5 to be coeval with each other and the oldest 
Galactic globular clusters, to within $1$ Gyr. This is inconsistent with the isochrones in Fig. 
\ref{f:hbmetexternal}, which suggest an 
age dispersion of $\sim 2$ Gyr between the four clusters, and that Fornax 1 should be $\sim 3$ Gyr younger 
than the Galactic clusters. Naturally, there are observational errors affecting the Buonanno et al. 
\shortcite{buonanno:98} measurements; however it is difficult to see how these could be of a magnitude 
large enough to account for the position of Fornax 1 in Fig. \ref{f:hbmetexternal}. This object appears
to be one of the most extreme second parameter clusters known. If the Buonanno et al. 
\shortcite{buonanno:98} ages are correct, then age may not be the sole second parameter operating in 
this cluster. In addition, Buonanno et al. \shortcite{buonanno:99} perform a similar analysis on Fornax 4, 
and find it to be significantly younger (by up to $\sim 3$ Gyr) than the oldest Galactic clusters 
(which is consistent with its position on the diagram if its photometric metallicity is correct), and also 
clusters 1, 2, 3, and 5 (which is inconsistent with its position on the diagram unless its spectroscopic 
metallicity is correct, and even then it would not be coeval with Fornax 1 according to the isochrones). 
Furthermore, Strader et al. \shortcite{strader:03} find clusters 2, 3, and 4 to be coeval but cluster
5 to be $2-3$ Gyr younger. Again, this is inconsistent with the position of the Fornax clusters in
Fig. \ref{f:hbmetexternal} if their metallicities are correct and age is the sole second parameter.
Further detailed study of the five Fornax clusters is clearly warranted.

The LMC is the most complex of the external systems, with a relatively large number of clusters. 
The majority of these are young-halo analogues. ESO121-SC03 stands out as having the reddest
HB morphology, but this cluster is known to be considerably younger than the rest of the plotted
clusters (e.g., Mateo et al. \shortcite{mateo:86}) -- it is the only known object to lie in the 
LMC age gap. The remaining clusters are interesting because according to the plotted isochrones 
several of them should be $1-2$ Gyr younger than typical old halo clusters (although the number
varies depending on whether the photometric or spectroscopic metallicities are used for the
clusters with uncertain measurements). Johnson et al. \shortcite{johnson:99} have studied Hodge 11, 
NGC 1466, and NGC 2257 (which do not have abundance discrepancies) and found them to be coeval with 
each other and the oldest Galactic globulars to $\pm 1.5$ Gyr. This is consistent with the positions 
of Hodge 11 and possibly NGC 2257 on the diagram, but not NGC 1466, which has a very red HB type for 
its metallicity. This may be another case where age is not the only second parameter. Similarly, 
Mackey \& Gilmore \shortcite{acs1} found NGC 1928, 1939 and Reticulum to be coeval with both a group
of old LMC clusters (NGC 1466, 2257 and Hodge 11) and the oldest Galactic globular clusters, to within
$2$ Gyr; while Brocato et al. \shortcite{brocato:96} measured NGC 1786, 1841 and 2210 to be coeval
with each other and the oldest Galactic globulars, to within $3$ Gyr. Because of the relatively large
quoted errors, these two studies do not place tight constraints on the HB-type vs. metallicity diagram;
however, we note that the positions of NGC 1786, 1841, 2210 and Reticulum on the diagram are 
inconsistent with the two studies' main conclusion -- that these clusters are coeval with the Galactic 
old-halo objects. Finally, if the spectroscopic metallicities are assumed for the five bar clusters 
discussed in Section \ref{sss:extramet} (NGC 1754, 1835, 1898, 2005, 2019) then their ages as derived 
by Olsen et al. \shortcite{olsen:98} are marginally inconsistent with the isochrones. This is reconciled 
if their photometric abundances are used. As with the Fornax dSph clusters, it seems that the LMC clusters 
form a useful system in which to study the second parameter effect.

Fig. \ref{f:hbmetall} shows HB-type vs. metallicity diagrams with all the external clusters plotted. These 
show the overall contribution of globular clusters to the Galactic halo in the thought experiment where the 
four systems have been fully accreted by the Milky Way. Scenario $\# 1$ (left panel in Fig. \ref{f:hbmetall}) 
is the diagram if the more metal-poor measurements are adopted for the clusters with abundance discrepancies, 
while scenario $\# 2$ (right panel) is the diagram if the more metal-rich measurements are adopted. These 
illustrate the maximum expected variation due to such measurement uncertainties. Scenario $\# 1$ provides 
a sample with a far smaller overall dispersion than scenario $\# 2$. Nonetheless, the clear result from 
examination of either diagram is that in this thought experiment, {\em the overwhelming contribution of 
clusters from external galaxies would be to the Galactic young halo sub-system}. This would be true based 
not only on the HB-type vs. metallicity diagrams, but also on the ages of the accreted clusters -- the 
mean age and the age spread in the merged external cluster system matches well that measured for the 
already established Galactic young halo. This result is fully consistent with the hypothesis of Zinn
\shortcite{zinn:93a} that the young halo clusters have been accreted by the Milky Way through merger events.

We remark that some authors (e.g., Zinn \shortcite{zinn:93b}; Smith et al. \shortcite{smith:98}) have
previously examined diagrams similar to Fig. \ref{f:hbmetall} for smaller samples of external clusters,
and have claimed that the external objects don't match exactly the region of the diagram occupied by
the young halo clusters. Looking at Fig. \ref{f:hbmetall}, we see that this is possibly the case. Many of 
the young halo clusters seem to be positioned at slightly higher metallicities (and thus redder
HB types) than the main body of external clusters. Even so, we see that the large majority of both samples
fall about the $1.1$ Gyr isochrone, and the region bounded by both ensembles is essentially equivalent
(apart perhaps from Fornax 1, which does not seem to have any young halo counterparts). Even if we accept
that the bulk of young halo and external clusters occupy slightly different regions of the diagram,
this is not a problem for the accretion scenario, and does not alter our conclusion that the main
contribution of external clusters would be to the young halo sub-system. All it means is that the various 
galaxies which have been accreted so far, and which might be accreted in the future, have clusters of 
somewhat different metallicities. Given the independence of these systems, along with the range of 
abundances observed within several of the four external systems considered here, this fact is 
hardly surprising.

There is one additional result. Although the majority of clusters would be donated to the young halo 
sub-system, there would also be a non-zero contribution to the old halo system. In scenario $\# 1$ 
this would be $6-7$ clusters (Terzan 7 and Pal. 12 must be assigned to the young halo on the basis 
of their ages), while in scenario $\# 2$ it would be $\sim 11$ clusters. That is, were all four external 
systems to merge into the Galactic halo, $\mathit{20-30}$ {\em per cent of the clusters would contribute 
to the old halo sub-system}. This is fully consistent with the results described above in Section
\ref{ss:characteristics}, where we showed that some fraction of the old halo clusters have kinematics
and ages consistent with the young halo ensemble. 

\subsection{Core radius distributions}
\label{ss:rccompare}

\begin{figure}
\includegraphics[width=0.5\textwidth]{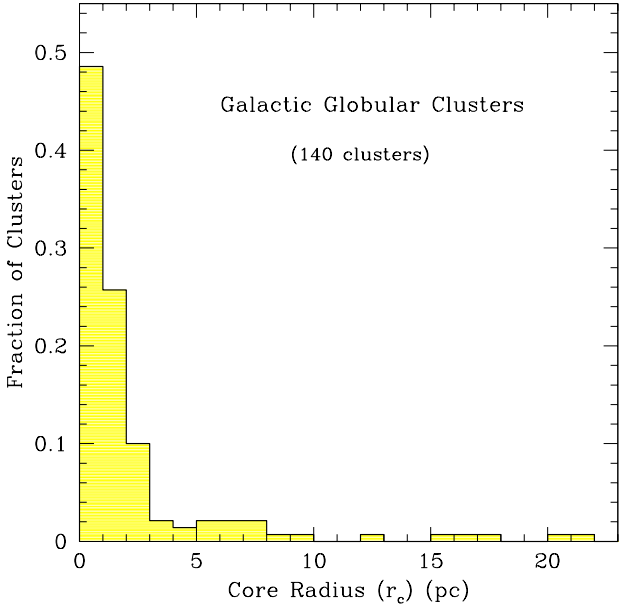}
\caption{The distribution of core radii for all $140$ Galactic globular clusters with suitable measurements.
The data are taken from the catalogue of Harris \shortcite{harris:96}, updated as described in Section 
\ref{ss:ggcupdated}. Note the long tail to very large core radii.}
\label{f:rchistall}
\end{figure}

The comparison between the Galactic sub-systems and the external cluster systems may be extended by
using the structural measurements from Table \ref{t:spfreq}. Mackey \& Gilmore \shortcite{sbp3}
demonstrated that the old clusters in the LMC and Fornax and Sagittarius dwarf galaxies possess very 
similar distributions of core radii. Any hypothetical merged external cluster system (such as that 
discussed in the previous Section) should therefore possess a characteristic distribution of core radii, 
matching those presented by Mackey \& Gilmore \shortcite{sbp3} for the individual external galaxies. This 
distribution can be compared directly to the distributions of core radii for the Galactic globular 
clusters sub-systems.

As described in Section \ref{ss:ggcupdated}, we have assembled core radius measurements for $140$ Galactic
globular clusters, of which we are able to classify $136$ into the three sub-systems. A histogram 
of core radius for our full sample of Galactic globulars is presented in Fig. \ref{f:rchistall}. The 
distribution is sharply peaked at small core radii (as should be expected, seeing as $\sim 20$ per cent of 
the Galactic population appears to be in a post core-collapse phase of evolution \cite{djorgovski:86}) but 
has a long tail to extremely large core radii -- significantly larger in fact, than any observed in the 
external cluster systems.

\begin{figure}
\includegraphics[width=0.5\textwidth]{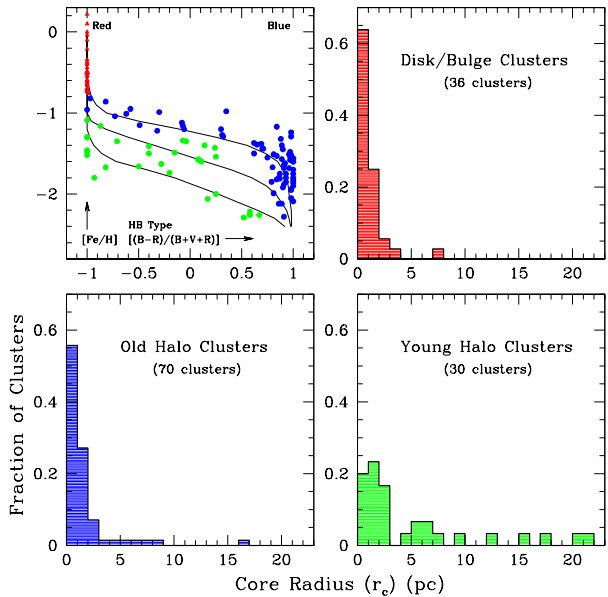}
\caption{The distribution of core radii for the three Galactic globular cluster sub-systems. The data are 
taken from the catalogue of Harris \shortcite{harris:96}, updated as described in Section 
\ref{ss:ggcupdated}. The upper left plot shows how the three systems have been defined, while the remaining 
three plots show the core radius distributions, as labelled. Note that the young halo system has a 
significantly distinct distribution -- almost all the clusters with very extended cores are members of this 
sub-system.}
\label{f:rchistsubsys}
\end{figure}

More enlightening is to plot similar histograms for each of the Galactic cluster sub-systems. These
distributions appear in Fig. \ref{f:rchistsubsys}. As with many of their other characteristics, 
the core radius distributions for these sub-systems are very distinct from one another. The bulge/disk
clusters are very concentrated at small radii, which makes sense because it is unlikely that a loosely 
bound cluster (i.e., with a large core radius) would survive long in the severe tidal environment
near the Galactic bulge. In contrast, almost all of the clusters with large core radii fall into
the young halo group, which possesses an utterly different core radius distribution to that for the
bulge/disk group. The old halo clusters exhibit a distribution somewhere in between. Like that for the
bulge/disk clusters, this distribution is concentrated at small core radii; but there is still a small
tail to large core radii. A simple K-S test highlights the differences between the three ensembles.
Using the K-S formulation in Press et al. \shortcite{press:92}, we reject the null hypothesis that the 
bulge/disk and young halo distributions were drawn from the same parent distribution at the $\sim 0.003$ 
per cent significance level (i.e., ${\rm Probability}(D>{\rm observed}) \sim 0.00003$, where $D$ is the 
maximum value of the absolute difference between the two cumulative distributions in $r_c$ -- in this case
$D = 0.5611$), while we reject the null hypothesis for the old halo and young halo distributions at the
$\sim 0.03$ per cent significance level ($P \sim 0.0003$ and $D = 0.4429$). The result for the bulge/disk
and old halo distributions is ambiguous however -- we find $P = 0.537$ ($D = 0.1603$) and are unable
to formally reject or accept the null hypothesis on this basis.

\begin{figure}
\includegraphics[width=0.5\textwidth]{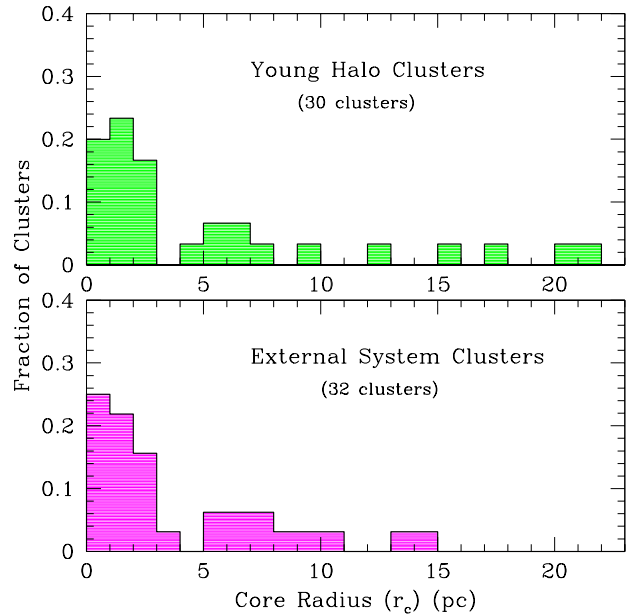}
\caption{The distribution of core radii for the Galactic young halo sub-system (upper panel) together with the merged external cluster system (lower panel). The two distributions appear to be very similar, except for the lack of extremely distended external clusters.}
\label{f:rchistexternal}
\end{figure}

Fig. \ref{f:rchistexternal} presents the histogram of core radii for the external clusters. It is
important to note several points about the construction of this histogram. First, as is highlighted in
Table \ref{t:spfreq}, we do not have core radius measurements for Lindsay 1 and Lindsay 38, so
these two clusters cannot be included in the histogram, or in any subsequent discussion. For NGC 1928
and 1939 we have only approximate upper limits for $r_c$. We place the most compact of the two clusters,
NGC 1928, in the $0-1$ pc bin, and NGC 1939 (which appears most similar to NGC 1835, which has $r_c = 1.16$ 
pc) in the $1-2$ pc bin. While it is possible that this cluster falls in the $0-1$ pc bin, switching it to 
this bin does not alter either the histogram or our subsequent analysis significantly. Finally, resolution 
effects must be accounted for. As discussed earlier, the core radius measurements for several of the clusters 
in Table \ref{t:spfreq} are upper limits. This is particularly so for the strong post core-collapse (PCC) 
candidate clusters (NGC 2005, 2019, and Fornax 5), which, for the construction of Fig. \ref{f:rchistexternal} 
and the remainder of our analysis, are set equal to the smallest measured core radius -- that for NGC 1916. 
This does not eliminate the resolution problem, but at least makes the effects visible and relatively easy 
to account for.

\begin{figure}
\includegraphics[width=0.5\textwidth]{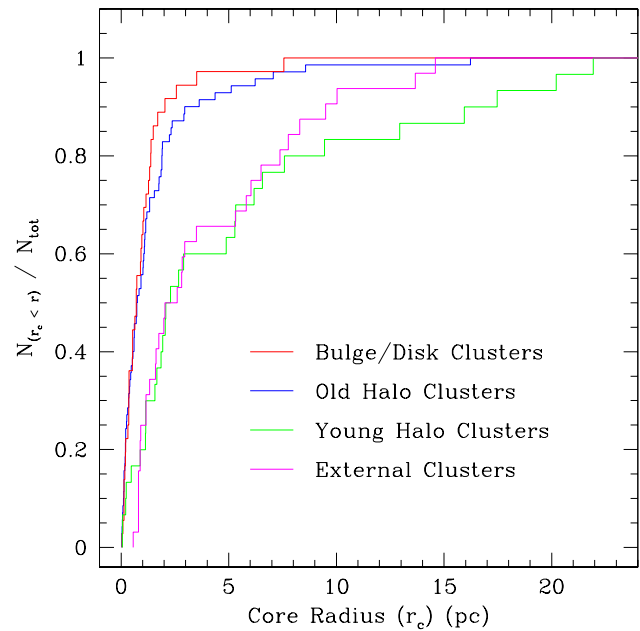}
\caption{Cumulative distributions in core radius for the three Galactic globular cluster sub-systems and all clusters in the merged external system, as labelled. The three Galactic sub-systems are clearly distinct from one another, as described in the text, while the young halo system and external ensemble match at high significance (ignoring the resolution problem at $r_c < 1$ pc).}
\label{f:cumdistall}
\end{figure}

Also shown in Fig. \ref{f:rchistexternal} is the histogram for the Galactic young halo sub-system. It is 
immediately apparent that this distribution is very similar to that for the external clusters, with the 
only major difference appearing at very large core radii. The young halo possesses several extremely extended 
clusters ($r_c \sim 20$ pc), while no analogues for these are found in the external galaxies. To highlight this
agreement, we have plotted the cumulative distributions of core radii for the external clusters and three
Galactic sub-systems in Fig. \ref{f:cumdistall}. It is clear that the young halo and 
external distributions trace each other very closely, only diverging at very large radii. At very small
radii the resolution limit of the $r_c$ measurements for the external clusters is evident. Given that
at least three of the external clusters are strong PCC candidates, it seems likely that the two
distributions would track each other closely even at very small radii if the resolution limit was
not present. A K-S test on the young halo and external distributions shows that we can accept the null
hypothesis that they were drawn from the same population at the $\sim 7$ per cent significance level
(i.e., $P = 0.928$). In conducting this test we ignored the differences between the two distributions at 
$r_c < 1$ pc to account for the resolution problem. Excluding this region, the greatest separation between 
the two curves occurs, as expected, at large core radii -- $D = 0.133$ at $r_c = 14.6$ pc. 

\begin{figure}
\includegraphics[width=0.5\textwidth]{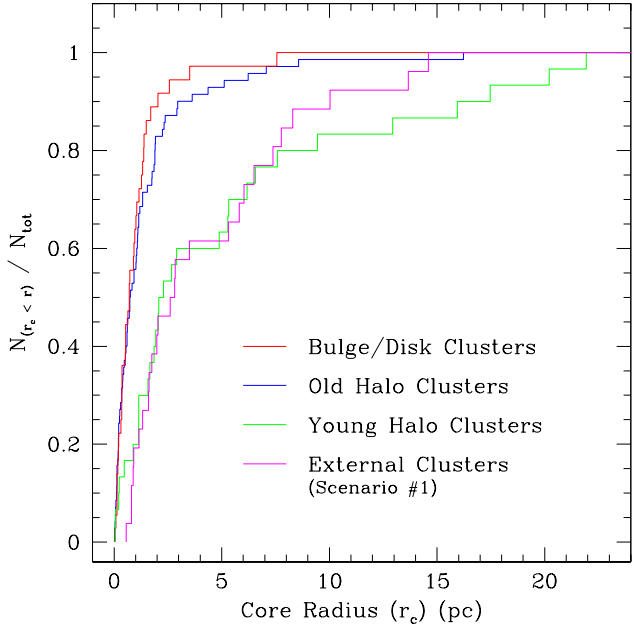}
\caption{Cumulative distributions in core radius for the three Galactic globular cluster sub-systems and the $26$ young halo analogues from the scenario $\# 1$ merged external system, as labelled. The significance of the match between the young halo and external systems is slightly increased (again ignoring the resolution problem at $r_c < 1$ pc).}
\label{f:cumdists1}
\end{figure}

In the previous Section, we demonstrated that in the hypothetical merger of the four external
galaxies with the Milky Way, the majority of the external clusters would contribute towards the young
halo sub-system. Therefore, it is perhaps not too surprising that the core radius distribution for
the young halo should so closely match that for the external clusters. However, we also demonstrated
that $20-30$ per cent of the external clusters would contribute to the old halo sub-system. It is
therefore perhaps not strictly correct to directly compare the Galactic young halo sub-system with the full
external cluster sample. Rather, we should select only the young halo analogues from this set of clusters.
The number of clusters in this subset depends on which metallicity measurements we adopt. For present
purposes we will stick with the so-called scenario $\#1$ set from Fig. \ref{f:hbmetall}, which produced the 
sample with least dispersion on the HB-type vs. metallicity plot. This sample consists of $26$ clusters,
since in scenario $\#1$ six clusters (NGC 1916, 1928, 1939, 2005, Hodge 11, and Terzan 8) are considered 
old halo analogues. We have plotted the cumulative distribution in $r_c$ for the scenario $\#1$ sample
in Fig. \ref{f:cumdists1}. Again, this distribution matches very well that for the Galactic young halo
sub-system. Repeating our K-S test (with differences for $r_c < 1$ pc ignored), the significance of the
match is slightly increased -- we accept the null hypothesis at the $\sim 5$ per cent significance level
(i.e., $P = 0.952$, with $D = 0.133$ at $r_c = 14.6$ pc as before).

\begin{figure}
\includegraphics[width=0.5\textwidth]{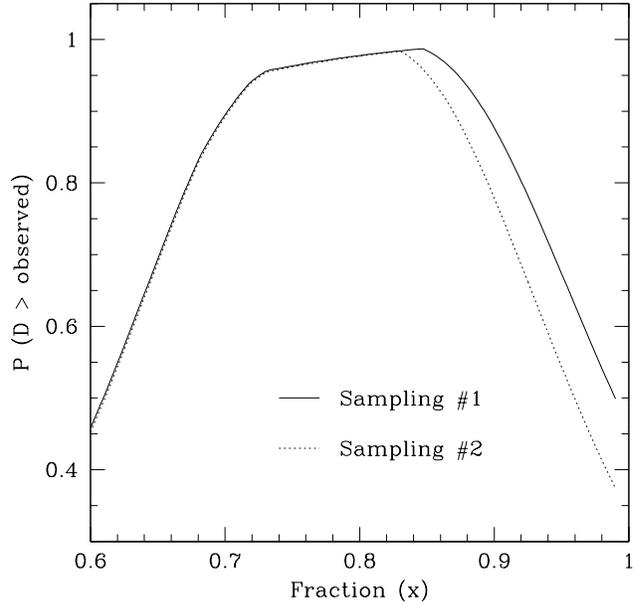}
\caption{Results of KS-tests between the core radius distribution for the old halo sub-system and composite 
distributions in core radius formed from the sum of a fraction $x$ due to the distribution of native Galactic 
clusters (e.g., the bulge/disk sample) and $(1-x)$ due to the distribution of accreted clusters (e.g., the 
external sample). The two curves relate to the two sampling techniques, as described in the text. Note that as
$x \rightarrow 1$ we expect the results of the KS-tests to approach the result of the formal test between the
bulge/disk and old halo samples described in the text ($P = 0.537$). Neither curve converges exactly to this
result however, due to the larger effective number in the two different samplings of the composite distribution
compared with the number of clusters in the bulge/disk group.}
\label{f:ksresults}
\end{figure}

This result supports our earlier claim that knowledge of the exact core radii for NGC 1928 and 1939 is
unimportant -- indeed, removing them from the analysis altogether only very slightly alters the
significance of the match. This also suggests that the core radius distribution of the external old halo 
analogues cannot be greatly different from the overall distribution of core radii for the full external 
cluster sample (although we note that the small number of clusters involved means that we cannot attach
too much significance to this result). Repeating the exercise for the (extreme) scenario $\#2$ sample of 
$21$ young-halo-type external clusters (we now add NGC 1754, 1898, 2019, ESO121-SC03, and M54 to the six 
old halo analogues listed above) somewhat degrades the formal significance of the match ($P \sim 0.75$) 
but does not alter our overall conclusion that {\it the distribution of core radii for the globular clusters in 
the four external dwarf galaxies is very similar to that for the Galactic young halo sub-system, and extremely 
different to those for the Galactic bulge/disk and old halo sub-systems.}

There is more interesting information to be gleaned from Figs. \ref{f:cumdistall} and \ref{f:cumdists1}. 
Although the K-S test comparing the Galactic bulge/disk and old halo $r_c$ distributions provided an ambiguous
result, by plotting these two cumulative distributions it is clear that the primary (and indeed only) 
difference between them occurs at $r_c > 1.5$ pc -- the old halo sub-system has, percentage-wise, more 
clusters with relatively large $r_c$ than does the bulge/disk system. Our main result from Section 
\ref{ss:hbcompare} was that if the four nearby cluster-bearing dwarf galaxies were to merge into the 
Milky Way, the majority of clusters would be donated to the young halo sub-system, but some would also go to 
the old halo sub-system. Let us therefore hypothesize that the present old halo sub-system is a composite 
system, containing primarily clusters native to the Galaxy, but also some clusters of external origin. We 
represent the distribution of core radii in this sub-system as the sum of a fraction $x$ due to the 
distribution of native Galactic clusters (e.g., the bulge/disk sample) and $(1-x)$ due to the distribution 
of accreted clusters (e.g., our external sample). For $0.6 \le x \le 1.0$, at increments of $0.25$ in $x$,
we calculated this composite $r_c$ distribution and compared it to the old halo distribution via 
a K-S test. Because of our resolution problem, for $r_c < 1$ pc we substituted the young halo distribution 
for the external cluster distribution -- as discussed earlier we expect the two to be very similar at small 
core radii. We devised two methods to sample the composite distribution at given $x$. In the first, we 
calculated the distribution at each step in the old halo distribution (i.e., at the core radius of each
cluster in the old halo sample), while in the second we calculated the distribution at each step in both
the bulge/disk and external cluster samples. 

\begin{figure*}
\begin{minipage}{175mm}
\begin{center}
\includegraphics[width=86mm]{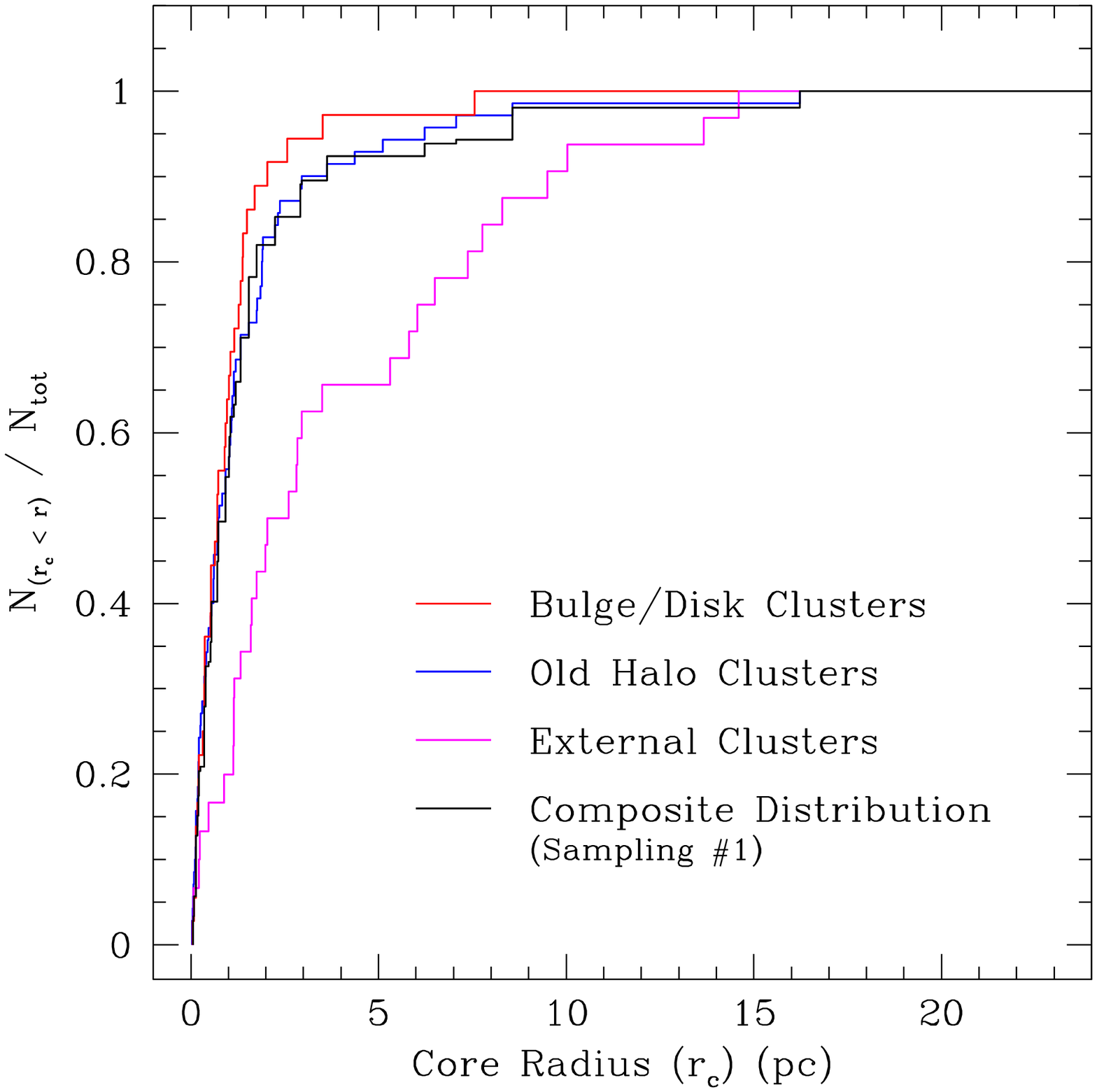}
\hspace{0mm}
\includegraphics[width=86mm]{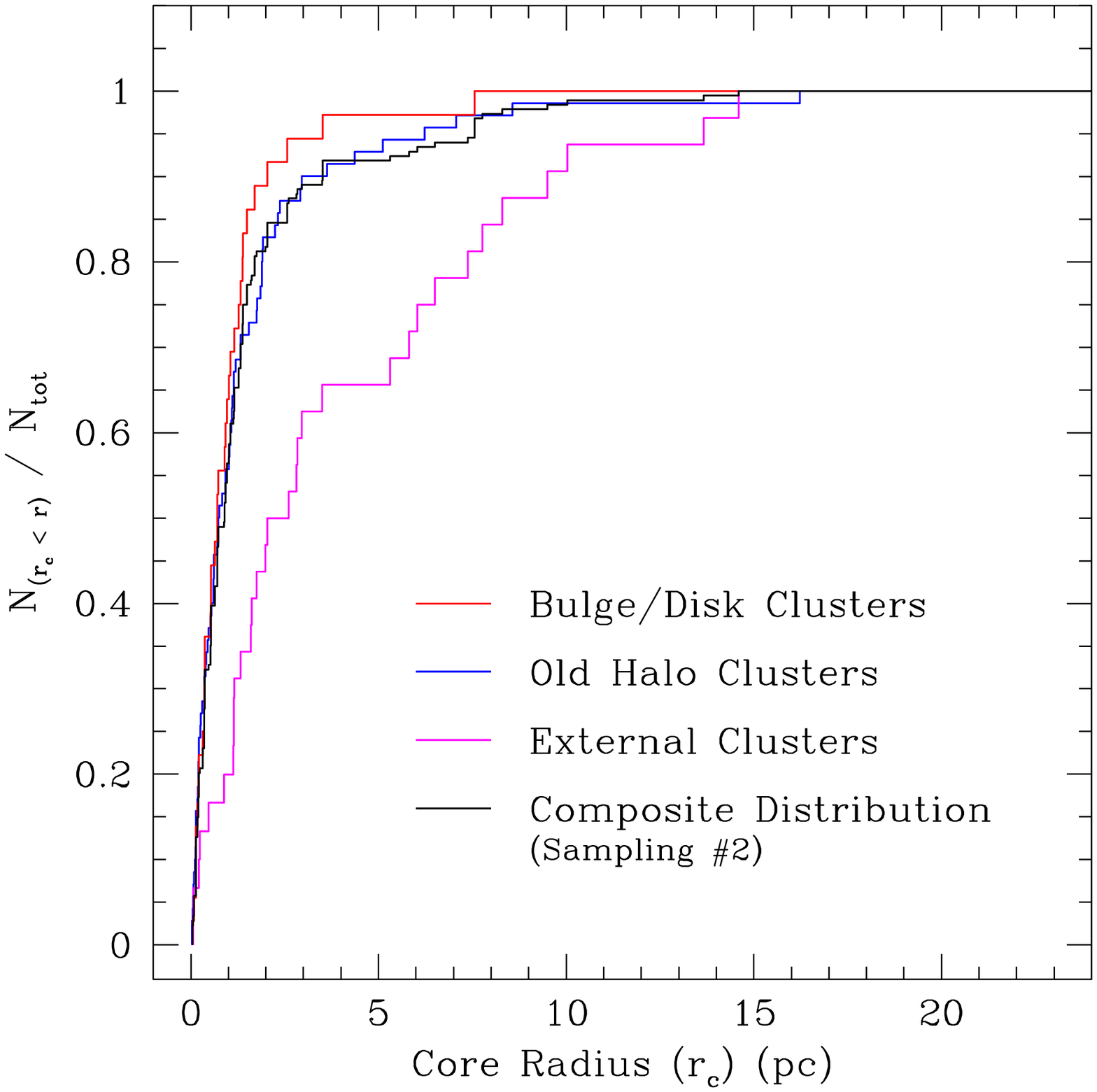}
\caption{Cumulative distributions in core radius for the Galactic bulge/disk and old halo sub-systems, plus the merged external sample and the two composite distributions with the highest significance matches to the old halo distribution (one for each sampling method). These occur for $x \approx 0.847$ ($P = 0.987$) (sampling method $1$ -- left) and $x \approx 0.830$ ($P = 0.983$) (sampling method $2$ -- right). Clearly these two composite distributions are both excellent representations of the old halo distribution.}
\label{f:cumdistcompo}
\end{center}
\end{minipage}
\end{figure*}

The results of our calculations are presented in Fig. \ref{f:ksresults}. This plot shows the output of the
K-S tests (i.e., $P(D>{\rm observed})$) over the chosen range of $x$ for each of the two sampling
techniques. Values of $P$ close to $1.0$ indicate that the null hypothesis (that the composite distribution
and the old halo distribution were drawn from the same parent distribution of core radii) can be accepted at 
high significance. The curve for the first sampling technique peaks at $x \approx 0.847$ ($P = 0.987$), while 
that for the second technique peaks at $x \approx 0.830$ ($P = 0.983$). That is, irrespective of how the 
composite distribution is sampled, it matches the distribution of core radii in the old halo sub-system at high
significance (better than $2$ per cent) when $x = 0.83 - 0.85$. The two curves in Fig. \ref{f:ksresults}
are above $P = 0.95$ ($5$ per cent significance) over the ranges $0.727 < x < 0.873$ and $0.727 < x < 0.853$
respectively, and above $P = 0.90$ ($10$ per cent significance) over the ranges $0.703 < x < 0.893$ and
$0.703 < x < 0.870$, respectively. Fig. \ref{f:cumdistcompo} shows the composite distributions for the two 
peak values of $x$, plotted alongside those for the other sub-samples. The composite distributions are each 
clearly an excellent match for the old halo distribution. Hence the distribution of core radii observed for 
the clusters in the old halo sub-system can be very well represented by assuming that $83-85$ per cent are 
native members of the Galaxy (and therefore possess a core radius distribution matching that of the bulge/disk 
clusters), while the remaining $15-17$ per cent are accreted objects (which possess a core radius distribution 
matching that of the external clusters). We note that as per the above results, accretion fractions covering 
the range $\sim 12 - 30$ per cent also provide adequate representations (i.e., matches at better than $10$ 
per cent significance). 

It is informative to note that $15-17$ per cent of the old halo sub-system numbers some $10-12$ clusters. 
Combined with the young halo system, this suggests that $\sim 40$ clusters in the Galactic halo might be 
of external origin -- $\sim 75$ per cent of which are of young-halo-type and $\sim 25$ per cent of which
are of old-halo-type. This is again entirely consistent with our conclusion from the previous Section, which 
stated that were the four cluster-bearing dwarf galaxies associated with the Milky Way to be accreted, the 
majority of new halo members would be of young-halo-type, while $20-30$ per cent would be of old-halo type.

\section{Discussion}
\label{s:discussion}
\subsection{Implications for cluster evolution and destruction}
\label{ss:clusterev}
What is the meaning of the comparisons presented above? There are two ways to view the results. Without 
assigning any physical explanation, we have clearly demonstrated that the native Milky Way globular clusters 
are very different from the majority of external globular clusters, both in terms of their HB morphology and
core radius distributions. On the other hand, the Galactic young halo globular clusters appear essentially
indistinguishable from the external globular clusters on the same basis. In addition, we have shown that
approximately $20-30$ per cent of the external globular clusters have HB morphologies consistent with
globular clusters in the Galactic old halo ensemble. Furthermore, we have demonstrated that the core radius
distribution of the old halo clusters can be very well represented as a composite distribution, consisting
of $\sim 83-85$ per cent objects that have a core radius distribution matching the bulge/disk clusters, and 
$\sim 15-17$ per cent objects that have a core radius distribution matching the external clusters. Overall,
we conclude that these results are fully consistent with the hypothesis (e.g., Zinn 
\shortcite{zinn:93a,zinn:96}) that all of the $\sim 30$ Galactic 
young halo clusters plus $10-12$ of the Galactic old halo clusters have been accreted via merger events
between the Milky Way and external dwarf galaxies. The remainder of the old halo clusters ($\sim 58-60$
objects) and the bulge/disk clusters ($37$ objects) are then native to the Galaxy.

Alternatively, given the weight of evidence we presented in Section \ref{ss:characteristics}, this conclusion 
can be accepted, {\em a priori}, as correct. In this case, it is possible to make some extra predictions from 
the above comparisons. On the basis of their HB types and core radius distributions, it appears that the 
external/young-halo clusters are a different class of cluster than the native Galactic clusters that are 
observed in the present epoch. But has this always been the case? Fall \& Rees \shortcite{fall:77} and 
Gnedin \& Ostriker \shortcite{gnedin:97} have considered destruction processes for globular clusters, such 
as evaporation (due to internal relaxation and external tides), and disk and bulge shocking. These studies 
have shown that loosely bound clusters (i.e., of low central concentration) are more susceptible to 
destruction from disk and bulge shocking than clusters of high concentration. Furthermore, Gnedin \& Ostriker 
\shortcite{gnedin:97} conclude that is it probable that a large fraction of the original population of 
Galactic globular clusters has already been destroyed. The lack of extended clusters in the present-day 
native Galactic population might therefore reflect such a process. 

It has been demonstrated on several occasions \cite{elson:89,elson:92,sbp1,sbp2} that the star clusters in 
the LMC and SMC exhibit a distinct trend in core radius with age -- specifically, that the spread in measured 
core radii is an increasing function of age. The core radius distributions observed here for the external 
cluster sample and the Galactic young halo clusters reflect the end point of this trend (i.e., the oldest
clusters and the maximum spread in core radius). Mackey \shortcite{thesis} has discussed at length the 
possible origins of the radius-age relationship. Assuming that the observed trend is a reflection of genuine 
variations in cluster structures rather than an observational artefact or sample selection effect, two classes 
of origin exist. The radius-age relationship can either be an evolutionary effect, where long-lived clusters 
are always formed compactly and some clusters grow large cores due to some unidentified expansion process, 
or it can be a formation effect, where in the past clusters were formed with a wide spread in core radius, but 
conditions have changed with time so that now only compact formation occurs. In either scenario, it seems
likely that it is the benign tidal environments of the dwarf galaxies in which the expanded clusters are 
observed that has allowed them to survive a Hubble time. What can these tell us about cluster formation and 
evolution in the Galaxy?

In the first case, where some clusters evolve large cores, it is easy to imagine that in the Galactic
environment any such expanding clusters would quickly be destroyed (via the processes outlined by Gnedin 
\& Ostriker \shortcite{gnedin:97}). This is similar for the scenario where clusters $\sim 10$ Gyr ago were 
formed with the core radius distribution exhibited by the external clusters -- any primordial expanded 
clusters would soon be disrupted in the Galactic environment. Both cases are equivalent to the core radius 
distribution of the external clusters evolving into the core radius distribution of the bulge/disk clusters. 
This hypothesis allows a simple calculation. The external clusters have $\sim 45$ per cent of members with 
core radii smaller than $2$ pc (reasonably concentrated), while the bulge/disk group possesses $\sim 90$ per
cent. It can be estimated that at present the observed population of native Milky Way globular clusters 
numbers approximately $100$ ($37$ in the bulge/disk, plus $\sim 83-85$ per cent of the $70$ old halo objects). 
Thus, $\sim 90$ have $r_c < 2$ pc, which would be $\sim 45$ per cent of the original population, assuming
no destruction of concentrated clusters. This population must therefore have numbered $\sim 200$ 
objects, {\it implying that $\mit{\sim 50}$ per cent of the original Galactic globular clusters have been 
destroyed up to the present epoch.} In fact, this is a lower limit, since it is not strictly true that 
concentrated clusters are not destroyed -- relaxation and evaporation processes predominantly affect these 
objects. Nonetheless, this estimate is consistent with that of Gnedin \& Ostriker \shortcite{gnedin:97} who 
predict that more than half of the present day clusters will be destroyed in the next Hubble time. 

It should be noted that an alternative possibility exists, whereby cluster formation conditions are
a function of environment. It is not difficult to imagine that harsh tidal conditions in the Galaxy
only ever allowed compact cluster formation (i.e., with a core radius distribution similar to that
of the bulge/disk clusters), while the more benign conditions in dwarf galaxies allowed clusters with
a wide range of core radii to be formed $10$ Gyr ago. Given that the details of cluster formation at 
very early epochs are presently unconstrained, little more can be said about such a scenario. 

Returning to the idea that destruction processes have played a large part in shaping the core radius 
distribution of native Galactic globular clusters, it is interesting to note that such processes have 
recently been observed in action. The extremely expanded young halo cluster Pal. 5 ($r_c = 21.93$ pc) has 
been found to possess massive tidal tails \cite{odenkirchen:01,odenkirchen:02,odenkirchen:03,rockosi:02}, 
and is hence in the process of being tidally disrupted. This could explain the origin of the tail to very 
large core radii in the distribution for the young halo clusters -- the only point where this distribution 
differs significantly from that for the external clusters. It will be useful to run $N$-body simulations 
for expanded clusters evolving in the Galactic potential, to investigate their survivability. These may 
begin in the Galactic tidal field (simulating formation) or may be introduced after some delay (simulating 
accretion). Such simulations will be able to place constraints both on whether the destruction scenario 
described above is viable (although the results of Gnedin \& Ostriker \shortcite{gnedin:97} suggest that 
it is) and on how long ago the young halo clusters can have been accreted. This in turn can help constrain 
the number of mergers the Milky Way has undergone, and how recently.

We note that one such set of simulations already exists \cite{dehnen:04}, although these are devoted to
modelling the specific case of the evolution of Pal. 5 rather than an investigation of any accretion
scenario. Nonetheless, these simulations are relevant here because they study the disruptive effects of 
the Galactic tidal field on a set of low-mass, low-concentration model clusters, similar to the more
extended objects observed in the Galactic halo and nearby external dwarf galaxies. The simulations
are run from $\sim 3$ Gyr ago to the present epoch, using the observed orbital parameters of Pal. 5
(peri-Galactic radius $\sim 5.5$ kpc and apo-Galactic radius $\sim 19$ kpc). The cluster suffers 
frequent strong disk shocks, and the models unambiguously predict the complete disruption of Pal. 5
after the next disk crossing in $\sim 110$ Myr. Irrespective of whether Pal. 5 is an accreted cluster
of not, these simulations clearly demonstrate that any accreted low-mass, low-concentration cluster 
unlucky enough to end up on an orbit which penetrates deeply and frequently into the Galactic potential, 
will only survive for a relatively short period of time (i.e., of the order of several Gyr). Again, this
shows the potential for $N$-body modelling of clusters to place at least partial constraints on the merger
history of the Galaxy. 

\subsection{Number of mergers}
\label{ss:mergers}
Our present results can provide a crude estimate of the number of mergers required to build up
the implied population of accreted halo clusters. In Section \ref{ss:hbcompare}
we showed two scenarios for the hypothetical accretion of the LMC, SMC, and Fornax and Sagittarius
dSph galaxies, based on the discrepancies in the measured abundances for the clusters in these systems.
In scenario $\# 1$, $6-7$ clusters were shown to match the Galactic old halo objects on the basis of 
their metallicities and HB morphologies, while in scenario $\#2$ this number was $\sim 11$. If we
therefore estimate that approximately $9$ of the external clusters are taken to be old-halo 
type objects, then the mean number of young-halo type clusters per external sub-system is $\sim 6$.
Comparing this number to the number of young halo clusters observed in the galaxy at present ($\sim 30$),
it seems that the Galaxy has probably experienced $\sim 5$ merger events with dwarf galaxies containing
clusters. An equivalent result is obtained if all external clusters (both young and old halo analogues)
are considered. However, we remark that a number of observations render it unlikely that the Milky Way 
has experienced a merger with a galaxy the size of the LMC or SMC in at least the last few Gyr, if ever.
In particular, these objects have a luminous mass of $\sim 10^{10} M_{\odot}$ and $\sim 10^9 M_{\odot}$
respectively (see e.g., Forbes et al. \shortcite{forbes:00}), which are equivalent or larger than
the observed luminous mass of the halo ($\sim 10^9 M_{\odot}$ -- see e.g., Freeman \& Bland-Hawthorn
\shortcite{freeman:02}). Such an event would also have delivered metal rich stellar populations and 
young clusters to the halo, which are not observed, and likely disrupted the Galactic thin disk
(again see e.g., Freeman \& Bland-Hawthorn \shortcite{freeman:02}). Therefore, it seems more sensible
to consider only the mean number of clusters in the Fornax and Sagittarius dwarf galaxies.
Repeating the calculation then suggests that the Galaxy has probably experienced $\sim 7$ merger events 
with dwarf-spheroidal-type galaxies containing clusters. This number is slightly larger than the
range suggested by van den Bergh \shortcite{vdb:00}.

This argument assumes that such galaxies contained the same number of globular clusters in the past 
as today (which seems reasonable, since destruction rates are significantly reduced in the weak tidal
field of a dwarf galaxy), and that not too many clusters have been destroyed once accreted 
into the Milky Way halo. If the destruction rate is $\sim 20$ per cent then the number of mergers increases
to $\sim 6$ in the first instance and $\sim 8$ in the second. In this context, it is interesting to note 
that the very luminous cluster $\omega$ Centauri seems likely to be the remaining core of a now disrupted 
dwarf galaxy (see e.g., Bekki \& Freeman \shortcite{bekki:03} and the references therein), just as M54 is 
apparently associated with the core of the Sagittarius dwarf \cite{layden:00}. In addition, evidence for a 
disrupted dwarf galaxy in Canis Major has recently been presented \cite{martin:04}, and it has been suggested 
\cite{vdb:04} that the very large and remote cluster NGC 2419 could also be the surviving core of a defunct 
dwarf galaxy.

The number of mergers with galaxies not containing clusters cannot, of course, be constrained by the
above argument. It is noted however, that such galaxies are invariably significantly
less massive than the satellite galaxies which do contain clusters. Forbes et al. \shortcite{forbes:00} 
list the dwarf galaxy companions to the Milky Way -- there are a further $9$ in addition to the four 
described here. Of these, only $5$ have Galactocentric radii smaller than Fornax (the most distant of the 
four in this study). Thus, a naive extrapolation suggests that the Galaxy might have undergone an additional
$6-11$ mergers with non-cluster-bearing dwarf galaxies. This is, of course, subject to the initial mass 
function and original spatial distribution of dwarf galaxies in the newly formed Local Group -- 
quantities which cannot be constrained here.

We remark that the numbers of mergers postulated here are not inconsistent with the mass of the luminous
Galactic halo. If we adopt the integrated luminosities listed by Forbes et al. \shortcite{forbes:00}
and a mass-to-light ratio $\sim 3$ (at most), we see that the mean luminous mass of local
cluster-bearing dwarf galaxies (i.e., Fornax and Sagittarius) is $\sim 10^{7.8} M_{\odot}$, and that
of local dwarf galaxies without clusters is $\sim 10^{6.6} M_{\odot}$. Thus, $\sim 7$ merger events
with cluster-bearing dwarf galaxies would have contributed $4.4\times 10^{8} M_{\odot}$ to the halo,
while $\sim 6-11$ mergers with dwarf galaxies bereft of clusters would have contributed
$2.4-4.4 \times 10^7 M_{\odot}$. If our hypothesized merger numbers are correct, the implication
is that $\sim 44-50$ per cent of the mass of the Galactic stellar halo may have been built up through
merger events. Da Costa \shortcite{dacosta:03} makes an important point regarding such accretion.
The mean abundance of halo field stars is, at $[$Fe$/$H$] \sim -1.7$, considerably lower than that
of dwarf galaxies like Fornax and Sagittarius. In order to reconcile these abundances, the majority
of mergers must have occurred at early epochs so that accreted dwarf galaxies were not too far along 
the path of chemical enrichment. It is not a viable hypothesis that only metal-poor dwarfs have been
accreted, since the evidence (at least locally) is that the most metal-poor dwarfs are also the least 
luminous, and therefore do not contain globular clusters.

The results described above are also consistent with more detailed studies of field halo stars.
Unavane, Wyse \& Gilmore \shortcite{unavane:96} examined the fraction of stars in the halo which have colours 
consistent with a metal-poor, intermediate-age population matching those typically observed in Local Group 
dwarf spheroidal galaxies, and concluded that the star counts imply an upper limit of $\sim 60$ mergers 
with low luminosity dwarf spheroidals (i.e., Carina-like objects), or $\leq 6$ mergers with more luminous 
Fornax-like objects within the last $\leq 10$ Gyr. The present study of the Galactic globular cluster
system seems to favour the latter scenario, and indeed our implied number of mergers is close to
that of Unavane et al. \shortcite{unavane:96}. Furthermore, Venn et al. \shortcite{venn:04} have 
recently demonstrated that the chemical signatures of most of the stars in local low mass dSph systems
(such as Draco) are distinct from the stars in each of the components of the Galaxy, implying that 
no Galactic component can have been formed primarily via the accretion of low mass dSph systems at late 
epochs. However, they do not rule out very early mergers with such objects before significant chemical 
enrichment can have occurred (a conclusion consistent with that of Da Costa \shortcite{dacosta:03}), nor 
mergers with larger systems like the Sagittarius dwarf or the LMC. As mentioned above, it seems clear
that the Milky Way has not merged with an LMC or SMC mass object (at least, not since very early epochs).
It is therefore apparent that a handful of mergers with Sagittarius or Fornax mass objects, mostly at
relatively early epochs, seems most consistent with the presently available data. This is just as argued
by the properties of the Galactic globular cluster system. We remark that if the young halo sub-system
is a good tracer of accreted material (and we see no reason why it shouldn't be), the majority of
field halo stars which have been donated to the Galactic halo via merger events should reside at relatively 
large Galactocentric radii (i.e., $R_{\rm{gc}} > 10-15$ kpc). That is, one might expect to observe a
transition between the representative kinematic and chemical properties of field halo stars, moving from
the inner to outer Galactic halo.

Finally, in the previous Section we noted that the observed distribution of cluster structures
in the present epoch implies that $\sim 100$ native globular clusters may have already been
destroyed since the formation of the Galaxy. Assuming a typical globular cluster mass equal to the
peak of the observed mass function -- that is, $2 \times 10^5 M_{\odot}$ \cite{harris:91} -- such
destruction of clusters may have contributed a further $2 \times 10^7 M_{\odot}$ to the stellar halo
(i.e., $\sim 2$ per cent).

\subsection{Implications for the second parameter}
\label{ss:secondparam}
How can one explain the difference in HB
morphology between the old halo clusters and the external/young-halo clusters? It is possible that several 
potential second parameters can come to the rescue. It has been demonstrated in the literature that both 
age (e.g., Lee et al. \shortcite{lee:94}) and central concentration (e.g., Fusi Pecci et al. 
\shortcite{fusipecci:93}) are, for many clusters, viable second parameters. A combination of these could 
reconcile the discrepancy between the old and young halo clusters. The isochrones of Rey et al. 
\shortcite{rey:01} indicate that an age difference of only $\sim 1-2$ Gyr would be consistent with most of 
the data (see Fig. \ref{f:hbmetall}). None of the studies demonstrating that specific external clusters are 
coeval with the oldest Galactic clusters have precision better than this. Furthermore, the mean ages for 
the young and old halo samples from the Salaris \& Weiss \shortcite{salaris:02} measurements show just 
such an offset ($\langle \Delta$\ Age$\rangle = -0.4$ Gyr for the oldest old halo clusters and 
$\langle \Delta$\ Age$\rangle = -1.9$ Gyr for the young halo ensemble).
Furthermore, in the scenario we described above it is the concentrated clusters which preferentially
survive in the Galactic population. Several authors have shown that more concentrated clusters
apparently have systematically bluer horizontal branches (see e.g., Fusi Pecci et al. \shortcite{fusipecci:93},
Buonanno et al. \shortcite{buonanno:97}, and references therein). In combination with the age effect
described above, it seems possible that this could explain the at least the global differences in HB 
morphology between Galactic old halo clusters and clusters of external origin. Nonetheless, as we demonstrated
in Section \ref{ss:hbcompare}, a number of unanswered questions remain regarding the variation of
HB morphology within the external sub-systems, in particular for the Fornax clusters. It is not clear
that age or cluster structure as second parameters are able to explain these adequately.

\subsection{Identifying accretion candidates among the old halo clusters}
\label{ss:oldhalocand}
Finally, we turn to the question of the identification of old halo clusters which have been accreted. It has
been estimated above that $\sim 15-17$ per cent (or $\sim 10-12$) of the old halo clusters are of 
extra-Galactic origin. By definition these will be indistinguishable from the Galactic native clusters
in terms of HB morphology. Nonetheless, it should be possible to identify those old halo clusters
which are likely to be of external origin. There are at least four possible distinguishing features 
which might identify these clusters -- chemical abundance patterns, ages, spatial motions (and distribution), 
and cluster structures. We concentrate here on the latter two in this list. 

We demonstrated above that the distribution of core radii for clusters in the old halo sub-system can be 
accurately reproduced by forming a composite distribution constructed from that for the bulge/disk clusters 
and that for the external clusters. Given that the bulge/disk clusters are, almost exclusively, very compact 
objects ($35$ of the $36$ clusters with structural measurements in this sub-system have $r_c < 3.5$ pc), 
it stands to reason that at least some of the accreted old halo clusters must be expanded objects. Of course, 
the old halo system extends to much greater distances from the main body of the Galaxy than does the 
bulge/disk system, so the mean old halo cluster size might be expected to be somewhat larger than the mean 
bulge/disk cluster size (see e.g., van den Bergh, Morbey \& Pazder \shortcite{vdb:91}). 
Nonetheless, it is difficult to explain the presence of the six old halo clusters 
with $r_c \ge 4.4$ pc. These are: M55 (NGC 6809) ($r_c = 4.4$ pc); NGC 6101 ($r_c = 5.1$ pc); NGC 7492 
($r_c = 6.2$ pc); NGC 5897 ($r_c = 7.1$ pc); NGC 2419 ($r_c = 8.6$ pc); and Pal. 15 ($r_c = 16.2$ pc). It 
is instructive to examine the positions of these objects in the Galactic halo. Three are very remote 
objects: NGC 7492 ($R_{\rm{gc}} = 25$ kpc), Pal. 15 ($R_{\rm{gc}} = 38$ kpc), and NGC 2419 
($R_{\rm{gc}} = 92$ kpc); while the other three are relatively central: M55 ($R_{\rm{gc}} = 3.9$ kpc), 
NGC 5897 ($R_{\rm{gc}} = 7.3$ kpc), and NGC 6101 ($R_{\rm{gc}} = 11$ kpc). Both groups are unusual. The 
remote clusters are more distant than the majority of the old halo ensemble -- only two other clusters in 
this $70$ strong group have $R_{\rm{gc}} > 25$ kpc and neither of these have $R_{\rm{gc}} > 30$ kpc. The 
young halo clusters on the other hand have $9$ (of $30$) clusters with $R_{\rm{gc}} > 25$ kpc, and four 
of these are objects with $R_{\rm{gc}} > 90$ kpc. In both this respect and that of their expanded 
structures, the three distant old halo clusters identified here more closely resemble accreted clusters 
rather than native Galactic objects.

In contrast, the three extended old halo clusters identified above which lie in the inner halo 
($R_{\rm{gc}} < 11$ kpc), are interesting because of their proximity to the main luminous body of the Galaxy.
Although direct $N$-body simulations are required to fully explore how the tidal influence of the Galaxy
would affect the destruction of loosely bound clusters, it seems difficult to understand how these 
three objects can have survived a Hubble time with extended structures in a comparatively harsh tidal 
environment. This problem can be circumvented in two ways. Possibly, these clusters are on the inner
parts of large orbits which carry them far out into the Galactic halo for long periods. The results
of Dinescu et al. \shortcite{dinescu:99} for M55 and NGC 5897 (NGC 6101 is not in their sample)
are ambiguous -- both orbits are of intermediate energy ($E_{\rm{tot}} \sim -1 \times 10^5$ km$^2$ s$^{-2}$),
eccentricity ($e \sim 0.5-0.6$, and inclination ($\Psi \sim 50\degr$), and have small prograde rotation
($L_{\rm{z}} \sim 200$ kpc km s$^{-1}$). These are not typical young-halo-like orbits; however it is
not difficult to imagine a scenario where the clusters were released into the Galactic halo only once 
their parent galaxies had been reduced to small orbits via dynamical friction with the Galactic dark 
matter halo. After all, Dinescu et al. \shortcite{dinescu:99} show $\omega$ Cen, a prime candidate for 
the remnant nucleus of a disrupted dwarf, to have an even smaller orbit than M55. A second option 
is that these three clusters arrived into the inner Galactic halo only recently (as, for example, the 
extended Sagittarius clusters Terzan 8 and Arp 2 have done), meaning they have not yet had time to 
disperse due to tidal effects. Certainly these objects deserve further attention as possible 
accretion candidates.

It is also interesting to examine the one bulge/disk cluster with a large core radius -- Pal. 11 
($r_c = 7.6$ pc). Although metal rich, this cluster is unlike any of the other bulge/disk objects 
because of its extended nature. Its position is also unusual for this ensemble --  it lies outside 
the bulge at a comparatively large Galactocentric radius ($R_{\rm{gc}} = 7.9$ kpc). While several other
bulge/disk objects lie at such radii, these (with a couple of possible exceptions) are never
more than $\sim 1.5$ kpc above or below the disk. In contrast, Pal. 11 lies $3.5$ kpc below the disk,
essentially in the inner halo. As with the six old halo clusters discussed above, this unusual object
deserves further attention.

Finally, in Section \ref{ss:characteristics} we identified seven old halo clusters from the sample of 
Dinescu et al. \shortcite{dinescu:99} with kinematics more characteristic of young halo objects.
Although this is not a definite indicator of extra-Galactic origin (as indeed an extended structure 
is not), such clusters must also be considered possible accretion candidates. The seven clusters of
interest are NGC 1851, 1904, 2298, 5024, 5904, 6205, and 7089. As we noted in Section \ref{ss:characteristics},
three of these seven clusters (NGC 1851, 1904, and 2298) have been suggested as former members of the 
proposed disrupted dwarf galaxy in Canis Major \cite{martin:04}. We also noted in Section 
\ref{ss:characteristics} that NGC 1851 (along with another proposed CMa member, NGC 2808) has an anomalously 
young age for an old halo cluster. The presence of these three clusters in this sample of kinematically 
interesting objects adds credibility to the identification of NGC 5024, 5904, 6205 and 7089 as accretion 
candidates. In passing, we note that none of these four objects is extended in structure. Two (NGC 5904 
and 6205) are in the sample of Salaris \& Weiss \shortcite{salaris:02}; however neither is exceedingly young
($\Delta$Age\ $= -1.3$ and $-0.1$ Gyr, respectively).

Returning to the question of possible CMa members, four more globular clusters have been linked with the 
tidal debris of this disrupted dwarf (the so-called ``Monoceros Ring'' or Galactic anti-centre stellar 
structure (GASS) \cite{newberg:02}) -- NGC 2808 (see e.g., Martin et al. \shortcite{martin:04}), 
NGC 5286, Pal. 1, and BH 176 (see e.g., Frinchaboy et al. \shortcite{frinchaboy:04}). None of these 
objects are in the sample of Dinescu et al. \shortcite{dinescu:99} and only NGC 2808 is in the sample
of Salaris \& Weiss \shortcite{salaris:02} (we have already remarked on the young age of this object).
In terms of HB type, NGC 2808 is an old halo cluster with a HB morphology lying on the old/young halo 
dividing line, while NGC 5286 is also an old halo cluster. Pal. 1 and BH 176 are in our bulge/disk 
sub-system. Both these clusters have large Galactocentric radii compared with the rest of the bulge/disk 
ensemble ($R_{\rm{gc}} = 17$ kpc and $9.7$ kpc respectively). In this respect they resemble the other anomalous 
bulge/disk object, Pal. 11, which we discussed above. In the same vein, van den Bergh \& Mackey 
\shortcite{vdb:04} have noted that Pal. 1 lies at an unusually large Galactocentric radius for its 
metallicity, similarly to Pal. 12 and Terzan 7, both of which are associated with the Sagittarius dwarf. 
These authors argue that this suggests Pal. 1 might have a similar evolutionary history to these two objects. 
We also note that Forbes et al. \shortcite{forbes:04} suggest BH 176 cannot be associated with the Canis Major 
dwarf because it does not seem to fit their derived age-metallicity relation for this galaxy.
Nonetheless, it is important to remark that the apparent in-plane accretion of the CMa dwarf might have
contributed clusters not only to the Galactic old halo system but also to the bulge/disk system.

The above results have important implications for studies which use globular clusters to help constrain
the properties and epoch of the formation of the Galaxy. In any such study it is vital to be clear about
the origin of the globular clusters under consideration, otherwise the results may evidently tell 
one more about the formation of a satellite galaxy rather than the Milky Way itself, and lead to 
erroneous conclusions.

\section{Summary}
\label{s:summary}
We have used high quality photometry from WFPC2 on the {\em Hubble Space Telescope} to examine the
horizontal branch morphologies of globular clusters in the LMC, SMC, and Fornax and Sagittarius dwarf
spheroidal galaxies. Adopting a uniform procedure allowed HB indices to be calculated for all the available 
clusters. Supplementing these with a few literature values provided us with measurements for complete samples
of LMC and Fornax globular clusters, and all known SMC and Sagittarius globular clusters. In addition, we
have compiled structural and metallicity measurements for these objects. Examining the HB morphologies
of the external globular clusters reveals them to be rich in RR Lyrae stars, with specific frequencies
much larger than those typical of Galactic globular clusters. Although several of these clusters have
been targetted by variability surveys, it seems likely that they possess a large number of RR Lyrae stars 
which remain to be discovered and characterized.

We have also filled out the database of HB indices, and structural and metallicity measurements provided 
by Harris \shortcite{harris:96} for the Galactic globular clusters, by using a combination of literature 
results and new calculations. Using these new data in combination with recent results from the literature 
concerning the kinematics and ages of Galactic globular clusters (e.g., Dinescu et al. \shortcite{dinescu:99};
Salaris \& Weiss \shortcite{salaris:02}), we have undertaken a detailed examination 
and comparison of the three globular cluster sub-systems defined by Zinn \shortcite{zinn:93a} (his so-called
``old halo,'' ``young halo,'' and ``bulge/disk'' systems). Our main results are as follows:
\begin{enumerate}
\item{The three groups possess very different spatial distributions. The metal-rich bulge/disk clusters 
are, as their monicker suggests, concentrated in the Galactic bulge and central disk region, and form a very
flattened system. The old halo clusters predominantly occupy the inner halo, while the young halo objects
extend to very large Galactocentric radii. Both halo systems extend inwards close to the Galactic centre. We
also note asymmetric distributions in the outer reaches of both halo sub-systems. For the old halo this
takes the form of an excess of clusters with large distances above the Galactic plane, while for the young halo
we observe a clear elongation of cluster positions projected onto the Galactic plane.}
\item{We observe a correlation between Galactocentric radius and cluster metallicity for the old halo clusters,
although their dispersion in $[$Fe$/$H$]$ is large and the correlation is not as strong as that noted by
Zinn \shortcite{zinn:93a} for his sample. The young halo clusters show no such trend.}
\item{The young halo clusters have large values of RR Lyrae specific frequency. In this they strongly
resemble the external globular clusters. In contrast the old halo and bulge/disk clusters generally
have small specific frequencies. This result is possibly related to cluster HB morphology -- by definition
the young halo clusters have red HB morphologies at given $[$Fe$/$H$]$.}
\item{As their name suggests, the young halo clusters are on average considerably younger than those in
the old halo sample (Fig. \ref{f:agehist}). The mean offset of young halo clusters is $\Delta$Age\ $= -1.9$ 
Gyr compared with M92. The old halo objects have a narrow peak in age at $\Delta$Age\ $= -0.4$ Gyr, plus a 
tail of $\sim 6$ clusters extending to much younger ages. The bulge/disk clusters exhibit a narrow peak in 
age at $\Delta$Age\ $\sim -2$ Gyr.}
\item{We do not observe any strong correlation between age and Galactocentric radius for any of the samples. 
There is however a hint that the innermost old halo objects may be slightly younger than the main body of 
this ensemble, with ages closer to the bulge/disk clusters. On a plot of $\Delta$Age vs. $[$Fe$/$H$]$,
the old halo clusters show a clear trend of higher abundances at younger ages. The bulge/disk clusters join
cleanly at the young end of this relationship. The young halo clusters, although possessing a large dispersion
in age, apparently show a considerably steeper age-metallicity relationship. The two Sagittarius clusters
Terzan 7 and Pal. 12 extend this trend to very young ages.}
\item{Examination of cluster kinematics reveals some of the bulge/disk objects to (unsurprisingly) 
possess disk-like orbits. The young halo clusters are characterized by large, energetic orbits of high 
eccentricity and intermediate inclination, which cover a very wide spread in orbital angular momentum,
including large retrograde motions. Some old halo clusters share similar characteristics, but most have
much smaller total energies and eccentricities. Dinescu et al \shortcite{dinescu:99,dinescu:03} have also
identified a number of old halo clusters with disk-like kinematics.}
\end{enumerate}

Taken together these results are similar to those presented by Zinn \shortcite{zinn:93a,zinn:96}; however
the new kinematic and age data extend the distinctions he observed between the Galactic cluster sub-systems.
These results are fully compatible with Zinn's \shortcite{zinn:93a,zinn:96} hypothesis that the majority
of the old halo and bulge/disk clusters are Galactic natives which formed in a dissipative collapse
like that originally envisioned by Eggen, Lynden-Bell \& Sandage \shortcite{eggen:62}, while the 
young halo clusters plus a few of the old halo objects have been accreted by the Galaxy through merger 
events such as that now observed for the Sagittarius dwarf galaxy (as originally proposed by
Searle \& Zinn \shortcite{searle:78}).

Using our new measurements for the external globular clusters, we have further tested this hypothesis 
via a detailed comparison between the HB morphologies and structural properties of these objects and those 
found in the Galactic sub-systems. Specifically, we imagined a scenario where the LMC, SMC, and Fornax 
and Sagittarius dSph galaxies have merged with the Galaxy, and investigated what sort of external clusters 
would subsequently inhabit the Galactic halo. The main results of our comparison are as follows:
\begin{enumerate}
\item{In terms of a plot of metallicity as a function of HB type, the majority of the external clusters
are essentially indistinguishable from those objects belonging to the Galactic young halo sub-system, 
but we also find a number of external clusters that have HB morphologies characteristic of old halo clusters. 
The fraction of old-halo type clusters in the external systems is $\sim 20-30$ per cent.}
\item{Examining the structures of clusters in the external systems and the Galactic sub-systems,
we find that the young halo, old halo and bulge/disk ensembles have very distinct distributions of cluster
core radii. The bulge/disk and old halo almost exclusively possess compact clusters, although the old halo
has a small tail to large core radii. In contrast, the young halo possesses a large fraction of very 
expanded clusters. The distribution of core radii for the external globular clusters matches the young
halo distribution very closely. The only major difference occurs for extremely distended objects, which
are not found in the external systems. These few young halo clusters (such as Pal. 5) are possibly in the
midst of final tidal disruption.}
\item{The old halo distribution of cluster core radii falls between that for bulge/disk sample 
and the external sample. We demonstrate that the old halo distribution can be well represented by a 
composite distribution formed by $\sim 83-85$ per cent of objects with structures typical of bulge/disk
clusters, and $\sim 15-17$ per cent of objects with structures typical of external clusters.} 
\end{enumerate}

Taken together these results fully support the accretion hypothesis outlined above. We conclude that
the $30$ young halo clusters and $15-17$ per cent of the old halo clusters ($10-12$ objects) are of 
extra-Galactic origin. The implied fraction of old-halo type accreted clusters ($\sim 25$ per cent
of accreted clusters) matches well the fraction of old-halo type clusters observed in the four external
galaxies ($20-30$ per cent). Based on the number of globular clusters counted in the Fornax and Sagittarius 
dwarf spheroidal galaxies, we estimate that the Milky Way may have experienced $\sim 7$ merger
events with cluster-bearing dwarf-spheroidal-type galaxies during its lifetime, building up $\sim 44-50$ 
per cent of the mass of the stellar halo. These mergers must have occurred at early epochs to ensure the
observed low mean abundance of field halo stars (see e.g., Da Costa \shortcite{dacosta:03}). 
We argue that the observed differences 
in structure between Galactic native clusters and those of external origin are due to destruction effects 
of the type outlined by Gnedin \& Ostriker \shortcite{gnedin:97}. In the future, large $N$-body simulations 
of the disruption of clusters in the Galactic tidal field will likely be able to place constraints on how 
long ago these mergers occurred. Examination of the present day
distribution of cluster structures implies that up to $\sim 50$ per cent of the original Galactic
population of globular clusters ($\sim 100$ clusters) may have been destroyed since the formation of the 
Galaxy. Such destruction would only contribute $\sim 2$ per cent of the mass of the stellar halo.

Finally, we try and locate old halo clusters which have properties characteristic
of accreted clusters. NGC 2419, 5897, 6101, 7492, M55, and Pal. 15 have very extended structures more typical
of young halo than old halo clusters, while NGC 5024, 5904, 6205 and 7089 have very energetic orbits,
with large eccentricities, which closely resemble typical young halo cluster orbits. Several of the
clusters associated with the proposed disrupted dwarf galaxy in Canis Major \cite{martin:04} also exhibit
kinematics and ages consistent with accretion. For example, NGC 1851, 1904, and 2298 have young-halo-like
orbits, while NGC 1851 and 2808 have anomalously young ages for old halo clusters. It is possible that
the in-plane accretion of the CMa dwarf has also contributed clusters to the Galactic disk system. The
identification of such objects has important implications for studies concerning the formation of the
Galaxy, since it is vital in such work to be clear about the origin of any clusters under consideration.

\section*{Acknowledgements}
This paper is based on observations made with the NASA/ESA 
{\em Hubble Space Telescope}, obtained from the data archive at the Space
Telescope Science Institute, which is operated by the Association
of Universities for Research in Astronomy, Inc. under NASA
contract NAS 5-26555. ADM is grateful for financial support from PPARC
in the form of a Postdoctoral Fellowship.



\bsp 

\label{lastpage}

\end{document}